# EXPLORING THE NRO OPPORTUNITY FOR A HUBBLE-SIZED WIDE-FIELD NEAR-IR SPACE TELESCOPE — *NEW WFIRST*

Alan Dressler[1], David Spergel[2], Matt Mountain[3], Marc Postman[3], Erin Elliott[3], Eduardo Bendek[4], David Bennett[5], Julianne Dalcanton[6], Scott Gaudi[7], Neil Gehrels[8], Olivier Guyon[4], Christopher Hirata[9], Jason Kalirai[3], N. Jeremy Kasdin[2], Jeff Kruk[8], Bruce Macintosh[10], Sangeeta Malhotra[11], Matthew Penny[7], Saul Perlmutter[10], George Rieke[4], Adam Riess[3], James Rhoads[11], Stuart Shaklan[12], Rachel Somerville[13], Daniel Stern[12], Rodger Thompson[4], and David Weinberg[7]

[1]The Observatories of the Carnegie Institution for Science
[2]Princeton University
[3]Space Telescope Science Institute
[4]University of Arizona
[5]University of Notre Dame
[6]University of Washington
[7]Ohio State University
[8]NASA: Goddard Space Flight Center
[9]California Institute of Technology
[10]Lawrence Livermore National Laboratories
[11]Arizona State University
[12]Jet Propulsion Laboratory, California Institute of Technology
[13]Rutgers University

## ABSTRACT

We discuss scientific, technical, and programmatic issues related to the use of an NRO 2.4m telescope for the WFIRST initiative of the 2010 Decadal Survey. We show that this implementation of WFIRST, which we call "NEW WFIRST," would achieve the goals of the *NWNH* Decadal Survey for the WFIRST core programs of Dark Energy and Microlensing Planet Finding, with the crucial benefit of deeper and/or wider near-IR surveys for GO science and a potentially Hubble-like Guest Observer program. NEW WFIRST could also include a coronagraphic imager for direct detection of dust disks and planets around neighboring stars, a high-priority science and technology precursor for future ambitious programs to image Earth-like planets around neighboring stars.

# 1. Introduction

NASA's acquisition of the two complete 2.4m aperture telescopes from the National Reconnaissance Organization (NRO) of the US Department of Defense is a significant and unexpected opportunity for the US scientific community. These are Hubble-sized telescopes that are judged suitable for astronomical science, telescopes which are in particular capable of delivering a field much larger than Hubble's, with optical systems that appear to be compatible with a large-area near-infrared camera. Some basic questions arise from this opportunity. Is an NRO telescope well suited to accomplish the science of the WFIRST, the top priority space initiative in the 2010 Astrophysics Decadal Survey *New Worlds, New Horizons* (*NWNH* — [http://sites.nationalacademies.org/BPA/BPA_048094](http://sites.nationalacademies.org/BPA/BPA_048094))? How does the larger aperture of an NRO telescope improve or limit the science program developed by the WFIRST Science Definition Team (SDT) or their Design Reference Mission 1 (DRM1), which is the descendant of the notional WFIRST telescope. Would the larger aperture allow for, and enable, additional science not included in the WFIRST program but high priority in *NWNH*, thus extending the key idea of WFIRST to cover the broadest possible program of high-priority science? Finally, would the availability of a nearly-completed 2.4m telescope improve or harm the chance of accomplishing the WFIRST mission in terms of cost and schedule?

The purpose of this paper is to provide preliminary answers to these and derivative questions. In particular, we attempt to evaluate the answers in the context of an NRO telescope without any significant alternations – "as is." With very limited resources and time, the answers will range from simple matters that can be easily assessed to complicated issues that will require months of study by teams of scientists and engineers. Our work here is to make a start to this process that will provide some justification for a much more detailed study. The goal of this white paper has been to provide information and talking points for the September, 2012 Princeton *NEW Telescope Meeting* and also as input to the new Science Definition Team assembled by NASA to "assess the possible scientific use of one set of the ex-NRO telescope assets for advancing the science priorities of the 2010 Astrophysics Decadal Survey."  The 'NEW' acronym stands for "New Worlds New Horizons Enabling Wide Field," so for the purposes of this paper we will call an NRO telescope configured to carry out the WFIRST program NEW WFIRST.

1.1 Context

*NWNH* was a multi-year effort involving many hundreds of astronomers and astrophysicists who prioritized the most compelling scientific opportunities in the 2011-2020 decade. The Survey considered a long and diverse list of potential programs and faced the usual challenge of matching a few of the best to the funding levels anticipated over this decade. Arguably, this Survey had both more compelling science programs and fewer potential resources than any previous astronomy survey.

In the field of space astronomy, this dilemma was exacerbated by the significant challenge of bringing the James Webb Space Telescope, the top priority of the previous survey, to a successful launch and operational phase; it was known from early in the Survey that most of the funding NASA would have applied to new missions in the 2011-2020 decade was already committed to finishing the Webb telescope. This meant that few of many excellent missions — and none of the most exciting "flagships'" — could be accommodated within the projected budget.

*NWNH* responded to this dilemma with a financially modest, but scientifically compelling, program for space: increase the frequency of the less-costly Explorer missions, and build a "junior flagship" — a 1.5m telescope capable of wide-field-imaging and spectroscopy, named WFIRST, that was an amalgamation of several diverse projects that shared common hardware. By combining two core programs — (1) dark energy surveys, and (2) microlensing planet finding — with other broad-purpose astrophysics surveys and a program of guest-investigator science, WFIRST was intended to accomplish as much of the high priority science as possible within the tightly constrained projected budget.

*NWNH* envisioned WFIRST as a flagship mission with broad scientific reach, and certainly very different than a dedicated "Dark Energy Probe." Like LSST, the top large ground-based *NWNH* priority, WFIRST is capable of producing outstanding science across a wide swath of astronomy. However, WFIRST is not a flagship mission in the JWST sense. WFIRST is a much smaller scale mission that does not require the development of novel technologies — as JWST has. WFIRST should be in comparison to JWST simple to manage and not susceptible to a large percentage cost growth.

Soon after the release of the *NWNH* Survey Report, the cost of completing JWST construction was determined to be even larger than in the projected numbers. In addition, the Administration made less money available to Astrophysics (by approximately $1.3B over five years) than had been assumed in formulating the *NWNH* plan. The accomplishment of *any* of the *NWNH* goals for space science has been put in doubt. However, there are positive developments since the Survey that raise hope. (1) The planned cadence of the Explorer program has been increased. (2) Participation by US astronomers in European Space Agency's *Euclid* mission has been secured, in this way providing some participation in the dark energy research that ranked very highly in the Survey. (3) NASA constituted a WFIRST Science Definition Team to study, to evolve, and to validate the cost of, the proposed program. The SDT refined and improved the WFIRST design and

confirmed the cost of the improved version within that derived in the *NWNH* "CATE" (cost-and-technology-estimate).

## 1.2 Why WFIRST?

The unexpected availability of an essentially complete telescope is a new wrinkle in the process NASA has employed to underwrite decadal reviews of astronomy and astrophysics. These reviews inform the process of formulating and carrying out missions that pursue the scientific priorities of the scientific community. There are, of course, many channels by which the community communicates its scientific aspirations to NASA, including other advisory committees and responses to NASA's own process of selection through competitive proposals, but the National Research Council's Decadal Survey activity is the only one that involves the entire astronomy/astrophysics community. This essential feature, combined with the time, depth, and rigor applied in Decadal Surveys, is why they have a unique weight, and why NASA treats them as a roadmap of scientific aspirations to be folded in with other budgetary and programmatic considerations. For this reason, it would indeed be fortunate if an NRO telescope were to be well suited to carry out the WFIRST mission — the top priority of the most recent Survey.

Conversely, an implementation path other than using an NRO telescope to accomplish the *NWNH's* top priority, whether judged by the CAA, a mid-decadal review, the next Decadal survey, or other deliberative process, would face complex questions and Hobson's choices that are likely to divide the community and potentially derail the Decadal process altogether. Another issue is essentially the "shelf life" of the NRO telescopes, a time beyond which components, systems, and technical expertise cannot be supported. For these reasons, we believe that the application of an NRO telescope to the WFIRST mission is the only attractive option at this time. However, because a forced fit is worse than any other option, the suitability of an NRO telescope for WFIRST science is the key issue that must be settled before any decision about the next step for NASA astrophysics after the launch of JWST can be taken.

## 1.3 The Landscape of Astronomical Research in 2020

Looking across the future landscape of current and future major facilities for astrophysics now anticipated in a post NWNH world, it is instructive to see how a NEW WFIRST would fit into the program (see Figure 1).

For general astrophysics, NEW WFIRST can play an essential role in complementing future ground-based 20m — 40m telescopes. For example, as Table 1 shows, based on the performance of HST + WFC3, a 2.4m space telescope equipped with forefront IR detectors can achieve broad-band point-source sensitivity of ground-based 'Extremely Large Telescopes' (ELTs), thus allowing these 20m – 40m telescopes to devote more of their valuable resources to their unique capabilities in multi-object spectroscopy in the optical and near-infrared.

As was articulated in *NWNH*, WFIRST was to be much more than a Dark Energy mission or an exoplanet mission. Indeed, these exciting core programs were only a "nose cone" of the WFIRST mission, leading the way for a diverse and powerful program of astrophysical research. With an unprecedentedly large field and high sensitivity for near-IR observations, a WFIRST mission enables forefront astrophysics, in the context of current, future ground and space missions. For this primary goal, NEW WFIRST could do even more. Furthermore, NEW WFIRST could also serve as a bridge to ambitious future exoplanet missions, such as the potentially "life-finding" TPF-C — a fantastic opportunity all by itself, one very highly regarded in NWNH.

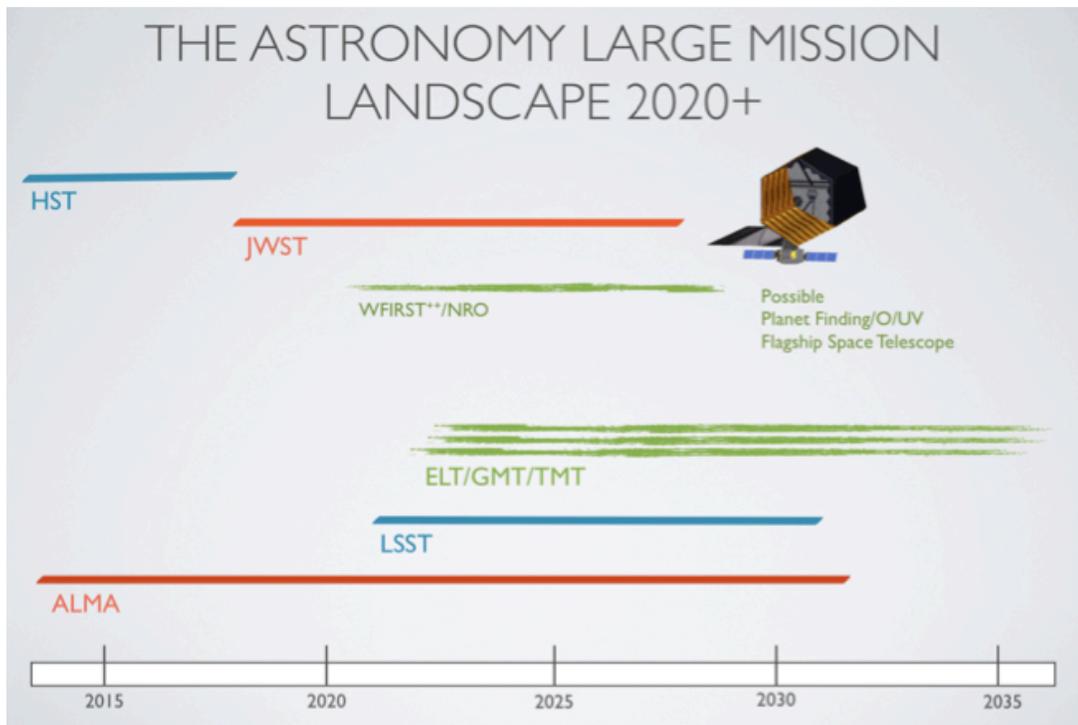

*Figure 1: Mission timelines for facilities especially relevant to NEW WFIRST.*

If launched soon after JWST, NEW WFIRST would significantly enhance the science return from JWST. With an imaging field of view at least 100 times the area of JWST's NIRCam, NEW WFIRST will discover numerous gravitationally-lensed galaxies, high redshift quasars, and other rare objects that will be superb targets for JWST follow-up. If operating concurrently with JWST, WFIRST will relieve JWST from some very expensive survey work that will be required its maturing mission to discover and study the first stars and galaxies, by performing surveys over far wider areas than JWST could ever accomplish. NEW WFIRST surveys could provide any number of valuable source catalogs, not only for JWST, but also for the Atacama Large Millimeter Array (ALMA) and future ELTs. In addition, such a space-based survey capability would provide a highly stable, infrared complement to *NWNH's* top priority ground-based initiative, the Large Synoptic Survey Telescope (LSST).

Table 1: Point-source Sensitivity

Sensitivity: HST and NRO-WFIRST Compared with Ground-based 30m-class Telescopes:
Imaging On-Source Exposure Time Required to Reach S/N=5 (hours)

|  |  | J=25.2 | H=24.5 | K=23.2 |
|---|---|---|---|---|
| **HST** |  | **F110W** | **F160/165W** | **F205W** |
| WFC3/IR | 123" x 136" | 0.15 | 0.30 | - |
| **30m ELT with AO** |  | **J** | **H** | **Ks** |
| assuming MCAO | 90" x 90" | ~0.1 | ~0.1 | ~0.01 |
|  |  | **F110W** | **F160/165W** |  |
| **NRO-WFIRST - estimated** | 3 x (1270" x 1270") | ~0.15 | ~0.3 |  |

*Notes:* Input SED: M0V-star, Vega magnitudes
HST/WFC3: 1 orbit = 2400 seconds; average zodiacal and earthshine background
HST/WFC3 and Gemini & VLT with AO: aperture radius = 0.3"
**30m assumes same Strehl performance as Gemini and VLT** -- in the background limited regime
30m: airmass = 1.2, seeing = best 70%-ile [J=0.66", H=0.64", K=0.61" without AO]

The burgeoning field of exoplanets, a high priority of *NWNH* for which no major program could be afforded, could be invigorated by NEW WFIRST. The increased light grasp and angular resolution of a 2.4m mirror would considerably strengthen the microlensing planet search compared to the original WFIRST mission (see Figure 2), as we detail in Section 5.3. More intriguingly, the combination of excellent on-axis performance and the large collecting area of the 2.4m NRO telescope would enable a coronagraphic imager that could answer crucial questions about proto-planetary disks and explore for the first time the habitable zones of about 30 nearby GFK stars. As described below, this would directly address one of the key exoplanet goals of *NWHH* not possible with the original WFIRST concept — direct imaging of exoplanets. This would as well complement new ground-based planet-imaging systems being readied for deployment on existing 8m – 10m telescopes to search for young planets several AU from the parent star. A coronagraphic imager on NEW WFIRST would importantly provide an essential technology pathfinder for future large space telescopes that can search for and characterize Earth-like worlds, including the search for life.

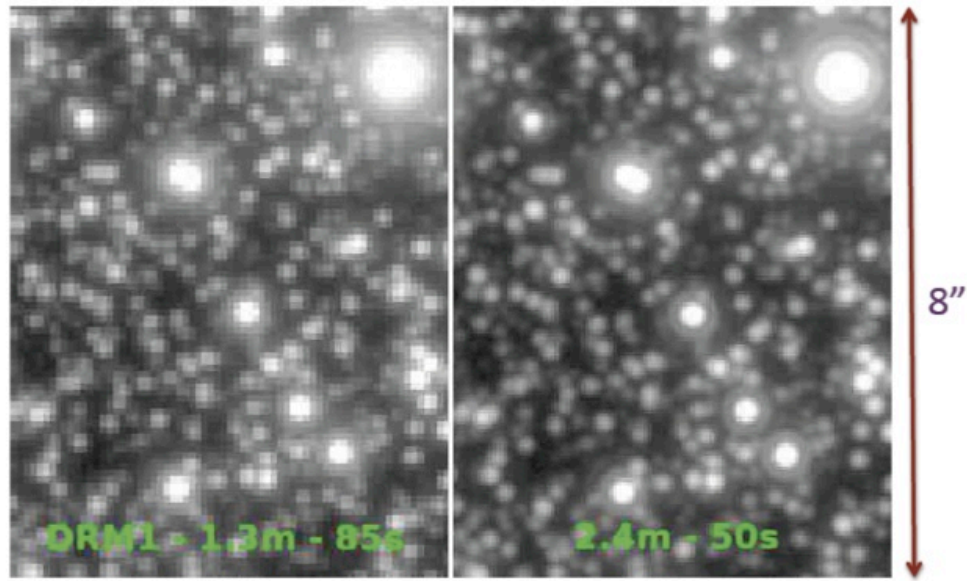

*Figure 2: The improved micro-lensing performance of a 2.4m NRO telescope (right) compared to SDT WFIRST (DRM1, left). The images compare an "equal-duration, equal-area" survey of a Galactic Bulge field.*

## 2. The NRO Telescopes – a Basic Description

NRO transferred ownership of two telescopes to NASA. The delivery includes 2 telescopes, each with the primary and secondary mirrors of a 3-mirror anastigmat design, associated support structures, outer thermal baffles, and various spares. A schematic of a telescope system that is similar to the delivered telescopes is shown in Figure 3. The outer thermal baffle (not shown) mounts to a separate interface and encloses the telescope assembly. The open volume — behind the primary mirror support structure to the base the spacecraft struts — can be used for instruments and supporting electronics.

The two-mirror telescope system has a 2.4m aperture and is F/8. The system's total wavefront error is less than 60 nm for a near-axis field point, supporting diffraction-limited imaging above about 840 nm. The telescope is designed to operate at room temperature, and the telescope structures use composites with low coefficients of thermal expansion (CTE) and invar materials.

The primary is made of ULE and has been lightweighted using an Exelis technology. The system mass is approximately 1200 kg, including the Outer Baffle Assembly (OBA); this is much less massive than the Hubble Space Telescope (HST). The secondary mirror (without any baffle) obscures less than 20% of the primary aperture. The secondary mirror is actuated.

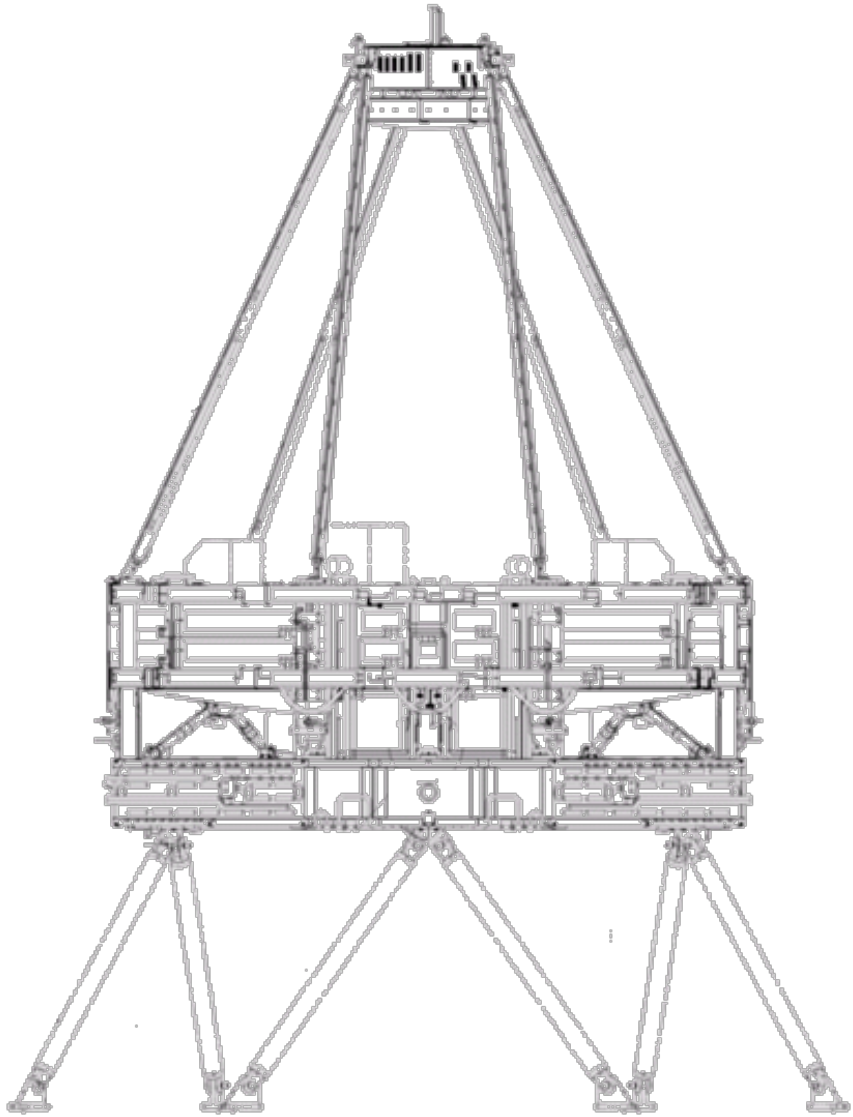

*Figure 3: A schematic of one of the delivered telescope assemblies.*

More information about the telescope, for example, the wavelength range and mirror coating material, and the operating temperature, will be shared as it becomes available.

2.1 Using the NRO telescope "as is"

In this white paper we investigate the application of an NRO telescope "as is" to the WFIRST mission. Here "as is" means no refiguring of the primary or secondary mirror, no mechanical modifications to the telescope structure, including mirror cells, actuators, thermal heaters and sensors, etc. At the Princeton Workshop we heard about the probable need to extend the OBA in order to provide a better light baffle for the telescope, and of course, there is no spacecraft supplied, so some modifications of structure might be necessary to accommodate its design. We imagine such modifications to have a very low-risk and, hence, they would not violate the "as is" principle.

The question of operating temperature, which will impact the scientific programs to be undertaken with NEW WFIRST, is more complicated, because the "as is" operating temperature is unknown. An upper bound is the demonstrated temperature of 293K where, assuming detectors cooled to 100K, the performance of NEW WFIRST in photometric bands Y, J, H is essentially uncompromised at a depth of $\approx$26 AB (5$\sigma$ point source detection) in an exposure of $\approx$ 800 s. This is about 2.5 times faster to the same depth as the Design Reference Mission 1 (DRM1) concept developed by the WFIRST SDT, discussed in Section 3. In contrast, Ks (long-wavelength cutoff TBD) observations to equivalent depth would be prohibitively long, due to the higher thermal background of a "room temperature" telescope. However, even in this case, with a limiting magnitude of Ks $\approx$ 24.6 AB in an 800 s exposure, the high image quality and wide field would make NEW WFIRST uniquely capable for an enormous range of science programs requiring Ks observations.

Performance in Ks would improve rapidly as the operating temperature is reduced from 293K: if lowered to the designed 'survival temperature' of the NRO telescope of 250K, the performance of NEW WFIRST would match the high sensitivity of the SDT DRM1, Ks $\approx$ 26 AB in an exposure of $\approx$ 2400 s, a time comparable to the DRM1 concept. However, it is not clear whether an NRO telescope could be operated at 250K without substantial compromise in optical performance or even increased risk to the overall system. For example, the degradation of the wavefront from stresses induced in the ULE primary mirror at colder temperatures is unknown. Further modeling could be done to determine the temperature at which such degradation would become unacceptable.[1] An example of increasing risk could be the integrity of adhesive bonds. Such factors will set a lower limit — as yet undetermined — for the "as is" operating temperature.

In summary, at this time we know only that the "as is" operating temperature of NEW WFIRST begins at 293K and extends to some undetermined lower temperature. If the science case for reaching DRM1-or-better sensitivity in Ks is judged to be compelling, it will become important, during this next year, to take the simplest, first steps to learn whether the NRO telescope could be operated colder

---

[1] We note that such degradation in image quality could be negligible at a wavelength $\lambda \approx$ 2$\mu$m, where the NEW WFIRST PSF would be sampled, even if the NRO specification for diffraction-limited performance in the visible had been seriously compromised.

than 293K — "as is."

## 3. The WFIRST Mission: Decadal Concept and SDT Evolution

3.1 The *New Worlds, New Horizons* concept for WFIRST

The Electromagnetic Observations from Space (EOS) panel of *NWNH* recognized that several of the more-than-fifty proposed 'activities' employed similar hardware, basically a space telescope of approximately 1.5m aperture with a field covering a good-fraction-of-a-degree, sampled by hundreds of millions of detector pixels sensitive to near-IR wavelengths. The EOS proposed, and the Decadal Survey Committee accepted, the idea of combining these activities to take advantage of this commonality, that is, to conduct a broad program of survey and targeted science with such a telescope. The result was WFIRST.

It is important to emphasize that it is the wide-field camera + detectors, and not the aperture size, that defines the WFIRST hardware. In particular, for the proposed activities of dark energy surveys, microlensing planet finding, and other near-infrared-sky surveys, the choice of a 1.5m aperture reflected the fact that these science goals could be achieved with a Kepler-sized telescope, a highly desirable feature in a constrained budget environment. However, all of these programs, and **especially** guest-investigator programs of targeted science, judged by *NWNH* as the **essential** component of the WFIRST mission, would benefit from a larger-aperture telescope. The defining feature of the WFIRST mission is a wide-field, near-IR camera, sampled at near the diffraction limit, as the scientific as well as literal focus of the optical system. The choice of 1.5m — the minimum required for these science programs — made sense in the context of *NWNH* because the construction of a more-capable, larger-aperture telescope would undoubtedly carry a much greater cost — a non-starter in the present budget climate. However, the availability — without cost to NASA — of a nearly-completed 2.4m telescope to carry out the same science programs (and likely more) could conceivably balance the augmented cost of other systems — for example, the larger spacecraft and the bigger camera – required for the larger telescope. If properly managed, this could result in a total investment in NEW WFIRST no larger than the $1.6B planned for the WFIRST in *NWNH.*

3.2 Evolution of the WFIRST concept

The WFIRST Science Definition Team (SDT) assembled by NASA, working with the Project team at GSFC and JPL, has submitted a report of its two-year-long study to better define the WFIRST mission and its hardware, and in particular to look for improvements, simplifications, and cost-and-schedule savings. The SDT membership and reports can be found at http://arXiv.org/pdf/1208.4012.pdf . This study was required to put to the test the notion in the EOS Panel Report that several science missions could be combined and carried out successfully. The JDEM-Omega space telescope was chosen as representative hardware to obtain a first-cut, reliable costing for WFIRST, a *NWNH* requirement for all of its recommended programs. However, this left an important opportunity for the SDT and WFIRST Project (GSFC) to explore the parameters of the hardware to see if, in greater detail and with greater certainty, these diverse scientific missions could all be supported with the common hardware, and how that hardware could be optimized for the task.

3.3 Comparing the SDT's DRM1 and DRM2

Two mission concepts, so-called Design Reference Missions (DRM), were considered by the SDT and Project. DRM1 is close in scale to JDEM-Omega, but included several innovations to improve performance. The intent was to meet all WFIRST objectives of the Decadal Survey. These include: (1) measurements of the acceleration of the expansion of the universe using the techniques of supernova Ia standard candles, weak lensing and baryon acoustic oscillations; (2) a census of exoplanets using the microlensing technique; and (3) near-IR surveys and targeted observations for the astronomical community. DRM2 is a smaller observatory designed to be less expensive, fitting approximately into the NASA $1B Probe class of missions. It has the capability to perform all observations required of WFIRST, but only in an extended mission lifetime.

DRM1 incorporates a 205K, 1.3m, unobstructed, three-mirror anastigmat telescope. It utilizes current-technology infrared H2RG sensors operating in the $\lambda$ = 0.7–2.4$\mu$m range, and fits within the fairing of a medium-lift-class launch vehicle. It has a single instrument with 36 H2RG detectors and has a field of view of 0.375 deg$^2$. There are two filter wheels in the beam with 6 filters, an R=75 grism for supernova spectroscopy, and an R = 600 prism for the galaxy redshift survey. DRM1 assumes that WFIRST will be deployed in an Earth-Sun L2 libration point orbit and have a minimum operational lifetime of 5 years. However, DRM1 will have no design elements (e.g., propellant supply) that intrinsically limit the lifetime to fewer than 10 years.

DRM1 changes relative to JDEM-Omega are as follows:

1) The telescope has an unobstructed design with the secondary mirror off-axis with respect to the primary mirror. This provides greater effective area for an equal-sized telescope, stiffer support of the secondary, and a cleaner point spread function due to the unobscured beam. The 1.3m DRM1 has nearly the same collecting area as the 1.5m obstructed JDEM-Omega, which had an unusually large central obscuration. Unobstructed telescopes of size similar to WFIRST have been flown in space and now have adequate technological heritage to be incorporated in the design.

2) The red wavelength cutoff of the observatory was increased from 2.0 to 2.4µm. This gives unique wavelength coverage compared to Euclid's stated red-wavelength cutoff of 2.0µm. Extending the operating wavelength increases the redshift limit of galaxy surveys to $z \approx 2.7$ (using the H$\alpha$ 656nm line), enabling in particular a better sampling of the history of the acceleration of the expansion of the universe and more complete coverage of the history of star formation in galaxies.

3) The detectors comprise a single instrument channel in the focal plane, compared with the three instrument channels in JDEM-Omega. For JDEM-Omega, two separate spectroscopy channels were incorporated for simultaneous imaging and spectroscopy surveys. For DRM1, all detectors are used in a single imaging channel, but with removable grisms on dedicated grism wheels for spectroscopy. This gives more flexibility in tailoring the observing program.

DRM2 was requested by NASA in response to the ESA's selection of the *Euclid* dark energy mission, which is scheduled to launch in 2020. DRM2 was designed to be non-duplicative with the capabilities of the Euclid mission and/or planned ground-based facilities such as LSST, and also to represent a lower cost alternative to DRM1, though likely with a reduced science return compared to DRM1. A recent committee of the National Research Council examined the potential overlap in the WFIRST and Euclid missions and found that there was limited duplication in their science capabilities ("Assessment of a Plan for U.S. Participation in Euclid" — http://www.nap.edu/catalog.php?record_id=13357 ). However, to reduce mission costs, DRM2 employs a 1.1m telescope (also a three-mirror anastigmat), utilizes a smaller number (14) of higher performance H4RG-10 infrared sensors operating from 0.7–2.4µm, uses an observatory design with only selective redundancy, and fits within a lower-cost, lower-throw-weight launch vehicle.

The field of view is 0.56 deg$^2$, larger than DRM1 due to the larger number of pixels with the H4RGs (240 megapixels for DRM2 compared to 144 megapixels for DRM1). The H4RG-10 detectors require further development to become space-qualified, but have a substantial advantage of 4 times as many pixels in a similar sized package. The orbit of DRM2 is the Earth-Sun L2 libration point, and the operational lifetime is 3 years. With the improved sensors, DRM2 has similar sensitivity to DRM1 for a fixed observation time, and can achieve many — but not all — of the Decadal Survey science objectives within its 3-year design lifetime. It is assumed that the actual observing plan for a WFIRST utilizing the DRM2 mission concept will be optimized based on the data available at that time.  All Decadal Survey objectives could be covered in an extended mission, although the time for guest investigator science would likely be significantly reduced compared to DRM1. Both DRM1 and DRM2 represent viable mission concepts that address the fundamental goals of the Decadal Survey recommendations.  DRM1 is more capable; DRM2 is less expensive.   The cost for DRM1 is estimated to be $1.6B, consistent with the Decadal Survey cost estimate for JDEM-Omega.  The cost for DRM2 is estimated to be ~1.0B.  The SDT report has a strong recommendation for near-term NASA investment to further develop H4RG detectors as a cost-effective way to improve the science return of any WFIRST design.

## 4. Feasibility study of an optic design imaging 0.375 deg$^2$

An attractive alternative to the SDT's DRM1 and DRM2 designs are the application of an NRO telescope to the science goals of WFIRST. The design of these telescopes provides a native wide-angle field-of-view, and in its current configuration could support a wide field imaging camera of at least 0.375 deg$^2$, the planned field-of-view (FOV) of DRM1.  As a proof-of-concept to using an NRO telescope "as is," we offer a wide-field instrument that provides a FOV of 0.375 deg$^2$, sketched in Figures 4 and 5.  The optical parameters of the unaltered two-mirror telescope were used, but the mirror surfaces were assumed to have zero wavefront error.

The instrument is designed to fit in the space behind the primary mirror and consists of two fold mirrors followed by tertiary and quaternary mirrors.  The tertiary and quaternary mirrors are both conic with an aspheric term that corrects coma. The respective sizes of the two mirrors are 0.58 x 0.38 meters and 0.4 x 0.26 meters.

An aspheric plate that corrects spherical is located near the pupil of the system. (Because it is thin and unpowered, it does not cause significant chromatic aberrations.)  The pupil is real and accessible to allow a good position for filters; two filter wheels can be accommodated. This design uses filters that are 5.4 inches in diameter.  A grism for R ≈ 600 spectroscopy, the dispersion chosen in DRM1 for the BAO redshift survey, is carried in the filter wheel.  A spectrum covering a wavelength range of 1.3–2.0μm (for an Hα redshift survey 1 < z < 2, see Section 5.2) extends about 3.3 mm at the detector, or ~300 pixels of the 4k array; each pixel covers about 0.0023μm in wavelength, or about 430 km s$^{-1}$.

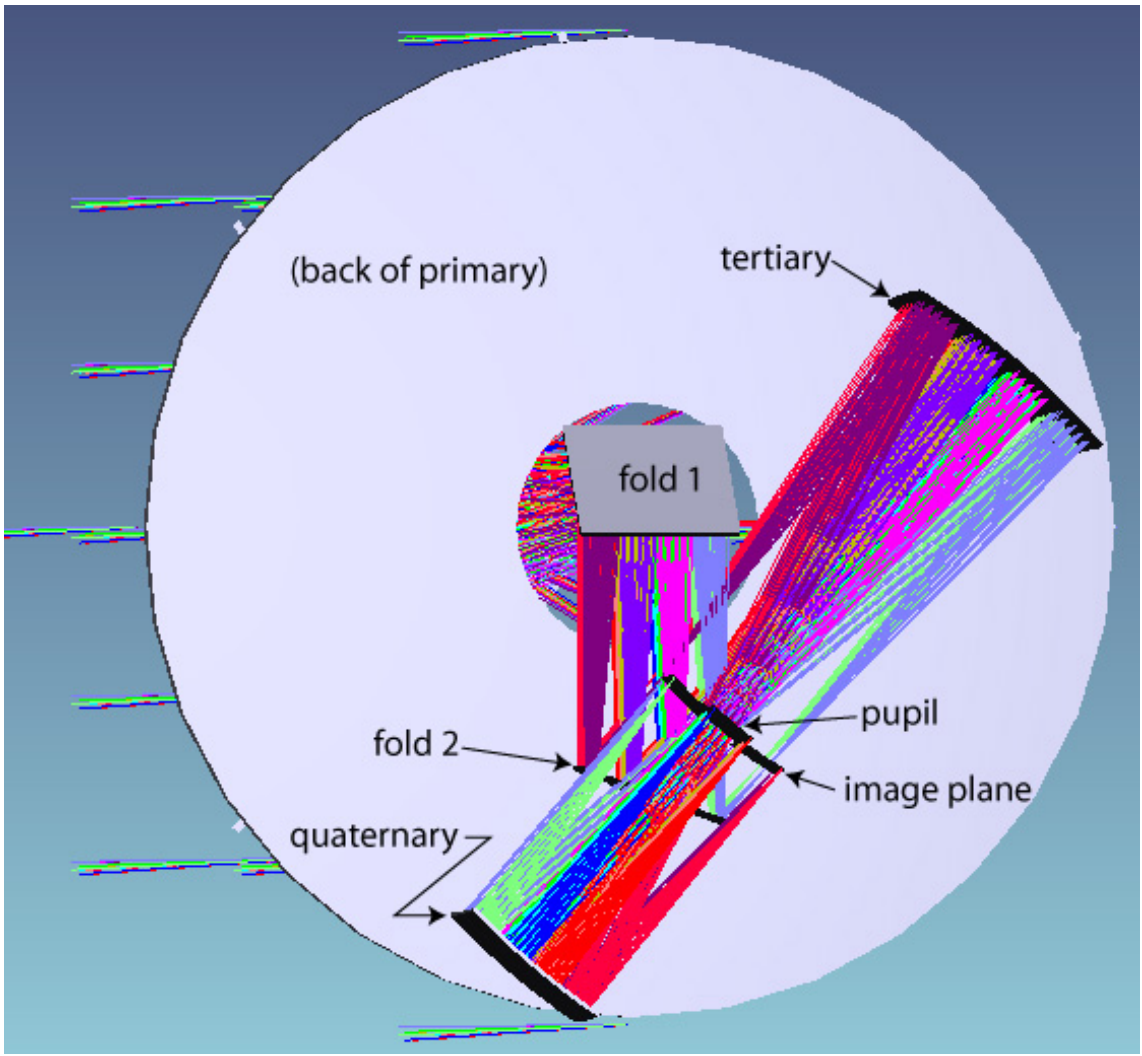

*Figure 4: An example of a possible wide-field instrument that can be added to the unaltered NRO telescope. The system has a total field of view of more than 0.375 deg$^2$, and fits into the accessible space behind the primary mirror. (Each ray bundle originates from a different field point in the system.)*

A sketch of the system's RMS wavefront error as a function of angle on the sky is shown in Figure 6. The RMS wavefront error is less than 50 nm over most of the FOV, allowing for diffraction-limited imaging (and thus a stable PSF) for wavelengths above approximately 0.7μm. The regions toward the bottom of the plot with very low wavefront error (toward the center field) are inaccessible due to the off-axis nature of this design; reflected beams need room to pass by the other optics in the system as shown in Figure 3.

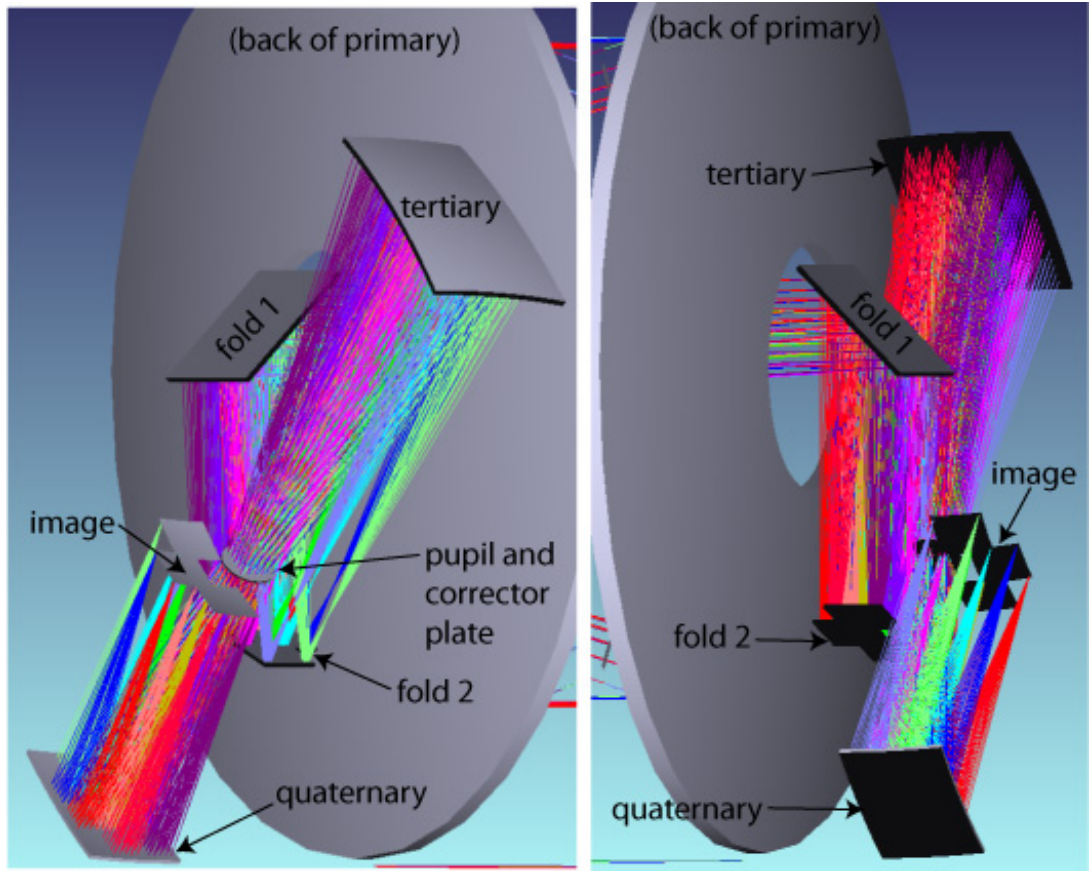

*Figure 5: Additional views of the wide-field instrument. The beam path order is fold 1, fold 2, tertiary, pupil, quaternary, and image plane. The field angles were chosen so that beams have adequate clearance from the edges of the other optics in the system.*

The image plane layout is designed in three squares, as shown on the sky in Figure 6 (superposed on the map of wavefront error). The configuration allows for easy tiling on the sky for observations that require large-scale surveys. Each of the three squares contains a 3x3 array of detectors; each detector is 4k x 4k pixels, with each pixel covering 0.11 arcsec on the sky and 10μm on the image plane.

This is a proof-of-concept system only, and as such, issues with the design remain. The issues are engineering challenges, though, not fundamental limitations. Among the issues:
- RMS wavefront error could be improved, to make room for alignment and surface figure errors.
- The instrument has uncorrected distortion.
- Additional clearance is needed between the beam and fold mirror 2.
- There is inadequate clearance between the back of the detector plane, where electronics must be located, and the pupil, where the filter wheel will be located.

- The filters must currently be 5.4 inches in diameter, which results in a large pupil wheel or sliding stage.
- The image plane configuration must be slightly reconfigured to allow for small gaps between the detectors.

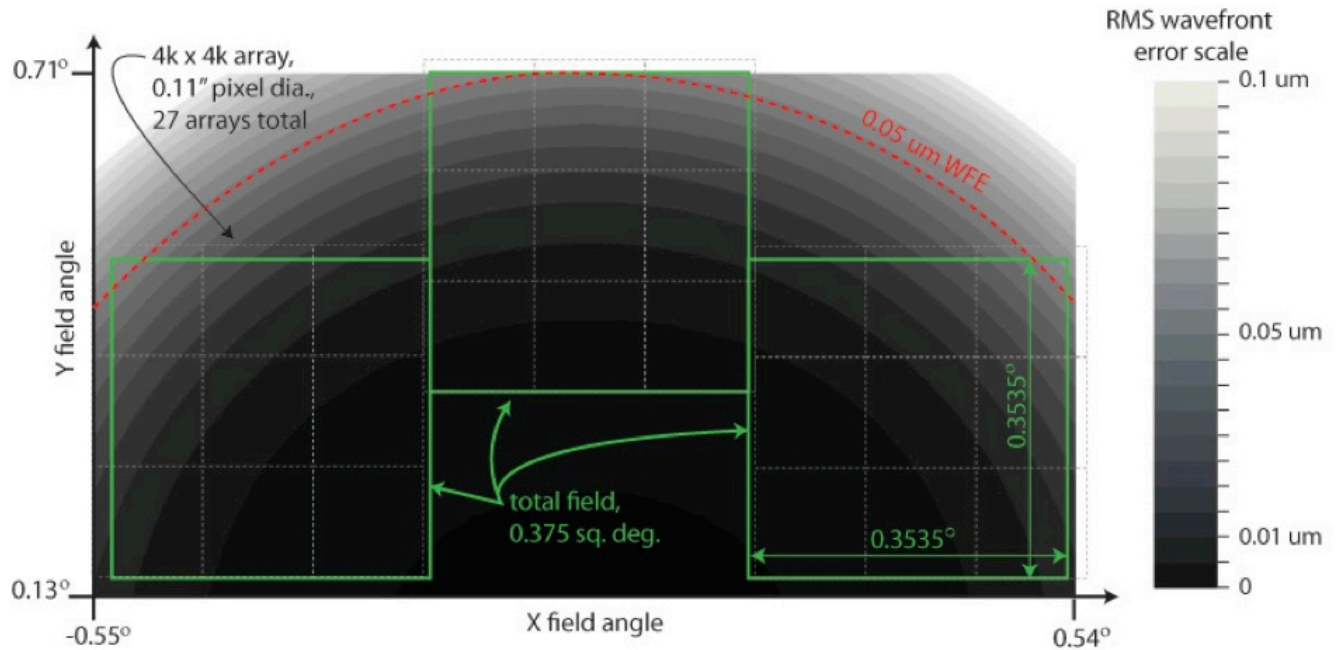

*Figure 6: A plot of the as-designed RMS wavefront error of the wide-field instrument as a function of angle on the sky. The range of the wavefront error contours is from 0 to 0.1μm. The line of 0.05μm wavefront error is shown in red, and indicates λ/20 performance for a wavelength of 1μm. The image plane outlines are shown in green and encompass a total of 0.375 deg$^2$ on the sky. Possible detector array locations are indicated by the gray dashed lines; 27 4k x 4k detectors with a pixel size of 0.11 arcsec would be required to cover the full field of 0.375 deg$^2$*

4.1 Alternate preliminary designs

The above design by E. Elliott at STScI was presented at the Princeton NEW Telescopes workshop by John MacKenty:
http://www.astro.princeton.edu/~dns/NRO_Meeting/mackenty.pptx

Another, more detailed proof-of-concept design, from the WFIRST Project office at GSFC, to use one of the NRO telescopes for WFIRST, was presented by Jeffrey Kruk:
http://www.astro.princeton.edu/~dns/NRO_Meeting/Kruk_NRO_TelWkshp_final.pdf
This design has instrumentation similar to that of the payload concepts described in the WFIRST SDT Final Report. Based on a thorough analysis, the study concludes that the performance of the NRO telescopes in conducting the WFIRST mission is significantly better in most respects than those contemplated in the SDT report for DRM1 and DRM2. The impacts of these performance differences on each WFIRST observing program is discussed, and key trade studies needing investigation by the next Science Definition Team are listed.

The payload design assumes that the telescope is used "as is" and that instruments must be modular and serviceable in-flight. The instrument volume is divided into three independent modules. A preliminary optical design for a three-mirror camera that fits within one such module provides diffraction-limited imaging over 0.28 deg$^2$. Filter and prism specifications are as given in the SDT Final Report. A duplicate instrument placed in a second instrument bay would double the FOV. The remaining bay can accommodate an auxiliary fine guidance sensor (needed during prism observations), a coronagraph, and perhaps additional small FOV instruments.

The NRO telescopes provide a narrower PSF and larger collecting area than the 1.1m and 1.3m telescopes studied by the SDT, but supports a somewhat smaller FOV in a single instrument. For each program, the optimization of area surveyed versus depth needs to be reassessed. In many cases, the increased read noise and observing overheads associated with short exposures tend to favor narrower and deeper surveys. The GSFC's design team presentation concluded that the improved spatial sampling and increased depth for a given exposure time are so great as to enable qualitative advances in the science program.

Rodger Thompson described a distinctively different conceptual design, called WNEW, for an NRO implementation of WFIRST:
http://www.princeton.edu/astro/news-events/public-events/new-telescope-meeting/program/NEW-Telescope-Meeting-Sept-2012-Program.pdf

The novel feature of WNEW is the use of dichroics to produce four 0.25 deg$^2$ focal planes all viewing the same position on the sky. In dichroic mode WNEW views the sky in four filters simultaneously and then switches to a second set of dichroics to view in four different filters. Only two observations are required to view in 8 different filters. In principle, the combination of the dichroic mode and the larger aperture of the 2.4m NRO telescope allows WNEW to complete all of the WFIRST DRM observations in roughly 25% of the time required by WFIRST. In addition, WNEW carries out the its imaging surveys in 8 filters rather than 4 with DRM1 or the GSFC design. By using WNEW in single focal plane or partial dichroic mode, observations with specialized filters, for example, narrow or very broad band would be possible. An example of an additional deep imaging science program is described that would image four 1 deg$^2$ fields in all 8 dichroic mode filters. It achieves 5$\sigma$ AB mag per pixel ranging from 32 to 31 for all but the longest photometric band, where telescope thermal emission limits the AB mag to 30.54.

Although a detailed design for implementing the WFIRST program on an NRO 2.4m is much needed, these preliminary exercises demonstrate that there is a feasible design for a wide-field camera with a FOV sufficient to carry out the WFIRST program

## 5. How well can an NRO telescope carry out the WFIRST mission?

5.1 Overview

WFIRST as defined by *NWNH* has several scientific programs:

i. **Dark Energy Program** (cf. GR modification), including general purpose high-Galactic-latitude "fractions of the sky" photometric and spectroscopic survey for weak lensing (WL) and BAO tests of cosmic expansion, respectively.
ii. **Microlensing Planet Finding**, a demographic survey of Mars-sized and larger planets that will complement and complete the Kepler survey by finding planets more than ~1AU from the parent star. Planet frequencies, orbits, and masses beyond the "snow-line" is essential information for developing a theory of planet formation.
iii. **General Astrophysics program**, including data from the dark energy near-IR surveys, other Guest Observer (GO) surveys, synoptic surveys, and targeted (single field) imaging proposed and carried out through the GO program.

In the following sections, we compare the performance of a NEW WFIRST with the science programs studied in the SDT report. We can most easily attempt to quantify the gain for dark energy science and microlensing science. For dark energy, the larger aperture enables a three times larger supernova survey, a significantly deeper and wider weak lensing survey, and a wider (but shallower) redshift survey. The larger aperture also enables strong lensing measurements of $\sim 10^4$ clusters in its 10,000 $\deg^2$ weak lensing survey. This provides a fourth approach for astronomical measurements of the effects of dark energy. For the microlensing science program, the improved sensitivity should nearly double the number of detected Earth mass planets, for example.

However, the greatest scientific gain of NEW WFIRST will be in the general astrophysics program, where the improved spatial resolution provides Hubble-class GO programs. We describe sample programs in the Milky Way Galaxy and its neighbors that focus on stellar populations, extragalactic programs that involve clusters of galaxies, the co-evolution of AGN and galaxies, and finally a series of programs, from QSOs to lensed galaxies to Lyα emitters, that probe the epoch of cosmic dawn. Much of the science described here will use data that come directly from the wide-field dark energy surveys or from surveys of similar depth in other parts of the sky. In this discussion we call out specifically a number of programs that require smaller-area, deeper surveys, or very deep single-field imaging. We expect that, over the lifetime of NEW WFIRST, a balanced program of all kinds will produce a science program of richness and depth approaching the unrivaled science bounty of HST.

The following science discussions all assume a "room temperature" telescope with detectors cooled to 100 K. In a few cases the advantages of a cooler operating temperature are noted, but the performance of NEW WFIRST discussed here is not dependent on this possibility.

5.2 The Dark Energy Program: NEW WFIRST Performance

The dark energy program of the WFIRST SDT DRM1 implementation has three distinct components: a synoptic survey to discover and monitor Type Ia supernovae at redshifts $0.2 < z < 1.7$, a wide area 4-band (YJHK) imaging survey for weak lensing measurements, and a slitless spectroscopic survey to obtain redshifts of emission line galaxies in the redshift range $1.3 < z < 2.7$ for measuring galaxy clustering — in particular, to constrain the expansion history using baryon acoustic oscillations (BAO) and the rate of structure growth using redshift-space distortions (RSD). These techniques are described in the SDT report and in more detail in Weinberg et al. (2012). In the 5-year primary mission of SDT DRM1, the SNe program would occupy 0.45 years of observation spread over a 1.8-year interval, returning to a set of fields for 33 hours out of every five days. The high-latitude survey (HLS) would occupy 2.4 years of observation, covering a total of 3,400 deg$^2$ with 4-band imaging and resolution $R \approx 600$ spectroscopy. More area could of course be covered in an extended mission after the 5-year primary mission.

We will assess the power of an NRO-based implementation of WFIRST by scaling to the capabilities of SDT DRM1. For purposes of this scaling analysis, we make the following assumptions:
1) the ratio of aperture areas is $0.85 \times (2.4/1.3)^2 \sim 3$, where the factor of 0.85 accounts for secondary obstruction in the NRO 2.4m telescopes.
2) the effective long wavelength cutoff is 2.0μm (vs. 2.4μm for SDT DRM1), enabling Y, J, H, and Ks-band imaging, and enabling H-alpha galaxy redshift measurements to $z = 2.0$ (as compared to $z = 2.7$).
3) the pixel scale is fine enough to fully sample (with a dithering strategy similar to that of SDT DRM1) the images at J and H band, but not at Y band; roughly speaking, this implies pixels scaled by the aperture ratio 1.3/2.4.

4) the active area of the focal plane would be the same in the two implementations, 0.375 deg$^2$.
5) the observing time devoted to dark energy would be the same as that in DRM1.

We further assume that the telescope and detectors will be cool enough that the sensitivity out to 2.0μm scales with the aperture ratio and is not degraded in the NRO implementation by higher thermal noise. A final design might or might not satisfy this condition, depending on details of the cooling mechanisms and the orbit. Conversely, imaging and spectroscopy beyond 2.0μm may be feasible if the detectors and optics can be kept cold enough. Achieving both conditions 3 and 4 would require a number of pixels on NEW WFIRST three times larger than that on SDT DRM1, which we assume would be possible with H4RG detectors. Of course, an SDT design with more pixels would also be more powerful than DRM1, but we do not consider this in our scaling. When measurements are limited by statistical errors rather than by systematic errors, their information content (in the sense of an inverse variance) should scale linearly with the number of pixels (at fixed pixel size) and with the amount of observing time.

We address each of the three dark energy programs in turn.

5.2.1 Supernovae (SNe)

SDT DRM1 would monitor 6.5 deg$^2$ with the depth needed to discover and measure Type Ia SNe to z = 0.8, and 1.8 deg$^2$ with the depth needed to measure Type Ia SNe to z = 1.7. This survey is expected to yield about 1900 confirmed Type Ia SNe detections, yielding measurements of the luminosity distance $D_L$ (z) in 16 redshift bins of width Δz = 0.1. The photometry and redshift measurements are accurate enough to yield statistical errors in distance modulus $\sigma_\mu$ ≤ 0.02 per redshift bin (1% in luminosity distance), after averaging over all supernovae in the bin. The SDT report considers two scenarios for systematic errors, a "conservative" one in which the systematic error per Δz = 0.1 redshift bin is 0.02 × (1+z)/1.8 mag, and an "optimistic" one in which it is 0.01 × (1+z)/1.8 mag. These systematic errors are assumed to be uncorrelated from one redshift bin to another, so the overall demand on systematics is tighter by a factor of ~ $(N_{bin})^{-1/2}$ = 4. The aggregate precision of the luminosity distance measurement, including the contribution of systematic errors, is $(\Sigma_i [\Delta \ln D_L(z_i)]^{-2})^{-1/2}$ = 0.32% for the conservative case and 0.23% for the optimistic case. The DRM1 SN survey incorporates JHK imaging, which we assume would become YJH in the case of NEW WFIRST.

To a first approximation, we expect that the survey speed for NEW WFIRST would be faster than that of SDT DRM1 by a factor of three because of the aperture ratio, neglecting increased overheads associated with more frequent repositioning of the telescope. For the conservative systematics case, the DRM1 SN survey is limited by systematic errors in most bins. In this scenario, therefore, an NRO implementation of WFIRST could do the SN program faster, but it might not do much better in the end, because increasing the number of supernovae would have only moderate impact on the total error. Indeed, a NEW WFIRST supernova survey might perform worse, because it would be limited (at each redshift) to bluer rest-frame wavelengths, where the intrinsic scatter of SNe peak luminosities is (slightly) larger, and where the systematics associated with dust extinction are larger.

For the optimistic systematics case, the DRM1 SN survey is limited by statistical errors. In this scenario, a NEW WFIRST survey could perform substantially better in the same amount of observing time simply by measuring more supernovae. If we assume that the *rms* statistical error in each redshift bin drops by $\sqrt{3}$, then the aggregate precision of the luminosity distance measurement improves from 0.23% to 0.17%. We expect the SN program contribution to performance measures like the Dark Energy Task Force (Albrecht et al. 2006) Figure-of-Merit (FoM) to scale approximately as $(\sigma_{SN})^{-2}$. Of course, the different characteristics of the NEW WFIRST telescope might lead to a different optimization strategy, allowing greater improvement than simply implementing the SDT DRM1 strategy over a 3× larger area. For example, a SN survey with NEW WFIRST might take advantage of the larger aperture by increasing the survey area for high-redshift ($z > 0.8$) supernovae.

The above analysis highlights the enormous importance of controlling systematics in a SN survey using NEW WFIRST. Because of the greater statistical power afforded by the larger aperture, the demands on systematics are tighter, but the loss of the longer wavelengths makes them more challenging to achieve. For example, evolution of the SN population over the redshift range of the survey could result in systematic errors that would ultimately limit the accuracy of this method. Another is the requirement for highly accurate relative photometry over a wide span of flux and redshift, with proper accounting for the redshifting of the complex SN spectral energy distribution. A reevaluation of how to control these potential sources of systematic error is needed because of the specific characteristics of a NEW WFIRST SN survey. Issues to be studied would likely include the resolution and wavelength cut-off of a prism or grism, the possibility of observing bright SNe in Ks band (even with a warm telescope), one or more spectrograph slits, and an IFU (providing separate clean spectra of host and SN, see Appendix B of the SDT Report). Spectroscopic observations of discovered SNe are one approach to reducing systematics — by measuring SN brightness over fixed rest-frame wavelength bands, thereby eliminating the need for k-corrections, and by addressing evolution systematics by matching SNe spectra. Another approach to taming systematics could be extending the red end of SN photometric measurements to reduce extinction and intrinsic variance.

5.2.2 Weak Lensing (WL)

The WFIRST SDT DRM1 survey would obtain galaxy shape measurements in three bands (JHK). The fourth band (Y) provides valuable leverage for photometric redshifts, but the pixel scale is too coarse to fully sample the Y-band images, precluding shape measurements with the accuracy required for WL. The DRM1 HLS would achieve an effective source surface density (i.e., a surface density weighted by the S/N of the shape measurements) of about 40 galaxies arcmin$^{-2}$, yielding 480 million shape measurements over its 3400 deg$^2$ area. These would allow precise measurements of the cosmic shear power spectrum in multiple "tomographic" bins of photometric redshift. The aggregate precision predicted for the shear power spectrum measurement (i.e., the precision on a constant factor multiplying the power spectrum at all redshifts) is 0.29%. This statistical precision imposes stringent demands on control of systematics, in particular the systematics on galaxy shape measurement.

The SDT report sets the requirement on WL systematics at a level that would allow the WL survey to remain statistics-dominated even if it were increased in area to $10^4$ deg$^2$, and detailed analysis indicates that the determination of the PSF from observations of stars should be accurate enough to control shape systematics at this extraordinarily stringent level, given the number of degrees of freedom in the optical system. Most WL galaxies would have shape measurements in all three bands, so the survey can independently measure three shear auto-correlation functions and three band-to-band cross-correlation functions. Since the true WL signal should be achromatic, this redundancy allows powerful internal tests for systematic errors, and informative diagnostics for any systematics that are detected.

From the point of view of weak lensing, an NRO implementation of WFIRST has two key advantages relative to SDT DRM1: faster survey speed because of its larger collecting area, and higher angular resolution because of its smaller diffraction limit. It also has two key disadvantages: the loss of K-band because of its higher operating temperature, and a more complex PSF of an obstructed as compared to an unobstructed optical system. One consequence of the latter is that the gain in angular resolution is smaller than the naive expectation from the ratio of mirror diameters; a second is that determination of the PSF at the required level of accuracy is more challenging.

Scaling with the assumptions above suggests that NEW WFIRST would be able to cover about three times the area of SDT DRM1 in the same amount of observing time, surveying $10^4$ deg$^2$ instead of 3400 deg$^2$. This would enable NEW WFIRST to cover a large fraction of the sky available at high Galactic and ecliptic latitudes. (Observations become less efficient at low Galactic latitude because of extinction and at low ecliptic latitude because of greater zodiacal background.) Furthermore, for the same limiting magnitude, the higher angular resolution would increase the effective source density. For plausible assumptions about the optical design, the half-light radius of the PSF would be about 1.3 times smaller for NEW FIRST, which should increase the effective source density by 30-50% if we assume that shape measurements can be made down to the same ratio of $r_{eff}/r_{PSF}$. We would thus expect the number of galaxies with shape measurements to increase by a factor of ~ 4, reducing *rms* statistical errors on the WL power spectrum by about a factor of two. Contributions to the DETF FoM should again scale roughly as the inverse variance of the power spectrum error, so this would represent a dramatic (approximately a factor of 4) increase in the power of the WL survey.

The greater statistical power of NEW WFIRST imposes stricter demands on the control of systematics relative to SDT DRM1, but the characteristics of the telescope impose greater challenges. We assume that the pixel scale on NEW WFIRST would not be fine enough for shape measurements in Y-band, but would be fine enough to allow J- and H-band shape measurements. This is a significant loss of redundancy relative to SDT DRM1, but it still allows measurement of two auto-correlations and one band-to-band cross-correlation (the same level of redundancy as in the WFIRST Interim Design Reference Mission). We regard this as a minimum requirement — there is little point in measuring WL with unprecedented statistical precision if one cannot check the measurement accuracy. The PSF for NEW WFIRST would be more complex than for the unobstructed SDT design, and a five-mirror optical path has more degrees-of-freedom to constrain than a three-mirror design. On the positive side, the NRO telescopes have active secondary mirrors that can be adjusted in response to wavefront sensors.

An NRO-based implementation would require careful simulations of survey strategies and calibration methods to ensure that the PSF can be determined with the necessary degree of accuracy. However, while we view keeping shape-measurement systematics below the statistical errors as a significant challenge, we do not see any show-stoppers. Another systematics challenge for WFIRST weak lensing is calibrating the distribution of photometric redshifts, for which one wants highly complete spectroscopic samples of $~10^5$ galaxies that sample the full magnitude and color range of the photometric sample. Compiling these calibration samples will require a combination of ground- and space-based observations, but the larger aperture of NEW WFIRST would be a significant asset for this purpose relative to SDT DRM1.

5.2.3 Galaxy Redshift Survey (BAO and RSD)

The spectroscopic component of the SDT DRM1 HLS would cover 3400 deg$^2$ to a line flux sensitivity of $1.0 \times 10^{-16}$ erg/cm$^2$/sec, sufficient to measure the Hα emission-line redshift of a 1.1 L* star forming galaxy at z=1.5 and a 1.7 L* star-forming galaxy at z = 2.6. Forecasts based on recent observational estimates of the Hα luminosity function and the clustering bias of high-redshift emission-line galaxies indicate that the survey would measure redshifts of about 17 million galaxies over 3400 deg$^2$ in the redshift range 1.3 < z < 2.7, yielding BAO determinations of the angular diameter distance $D_A(z)$ and Hubble parameter $H(z)$ in 14 Δz = 0.1 redshift bins. The forecast aggregate precision from BAO is $(\Sigma_i [\Delta \ln D_A(z_i)]^{-2})^{-1/2} \approx 0.5\%$ for $D_A$ and $(\Sigma_i [\Delta \ln H(z_i)]^{-2})^{-1/2} \approx 0.7\%$ for H, and analyses using the full galaxy power spectrum could improve the precision by a factor ~1.5-2, depending on the accuracy that can be achieved in modeling of non-linear evolution and non-linear galaxy bias. The precision of RSD measurements depends on how small a scale one can reach before systematic uncertainties in modeling non-linear effects exceed statistical errors: for assumptions that are probably conservative, the SDT report forecasts an aggregate error of 1.4% on the product $\sigma_{8m}(z)f(z)$ of the fluctuation amplitude and logarithmic growth rate.

From the point of view of BAO measurements, the critical properties of a redshift survey are its comoving volume and its sampling density, specifically the product nP of the mean galaxy space density n, and the amplitude of the galaxy power spectrum P(k) at the wavenumber characteristic of BAO, approximately k = 0.2 h Mpc$^{-1}$. Relative to a survey with perfect sampling (i.e., very large n), the BAO measurement from a sparsely sampled survey is degraded in variance by a factor of ~ (1+1/nP), equivalent to reducing the volume of a perfectly sampled survey by the same factor. Given recent estimates of emission line galaxy bias, the SDT DRM1 redshift survey is forecast to have nP > 1.2 from z = 1.3 - 2.3, so in this range it would be limited mainly by finite volume rather than galaxy shot noise. The predicted sampling density falls to nP = 0.5 by z = 2.65. With the same assumptions, the Euclid redshift survey has nP > 1 only for 0.7 < z < 1, falling to nP ≈ 0.3 by z = 1.3 and nP ≈ 0.15 at the outer redshift limit z = 2.0.

With a 2.0μm cutoff, a NEW WFIRST redshift survey would be limited to z < 2.0; the lower redshift cutoff would probably be dropped below z = 1.3, depending on the degree to which ground-based surveys had fully sampled this redshift range. However, NEW WFIRST should be able to map to the same flux limit about three times faster than SDT DRM1, so in the same observing time it could cover ~10$^4$ deg$^2$ instead of 3400 deg$^2$, with nP > 1 over the full redshift range of the survey. In contrast to SN and WL, the systematic errors for BAO are expected to be easily sub-dominant to statistical errors, so the likely gain (in the sense of inverse variance or contribution to DETF FoM) is a factor ~ 3 or more. RSD measures of structure growth should gain by a similar factor, as should other approaches using the galaxy power spectrum, up to the point that they become limited by systematic uncertainties in theoretical modeling.

Relative to SDT DRM1, a NEW WFIRST redshift survey would be less complementary to Euclid's because it would cover essentially the same redshift range. However, even if it mapped the same sky area as Euclid, it would not be redundant, because Euclid's BAO measurements at z > 1 will be severely degraded (relative to a perfect survey) by shot noise, while NEW WFIRST would not. The aggregate statistical precision of the NEW WFIRST measurements of BAO and RSD would be substantially higher than Euclid's (because of greater depth) or SDT DRM1's (because of greater area), yielding greater leverage on dark energy.

5.2.4 Dark Energy Performance: Summary

With the assumptions outlined above, an NRO-based implementation of WFIRST would have roughly three times the survey speed of SDT DRM1 to the same depth. It would have higher angular resolution, but it would likely have a shorter red wavelength cutoff or reduced sensitivity at its reddest wavelengths.

In the same amount of observing time, NEW WFIRST could carry out a substantially larger SN survey than SDT DRM1. However, the relative performance would depend critically on the ability of NEW WFIRST to achieve comparable or better control of systematics. The absence of K-band imaging is a significant loss in this regard. Using an IFU for SN spectroscopy and spectrophotometry becomes an especially attractive option for NEW WFIRST, perhaps enabling a SN program that is substantially better than that of DRM1 in both statistics and systematics.

For WL, NEW WFIRST could achieve a factor $\sim 3$ increase in survey area in the same observing time, covering much of the sky that is available at high Galactic and ecliptic latitudes. It should also achieve a higher effective source density because of higher angular resolution, and the combined effect could be a factor $\sim 4$ in the number of galaxies with good shape measurements. Achieving the required control of shape measurement systematics will be more difficult because of the greater PSF complexity, but it should be possible if sufficient attention is paid to these systematics when designing the hardware, survey strategy, and analysis software. The loss of K-band imaging reduces redundancy and the ability to carry out internal consistency checks, but making shape measurements in two bands (J, H) rather than three is acceptable. An additional band between H-band and K-band with a red cutoff of $2.0\mu m$ could be considered to mitigate the loss of K-band.

For BAO, RSD, and other applications of galaxy clustering, NEW WFIRST could again achieve a factor $\sim 3$ increase in survey area relative to SDT DRM1. The redshift range of the emission line galaxy survey would shift because of the wavelength cutoff, perhaps covering $1.0 < z < 2.0$ instead of $1.3 < z < 2.7$. While the resulting measurements would be less complementary to Euclid's, their statistical precision would be much better than either Euclid's or SDT DRM1's, with a clear net gain for dark energy constraints.

Note that our assumption of fixed focal plane area is crucial to these conclusions. If we instead assumed fixed pixel count, then the smaller area associated with finer pixels would give up most of the gains described above. An SDT-like WFIRST with H4RG detectors and a larger total pixel count could also outperform the SDT DRM1 design. The maximum focal plane area with good image quality in an SDT-like design is probably ~0.6 $deg^2$, so the maximum gain in survey speed relative to DRM1 would be a factor ~1.6. However, an extended mission (10 years rather than 5 years) could make up much of the remaining ground, relative to 5 years with NEW WFIRST.

5.3 The Exoplanet Microlensing Program: NEW WFIRST Performance

In broad-brush strokes, it is fairly straightforward to demonstrate that a suitable NRO implementation of WFIRST is likely to be more capable than any of the WFIRST SDT DRM designs, in the sense of allowing a higher planet yield per unit observing time for the same assumptions. Furthermore, the NRO implementation of the WFIRST design may allow one to substantially increase the sensitivity, and may provide dramatically more information about the properties of the individual planet detections. However, demonstrating these two ideas conclusively requires additional study.

We first provide the basic argument that suggests a more capable mission for the NEW WFIRST, and then provide a somewhat more detailed estimate of the improved yields.

5.3.1 General considerations

The qualitative science requirements for a space-based microlensing survey are (1) sensitivity to a large number of planets over a broad region of mass and semi-major axis, (2) sensitivity to planets with mass down to that of Mars, (3) sensitivity to free-floating planets down to the mass of the Earth, and (4) ability to obtain mass measurements for a significant fraction of the primary lens stars.

Planetary companions to stars are detected via short-duration (hour to day), relatively low-amplitude (one to tens of percent), planetary deviations from the longer timescale (weeks to years) microlensing events caused by their host stars. The planetary perturbations become briefer and more rare as the planet mass decreases, with Earth-mass planets having detection probabilities of <1% and durations of order an hour. Free-floating planets are detected as isolated microlensing events; the durations and probabilities of these events also decrease as the planet mass decreases. For both bound and free-floating planets, the minimum detectable mass is set of by the angular size of the source star: planets with the mass of Mars or less give rise to relatively low-amplitude perturbations (<~10%) due to suppression and smoothing by the finite size of the typical main-sequence stars. Finally, measuring the mass of luminous primary stars requires measuring the relative proper motion between the lens and source, which are 2-10 milliarcseconds (mas) per year.

For a typical microlensing event, the detection probability for an Earth-mass planet at 2 AU is ~1.5%; therefore, in order to detect of order 200 such planets, a total of ~200/0.015 ~$10^4$ microlensing events must be found and monitored for planetary perturbations. The microlensing event rate toward the Galactic bulge is $5 \times 10^{-5}$/star/year, and thus ~$10^4$/$5 \times 10^{-5}$/stars/year ≈ 200 million star-years must be monitored. The stars must be monitored with the cadence needed to detect and resolve the perturbations of the lowest-mass planets of interest, which translates into a cadence of ~15 minutes. The photometric precision must be sufficient to detect the 5–10% perturbations at high signal-to-noise ratio, so photometry with a precision of a few percent is required for the faintest monitored stars. The typical stellar density in the Galactic bulge is ~$10^8$ stars per square degree down to J=23. Thus, areas of more than 2 square degrees must be monitored with cadences of ~15 minutes, and the total photon rate must be sufficient that photometric precisions of a few percent are obtained for the faintest sources. This latter requirement can be met by mirror apertures of more than 1 meter, wide-band near-IR observations, and high angular resolution — needed in order to resolve the dense background of blended stars. Given the areal density of ~$10^8$ stars per square degree, this implies angular resolutions of <0.4", commensurate with an aperture of 1m or larger and observations at $\lambda \sim 1.5 \mu m$.

Thus the primary requirements for planet detection are (1) a relatively large (substantial fraction of a degree), near-IR focal plane, (2) effective aperture of more than 1m with a wide near-IR bandpass, (3) angular resolution of <0.4". Once the third requirement is met, the total yield of planet detections scales primarily with the total amount of observing time (nearly linearly), the area of the detector, and the total photon collection rate for typical source. In addition, these three factors can also be traded off of one another. Therefore, given that the NEW WFIRST has a larger total photon collection rate, and provided that the detector will have a similar effective area as in the SDT DRM designs, then NEW WFIRST will be more intrinsically capable, i.e., will yield more planet detections per unit observing time.

There are additional, secondary requirements that may bear on the NEW WFIRST design and its ability to carry out a microlensing mission. First, the spacecraft must be able to point continuously or nearly continuously at the Galactic bulge for at least the duration of the typical primary microlensing events, or more than 40 days. Second, the storage capacity and downlink rate must be sufficient to store and transmit the ~700 full-frame images per day during the bulge observations. Third, there are requirements on the pointing accuracy and slew-and-settle times. Finally, a broad near-IR filter and a narrower bluer filter are required, the first for the primary science data, with observations in the second filter acquired every ~12 hours for characterization of the source stars.  This, in turn, leads to a requirement on the total number of filter changes during the microlensing experiment of >1000.

In addition to a large yield of planets over a broad region of parameter space, the other dramatic improvement of a space-based microlensing mission over ground-based surveys is the ability to automatically estimate the masses of the majority of the planetary host stars and thus planetary companions.  This qualitatively new capability is enabled by the small and stable PSFs afforded in space.  PSFs of angular size <0.4" essentially resolve out most of the background stars, and thus allow one to isolate the time-variable flux of the microlensed source from any flux from the lens, or companions to the lens or source.  Light from companions to the lens and source can be further distinguished from light from the lens based on observations taken over a very wide time baseline from the event, which will allow one to infer the relative lens-source proper motion as the source and lens move apart.  The number of events for which this measurement can be made depends quite sensitively on the PSF size and stability (Bennett et al. 2007).  For this application, therefore, NEW WFIRST may prove substantially more capable than any of the SDT DRM1 designs.

A quantitative assessment of the potential of NEW WFIRST to measure host star masses requires detailed simulations, which have been done using only one of our simulation codes for only one of several suggested NEW WFIRST optical layouts.  However, we can make some general statements.  First, at the angular resolution of the NRO telescope of 0.16 arcsec for a typical wavelength of 1.5$\mu$m, we expect nearly all unrelated background stars to be resolved.  Furthermore, NEW WFIRST will be able to resolve sufficiently luminous companions to the lens or source with projected separations of >1300 AU, although these will constitute a relatively small fraction of binaries.

Finally, NEW WFIRST will likely also be able to measure the relative lens-source proper motion for a majority of the luminous lenses.  Figure 7 shows the distribution of lens-source proper motions for a sample of microlensing from a WFIRST simulation. The peak of the distribution occurs near 5 mas/yr, and thus over the course of a ~5 year mission, the separation between the lens and source will have changed by ~0.025 arcsec.  For typical lens fluxes relative to the source, images taken well before or after the microlensing event will exhibit a detectable elongation, and thus yield a measurement of the lens-source proper motion (Bennett et al. 2007).  Figure 7 shows that DRM1 will be able to measure proper motions for a little over half of the microlensing events with luminous lenses. NEW WFIRST will perform better, allowing measurements of the proper motion of most events.   One potential complication with these measurements is the more complicated PSF shape of the NRO telescope due to the obstruction of the secondary mirror.  This is an important topic for future study, although, it seems likely that the weak lensing requirements on the PSF shape will be more stringent than are required for the mircolensing planet search.

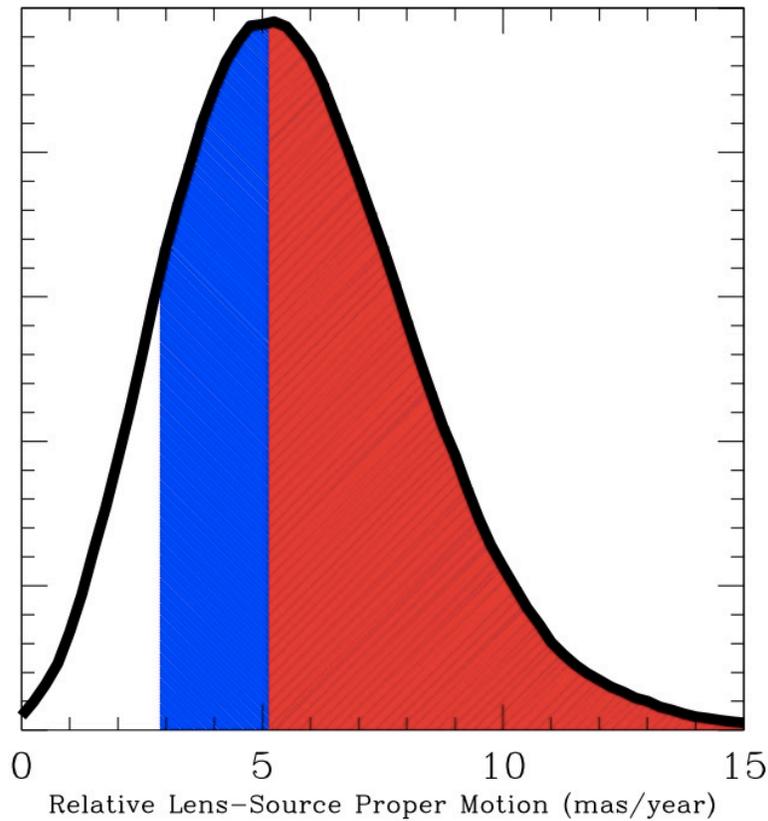

*Figure 7: The solid black curve shows the distribution of relative lens-source proper motions from a Galactic simulation based on a model by Han & Gould (2003). The shaded red region shows a rough estimate for the range of proper motions that will be detectable with the SDT DRM1 telescope (with resolution of 0.29 arcsec), whereas the blue shaded region shows the added range for NEW WFIRST telescope (with resolution 0.16 arcsec).*

5.3.2 Comparison Between WFIRST SDT DRM1 and NEW WFIRST

The baseline WFIRST SDT DRM1 microlensing survey consisted of 6 uninterrupted 72-day seasons for a total of 432 days. A total of 7 fields were monitored, for a total of 2.625 deg$^2$. The exposure time per field was 85 s, with a cadence of 15 minutes. The majority of the observations were taken in a wide filter "W" with 0.92-2.40$\mu$m, with observations every 12 hours taken in a "Y" filter with 0.92-1.21$\mu$m.

The estimated yields for SDT DRM1 are shown in Table 2. The simulation methodology and assumptions are described in Section 2.2.4 and Appendix A of Green et al. (2012). The "best estimate" yields are given in Table 3 of that report, which ultimately adopts a planet distribution function for cold exoplanets as measured from ground-based microlensing surveys by Cassan et al. (2012), as well as microlensing event rates extrapolated from observed rates for bright clump giant sources. Very briefly, DRM1 is expected to detect ~2200 bound exoplanets with masses in the range of 0.1-10,000 $M_{Earth}$ and semi-major axes in the range of 0.03-30 AU. Of these, ~240 will have mass similar that of the Earth, and ~30 will have mass similar to that of Mars. In addition, if there is one free-floating Earth-mass planet per star in the Galaxy, SDT DRM1 will detect ~30.

While we have not performed detailed simulations of a NEW WFIRST microlensing survey, we can very roughly scale the results from the SDT DRM1 to obtain a rough idea of the expected planet yield. To do so, we make the same set of assumptions adopted in Section 5.2 for evaluating the expected results of the dark energy experiments. Specifically, we assume:

1. A factor of $0.85 \times (2.4/1.3)^2 = 2.9$ larger photon rate from the larger aperture, where the factor of 0.85 accounts for the obstructed aperture.
2. An effective wavelength cutoff of $2.0\mu m$, corresponding to a 1.27 smaller photon rate relative to the SDT DRM1. The factor of 1.27 is appropriate for a flat-spectrum source, but we adopt it uniformly for simplicity.
3. An active field-of-view for the detector of 0.375 $deg^2$, identical to that for the SDT DRM1.
4. The same total observing time for the microlensing observations as DRM1, 432 days.

    We also make the following additional assumptions:

5. We assume the same exposure time, cadence, number of fields, and field locations as for SDT DRM1.
6. We assume the contribution to the background flux due to all sources is the same for both NEW WFIRST and SDT DRM1 designs.

We discuss the appropriateness of the last two assumptions below.
With these assumptions, the only effect on the yields is due to the change in the total number of photons per exposure. The net increase in the photon collection rate is $\sim 0.85 \times (2.4/1.3)^2/1.27 \sim 2.28$. The effect of the increased photon rate on the yields depends primarily on the form of the cumulative number of detections as a function of the minimum threshold for detection, which roughly follows a power law of the form $N(>\Delta\chi^2) \propto (\Delta\chi^2)^\alpha$, with the exponent $\alpha$ depending on the mass and separation of the planets of interest. Averaged over separations between 0.03-30 AU, the exponent varies from $\alpha = -1.2$ for low-mass planets of $\sim 0.1 M_{Earth}$, to $\alpha = -0.3$ for high-mass planets of mass >100 $M_{Earth}$. Since $\Delta\chi^2 \propto N_{phot}$, this number of detections scales with the photon rate in the same way. We have calibrated this scaling explicitly using simulations that adopted different photon rates, but were otherwise identical.

The yields estimated in this way are compared to the yields from DRM1 in Table 2. Overall, we estimate that NEW WFIRST will detect ~1.6 times more planets, with yields that are a factor of 30-40% larger for the high-mass planets increasing up to a factor of ~3 times larger for Mars-mass planets. The dramatic increase in the yields of the lowest-mass planets simply reflects the fact that these planets are near the edge of the mission sensitivity for DRM1. This further suggests that NEW WFIRST may have significant sensitivity to other classes of planets that were just beyond the region of primary sensitivity for DRM1, e.g., very small separation planets and habitable planets.

Table 2: Comparison of planet yield with SDT DRM1 and NEW WFIRST

| $M/M_{Earth}$ | SDT DRM1 | NEW WFIRST |
|---|---|---|
| 0.1 | 30 | 82 |
| 1 | 239 | 379 |
| 10 | 794 | 1322 |
| 100 | 630 | 1067 |
| 1000 | 367 | 509 |
| 10,000 | 160 | 205 |
| Total | 2221 | 3564 |

*Table 2: Predicted yields for bound planets for the NEW WFIRST implementation, compared to the predicted yields for the SDT DRM1 mission (from Green et al. 2012). Both assume the same number of fields, same area of the detector, and the same total observing time of 432 days. For these yields, we have assumed the planet distribution function for cold exoplanets as measured from ground-based microlensing surveys by Cassan et al. (2012). The effects of the changes in the background flux for the NEW WFIRST design relative to the SDT DRM1 design have not been considered.*

5.3.3 New Microlensing Event Rate Measurement

Since 2004, the main focus of gravitational microlensing surveys of Local Group galaxies has been the study of extrasolar planets, and as a result, there have been no papers on the microlensing rate or optical depth toward the Galactic bulge using events observed more recently than 2002. However, the more recent OGLE-III, MOA-II, and OGLE-IV surveys have discovered more than 10 times as many microlensing events as were known in 2002. As a result, these analyses did not measure the microlensing event rate in the central Galactic bulge, and so the event rate estimates used for Table 2 are based upon an extrapolation from published measurements in fields several degrees away. However, a new microlensing rate measurement from the MOA-II survey (Sumi et al., in preparation) does overlap the with the WFIRST fields and is based on a much larger sample of events. This new microlensing rate measurement implies an event rate that is a factor of 1.77 larger than the results listed in Table 2, so these estimates should be regarded as conservative.

5.3.4 A Preliminary NEW WFIRST Microlensing Simulation

We have done a preliminary investigation of the GSFC NEW WFIRST proof-of-concept design, which has a 25% smaller FOV than the SDT DRM1 and the baseline STScI NEW WFIRST proof-of-concept design presented in section 4. The improvement over the DRM1 exoplanet yield with this telescope design is somewhat smaller than the improvement given in Table 2, due to the smaller area of the focal plane. When the observing strategy is optimized for this FOV, the NEW WFIRST survey will detect Earth-mass planets at a 35% higher rate than the DRM1 design, instead of the 59% improvement listed in Table 2. However, both of these numbers assume a higher data rate than was possible with the SDT DRM1 design. This is not a problem for the GEO and HEO orbits being considered for NEW WFIRST, but it could be a problem for NEW WFIRST if it flew in an L2 orbit. In fact, if NEW WFIRST flies in a L2 orbit with the same data rate assumed for the SDT DRM1 design, the NEW WFIRST Earth-mass planet detection rate would be 24% lower than the SDT DRM1 discovery rate. The reason for this is that the better image sampling of NEW WFIRST means that there are more pixels per star, so fewer stellar photometry measurements can be made with a fixed pixel download rate. However, there are likely to be data compression and/or data system hardware solutions to this problem.

5.3.5 Caveats

Of course, the rough estimates for the yields presented here should be considered only illustrative of the improved capability of NEW WFIRST, a much more careful assessment using a full simulation is needed to provide robust estimates.

Such a simulation must in particular include a detailed accounting of the various background noise sources, whose magnitudes are likely to be significantly different for NEW WFIRST relative to SDT DRM1. In particular, the ~1.8 times better resolution of an NRO telescope at fixed wavelength means that a larger fraction of the unrelated bulge stars will be resolved, thereby lowering the background. Furthermore, although a shorter wavelength cutoff would result in a smaller photon flux, this is partially compensated by the smaller effective wavelength for the broad microlensing filter and the resulting better resolution. Of course, the improved resolution also means that the noise contribution due to the truly smooth backgrounds (i.e., zodiacal light) will be smaller. When comparing the yields of the 1.3m DRM1 to the 1.1m DRM2 missions, the SDT found that the net effect of these various factors was relatively small (~10%). However, the angular resolution of NEW WFIRST is substantially better than SDT DRM1, and therefore these effects are likely to be more important. In addition, one must consider the larger thermal background due to the assumed higher operating temperature of the NEW WFIRST telescope.

Another effect we have not considered, but which is unlikely to be important for the overall yields, is the effect of systematics on the photometry. Due to the larger photon rates, a larger number of stars in the NRO design will have photometry limited by systematics, rather than photon noise statistics. However, the planetary perturbations of interest are a few to tens of percent, substantially larger than the expected systematics limit of a few millimagnitudes. Indeed, the overwhelming majority of stars are expected to be photon-noise dominated, for all conceivable designs. This is qualitatively different and in stark contrast to the dark energy experiments, many of which operate near the systematics limit. In such cases, a more sophisticated analysis is required to see if the larger aperture of NEW WFIRST translates into a net improvement in the science metrics.

5.3.6 Future Work

The primary future work that is needed is a full simulation of a NEW WFIRST mission, in order to validate the rough scaling relations we have provided, and to determine the effects of the different backgrounds on the yields. Furthermore, optimization of a microlensing survey with NEW WFIRST should be carefully considered; it may well be that the optimal location, number of fields, and field cadence will be very different than that assumed here. Most importantly, the sensitivity of NEW WFIRST to planets near the edge of the sensitivity for the SDT DRM1 design should be carefully considered — in particular, for habitable planets. Finally, a quantitative and detailed estimate of fraction of events for which primary mass measurements are possible should be performed.

5.4 The Stellar Populations and Properties of Local Galaxies

5.4.1 Local Stellar Populations

Since the first studies by Hertzsprung and Russell over 100 years ago, the Milky Way's stellar populations have represented a unique and fundamental data set for astronomy. Exploration of these nearby stars and star clusters provides exquisite details on many astrophysical topics, including star formation, stellar evolution, and the initial mass function. These studies directly enable detailed tests of stellar models, and thereby provide the crucial calibration to link the color and luminosity of a distant and unresolved stellar population to its age and metallicity (Bruzual and Charlot 2003). More generally, the distribution of nearby stellar populations within the Milky Way provides unique insights on the formation and assembly of large galaxies.

Over the past few years, large wide-field surveys such as SDSS, 2MASS, WISE, RAVE, and UKIDSS have provided a wealth of new data to investigate local stellar populations. Despite the many scientific rewards from these surveys (e.g., SDSS's discovery of those galaxies most-dominated by dark matter), astronomy still suffers from the lack of a wide-field, high-resolution IR map of the Galaxy. The NRO 2.4m telescope can provide such a survey, and transform several general scientific areas:

**1) Stretch the color-magnitude relation of stars**
Stellar populations such as globular clusters have been imaged with high-resolution UV and optical cameras. The sensitivity achieved with a 2.4m IR telescope will transform these data sets into a panchromatic nature, thereby establishing the luminosity functions of post main-sequence phases of stellar evolution in the IR bandpasses (e.g., the RGB and AGB), and stretching age-sensitive features (e.g., the turnoff) over a wider color baseline. These data will lead to an unprecedented calibration of IR stellar models and improved ages of star clusters that represent important calibrators to many astrophysical relations. As an example, Figure 8 illustrates the morphology of the predicted color-magnitude diagram for an old stellar population in the optical and IR baselines (models from Dotter et al. 2007), as well as the associated uncertainties in the distance and reddening at fixed age and metallicity. The panel on the right shows a recent 3-orbit observation of 47 Tuc with WFC3/IR on HST, and demonstrates the potential of future high-resolution IR probes of crowded fields (Kalirai et al. 2012).

**2) Towards a complete stellar and substellar census**
Low mass stars dominate the stellar luminosity function. At an age of 1 Gyr, a 0.08 $M_\odot$ star has $M_V = 19$, but a color of V-H = 8. The resolution and sensitivity of NEW WFIRST in IR bandpasses will enable the most complete stellar luminosity function to date, providing characterization from the brightest giants to the faintest dwarfs in stellar systems out to several kiloparsecs. This survey will directly inform the Galactic mass budget, drastically improve the color-magnitude relation of cool stars, and provide resolved estimates of the initial mass function and its variation with environment. These quantities in turn relate to the physics governing the internal and atmospheric structure of stars, and have widespread consequences for many astrophysical applications, including the measurement of star formation mechanisms and masses of galaxies. This survey will also provide a global mapping of objects cooler than T dwarfs ($T_{EFF} < 700$ K) — so called "Y dwarfs". Such objects must exist: objects with inferred masses down to 5 $M_{JUP}$ have been identified in pencil beam star-forming regions. These observations can provide a crucial test for the structure and evolution of both low-mass hydrogen burning stars and brown dwarfs, and inform the threshold mass signifying hydrogen versus deuterium burning evolution.

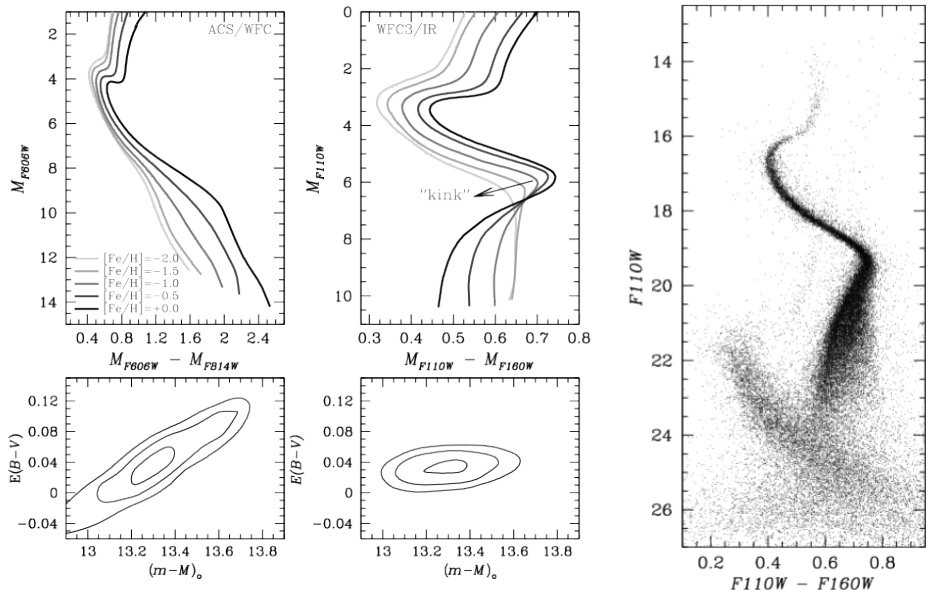

*Figure 8: Using the same stellar evolution models at a fixed age and metallicity (Dotter et al. 2007), we fit newly obtained WFC3/IR observations of 47 Tuc (right panel) and compare the resulting constraints on distance and reddening to a fit of the visual CMD of the star cluster from the ACS Survey of GCs (left panel) and the constraints derived from WFC3/IR data (center panel). The contours represent 1, 2, and 3σ best-fit values. As expected, with all other constraints fixed, degeneracies in the fundamental parameters of the GC are largely lifted in the WFC3/IR bandpasses given the "kink" on the lower main sequence caused by H2 (see Kalirai et al. 2012). By combining the constraints from the visible and (more accurate) IR data in quadrature, the "fitting" error in the age of 47 Tuc is improved by a factor of ~2.*

## 3) Map substructure in the Milky Way halo

Deep, wide-field IR imaging with NEW WFIRST will enable a new breakthrough in the characterization of the Milky Way's stellar halo. The stellar halo is old, and therefore the brightest sources in both substructure and dwarf satellites are red giants. The survey will be sensitive to the complete red giant branch in systems out to the edge of the Milky Way halo. This will lead to a new generation of high-resolution star-count maps from which the Galactic stellar density, surface brightness profile, chemical abundance gradients — through optical-near IR photometry and follow up spectroscopy — and level of substructure will be measured. These observed properties of the stellar halo are intimately linked to the accretion processes that built up the Milky Way. A NEW WFIRST IR survey will also enable the detection of a large population of low-luminosity dwarf satellites that are known to exist (Tollerud et al. 2008). The luminosity function of these dwarf spheroidals will provide the best test to date of the missing satellite problem

In addition to these general science themes, a NEW WFIRST IR survey of the Milky Way can enable a host of other transformational studies. For example, lower latitude investigations in the Galactic plane can provide the highest resolution mapping of the Milky Way disk and bulge to date. The IR imaging and spectroscopic sensitivity will penetrate dust-obscured regions and characterize the temperature and surface gravity of star-forming complexes out to the edge of the Milky Way. These surveys will also yield new insights on the structure of the Galactic disk, including the spiral structure. Surveys of the Milky Way can also extend SDSS studies of the remnant white dwarf luminosity function from the Galactic disk to the halo (e.g., Harris et al. 2006). The truncation of the remnant luminosity function is shaped by the formation time of the stars, and therefore such studies yield independent age constraints for the halo.

5.4.2 Extragalactic Stellar Populations

The scientific potential of extragalactic stellar population studies is large. Extragalactic stars have proven to be one of the most secure ways to probe the star formation history of a galaxy, and thus anchor the "ground truth" for constraining the astrophysical processes that drive galaxy formation, and for calibrating widely used techniques to derive star formation rates and histories over cosmic distances. Although main sequence stars remain the gold standard for such work, truly ancient main sequence stars are only routinely detectable within the Local Group. In contrast, luminous evolving stars are detectable to much larger distances, and probe stellar populations of all ages.

The main factor limiting stellar population studies is angular resolution. With ground-based telescopes, light from individual stars is blended together, prohibiting the study of stars beyond the local group. In contrast, the superb resolution available with a 2.4m space-based telescope allows resolution of individual red giant stars out to distances of ~5 Mpc in the main body of galaxies, or out to much larger distances in low surface brightness regions (halos, intracluster light, dwarf galaxies, etc). High luminosity stars, which more easily rise above the crowding limit, can also be detected out to distances greater than ~10 Mpc, allowing the detection of massive main sequence stars, blue and red supergiants, and asymptotic giant branch (AGB) stars (see Figure 9).

Over the last two decades, HST has transformed the study of extragalactic stellar populations, particularly in the optical. However, exploiting NIR stellar populations has largely been untapped, beyond an initial HST "snapshot" survey (Dalcanton et al 2012a, Melbourne et al 2012), the MCT survey of M31 (Dalcanton et al 2012b), 2MASS observations of the Magellanic Clouds (Nikolaev & Weinberg 2000) and small fields observed with adaptive optics from the ground (e.g., Melbourne et al 2010, Olsen et al 2006, Davidge et al 2005). Optical studies as well have been limited by the area accessible in a single ACS or WFC3/UVIS field of view. We now highlight just three of the many ways in which red optical and NIR wide-field imaging with WFIRST could prove transformative.

### *Tracing the History of Galaxy Accretion*

As Kathryn Johnston recently noted in her AAS keynote address, "Stars remember what gas forgets." In other words, once a star forms, it can retain an imprint of its initial kinematics for many dynamical times. Moreover, a star's color and luminosity allows it to be associated with a mass, metallicity, and evolutionary state, which in turn reveal the star's age and chemical enrichment.

The power of this approach has been realized most fully in photometric (and spectroscopic) studies of the extended halo of M31 (e.g., Guhathakurta et al. 2005; Ibata et al. 2007; McConnachie et al. 2009). Most recently, the PAndAS survey (McConnachie et al. 2009), has mapped red giant and AGB stars in a region spanning 300 kpc across M31. The colors of the individual stars allows them to be grouped in metallicity, thereby revealing rich substructure that traces the history of past accretion events. Unfortunately, this rich information is currently available for only three galaxies — M31, M33, and the Milky Way, the latter of which is complicated by our unique view from within the galaxy — making is difficult to derive generic constraints on the galaxy assembly process. Preliminary data is available within the halos of a larger number of nearby galaxies (Mouhcine, Ibata, and Rejkuba 2011; Radburn-Smith et al. 2011), but due to the limited field of view of HST, even a dozen sparse pointings sample only a small fraction of a typical halo. Comparable studies cannot be undertaken from the ground, because steeply rising galaxy number counts rapidly swamp the population of halo stars, requiring effective star-galaxy separation to faint magnitudes.

The wide field capabilities of DRM1 or NEW WFIRST have the potential to revolutionize this field. A series of 6 pointings with moderate exposure times (equivalent to 2-10 HST orbits) should reveal more than 1 magnitude of the red giant branch out to the virial radius for typical nearby galaxies, of which 100's are close enough to employ this technique (see Figure 10). This strategy would allow studies comparable to that in M31 to be carried out for galaxies spanning a nearly complete range in stellar mass, in a wide range of environments.

With a larger investment of time, comparable to typical Treasury programs on flagship missions, one could fully map the core of the Virgo cluster (D=16 Mpc), mapping the stellar populations in the intracluster light (ICL) across the central 2x2 deg$^2$ field (0.5 x 0.5 Mpc$^2$). Detection of individual stars in the ICL has already been demonstrated for individual pointings with HST (e.g., Williams et al. 2009), but scaling to a meaningful area is far beyond the capabilities of current capabilities.

### *Fully Characterizing the Local Volume*

There are ~450 galaxies cataloged within 12 Mpc (Karachentsev et al. 2004 – Catalog of Neighboring Galaxies), but the majority of them have no observations with HST resolution. They therefore lack secure distance measurements, in spite of the fact that many of these galaxies have been targeted by major flagship programs (SINGS, LVL, GALEX, S4G). Moreover, among the massive galaxies that do have HST observations, most lack sufficient areal coverage to fully characterize their stellar populations. A targeted key program by the NEW WFIRST could easily generate the definitive legacy data set on luminous evolving stars within this volume. These constraints on stellar populations would significantly augment the astrophysical processes that can be probed with existing multi-wavelength data. The same program could simultaneously measure tip-of-the-red-giant-branch distances, allowing a complete mapping of the flow field across a ~25 Mpc diameter volume.

### *Calibrating Models of Luminous Evolving Stars*

The interpretation of the integrated light of distant galaxies relies on spectrophotometric population synthesis models (e.g., Bruzual and Charlot 2003), and in turn, on the quality of the stellar models on which these are based. Among the major weaknesses in these models (e.g., binary star evolution, uncertain helium and metal-enrichment laws, etc.), the most obvious deficiency is in the description of the asymptotic giant branch (AGB) phase. At $z = 0$, the impact of AGB stars is most pronounced in the near-infrared (NIR), due to their red colors. However, at higher redshifts ($z \geq 1$), where galaxies are younger, AGB stars are more luminous and can potentially make up nearly half the luminosity at all restframe optical-to-NIR wavelengths (see Maraston et al. 2006; van der Wel et al. 2006). More recently, Dalcanton et al. 2012a and Melbourne et al. 2012 have pointed out the equally important role of massive red core Helium burning (RHeB) stars, which can contribute as much light as AGB stars in the NIR. These stars are sufficiently massive that they produce light on even shorter timescales than AGB stars, and can therefore produce rapidly varying NIR mass-to-light ratios.

Despite this important role, the present-day modeling of AGB and HeB stars is embarrassingly primitive and uncertain. The most basic aspects – the lifetimes and luminosities, the colors, the transition from O-rich to C-rich atmospheres for AGB stars – depend directly on processes like rotation, convective overshooting, mass loss, and convective "third dredge-up" episodes. These processes are far from being described by a theory based on first-principles (see e.g., Woitke 2006a,b; Herwig 2005), so currently, the best that one can do is assume simple parametric laws, and then calibrate them by means of synthetic models.

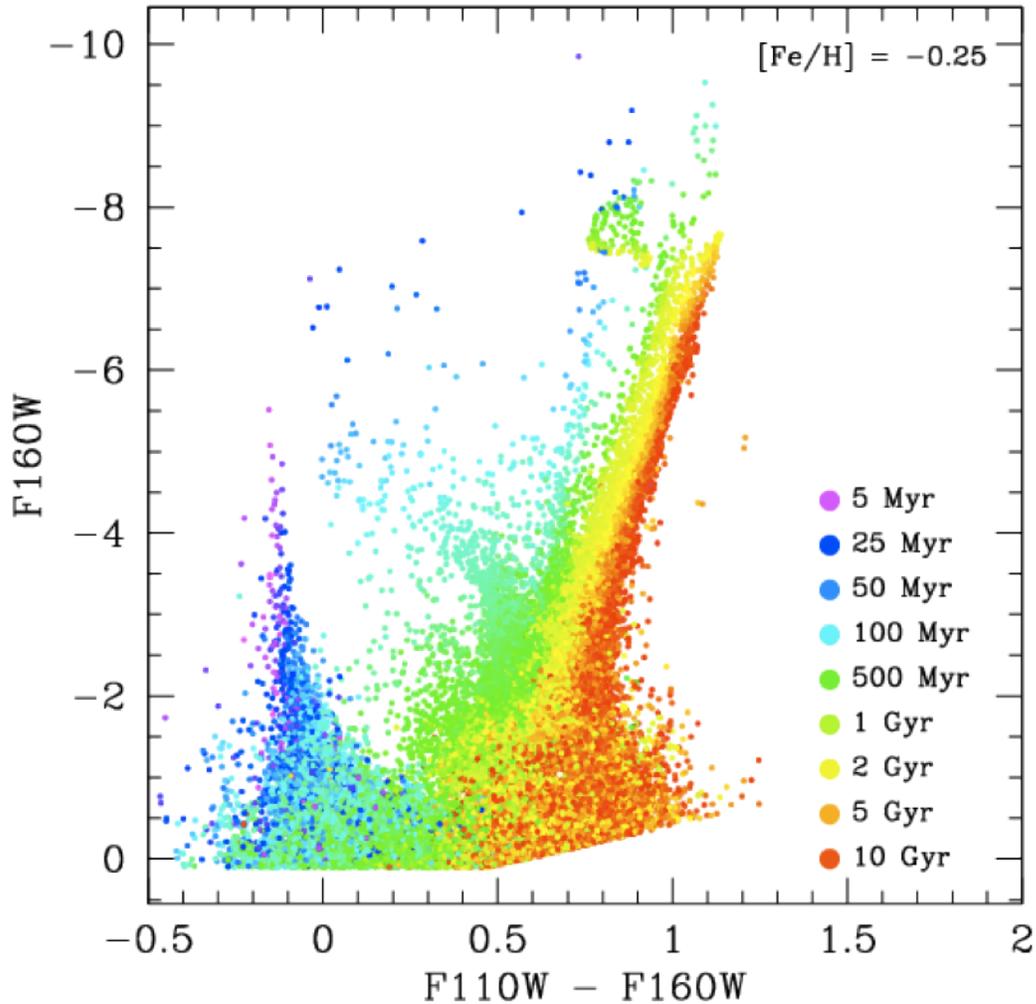

*Figure 9: Luminous NIR stellar populations as function of color and magnitude, color-coded by age for a continuous star formation history. Stars of all ages are detectable with modest exposure times for galaxies within 10-15 Mpc. Metallicity variations (not shown) affect the color of the red giant branch (1-13 Gyr), and the structure of the blue and red Helium burning sequences (25-500 Myr). AGB stars are found brighter than $M_{F160W} \approx -6$.*

Unfortunately, there are surprisingly few opportunities to actually calibrate these first-order models, in spite of their ability to make testable predictions. Because they are rapidly evolving, AGB and HeB stars are intrinsically rare, making them challenging to calibrate in individual clusters. Plenty of constraints can be derived from the stars in the Magellanic Clouds (extensively sampled by objective prism surveys, 2MASS, OGLE, MACHO, and Spitzer's SAGE and S3MC surveys), but other observational constraints at both lower and higher metallicities are inadequate; indeed, models of AGB stars have been tuned to work for the Magellanic Clouds only. Some progress should be made in the near future with the HST survey of M31, but this is only one system, and given the complexity of cool star atmospheres, the effects of chemical composition are expected to be significant. The limited calibration of these stellar models is therefore likely to be the dominant uncertainty in critical observational studies of galaxy evolution, which rely almost entirely on interpreting colors and luminosities of galaxies at high redshift. This problem increases in severity as NIR becomes the dominant mode of extragalactic imaging (e.g., WFC3/IR, JWST, MCAO).

With its wide field of view, NEW WFIRST has the potential to generate orders-of magnitude-increases in existing samples of NIR-bright AGB and HeB stars, thus providing essential calibration data for the rapidly evolving massive stars that dominate the light in the NIR. Moreover, a small pilot survey with WFC3/IR (Dalcanton et al. 2012) has shown that NIR stellar populations are far richer than previously believed, and thus should be as effective for deriving star formation histories as the better studied optical populations, once models on NIR-bright stars are properly calibrated.

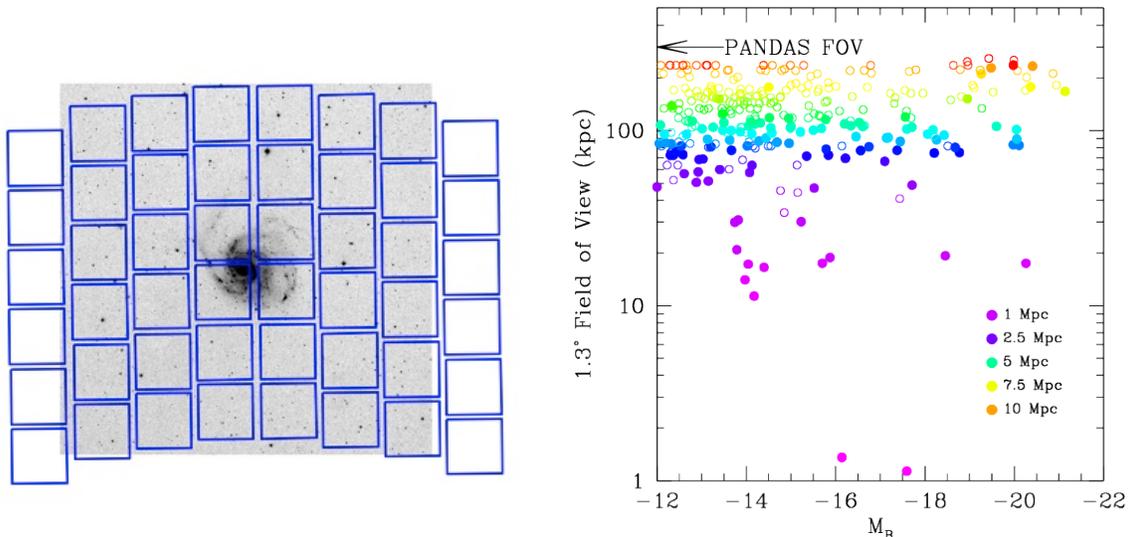

*Figure 10: Example mapping the extended structure of M101 with a 3-position pointing of the DRM2 FOV (left). The largest extent of the DRM2 FOV would cover a linear extent comparable to that of the PANDAS survey of M31's halo for galaxies at distances D ≥ 10 Mpc.*

5.5 Extragalactic Programs

5.5.1 Studying Galaxy Evolution with the Dark Energy Surveys of NEW WFIRST

Galaxies show remarkable diversity, yet at the same time they exhibit certain strong correlations and obey mysteriously tight scaling relations. Developing a detailed understanding of the physics that shapes observable galaxy properties within the framework of the dark-matter and dark-energy dominated universe remains one of the major challenges for astrophysics in the next decades. Large surveys such as the Sloan Digital Sky Survey (SDSS) have allowed us to exquisitely quantify these correlations for nearby galaxies, and by enabling us to "slice and dice" our sample into many bins in galaxy properties and environment, these surveys have provided insight into the physics by teasing out the 'main controlling variables' from those that are just along for the ride.

One of the most profound correlations seen in nearby galaxy populations is that between galaxies' *spectrophotometric* properties (probing their star formation history) and their *structural* and morphological} properties. For example, it is well known that early type galaxies locally tend to have more compact light profiles and also have red colors and little recent star formation, while late type, disk dominated galaxies have more extended light profiles and blue colors indicative of their ongoing star formation. A correlation with galaxy mass and environment is also seen, with red early types dominating at high masses and in rich environments, and blue late types dominating for lower mass systems and in less dense environments. SDSS revealed that galaxy mass (or internal density, or velocity dispersion) may actually be the more fundamental parameter, with much of the trend between galaxy type and environment coming indirectly from the correlation between mass and environment (massive galaxies are found preferentially in rich environments).

At high redshift, sub-arcsecond resolution is needed to study the internal structure of galaxies, which — for large samples over large fields — is only possible from space. In addition, studying galaxies in the rest-optical beyond $z \sim 1$ requires deep near-IR imaging over large fields, again, only feasible with a space telescope. Early results from Hubble's WFC3 have revealed that the galaxy population at $z > 1.5$ looks very different from today's — we commonly see extremely clumpy disks, ultra-compact and massive "red nuggets", and small blue 'tadpole' galaxies. Intriguingly, some of the correlations that are prevalent at $z = 0$ appear to be already in place at $z > 2$, namely there is still a strong correlation between specific star formation rate (sSFR) and galaxy structure (with compact galaxies in general having the lowest sSFRs). However, other correlations have disappeared or even reversed — for example, massive galaxies at $z > 2$ tend to be actively star-forming and disk-dominated. On the flip side, disks with prominent spiral or bar features, common locally, are extremely rare. There have been claims that the relationship between star formation rate and large scale environment also reverses at $z > 1$ (so that objects in denser environments have higher SFR than in the field), but these claims remain controversial, and it is impossible to tease out the 'driving parameters' with current samples.

The very large near-IR imaging surveys with NEW WFIRST that are planned for the Dark Energy program will make it possible to quantify — over most of the history of

galaxies — correlations between integrated galaxy properties (stellar mass, velocity dispersion), galaxy internal structure (size, light profile shape, morphology), and large scale environment, and to subdivide the samples into many bins in redshift, mass, and other quantities, in order to isolate the 'driving parameters.' In order to isolate the physical processes responsible for these correlations and to well populate these subdivisions at multiple redshifts, very large galaxy samples ($>10^7$) are required. In addition, wide fields are needed to collect large enough samples over the full range of environment, including even the rare, densest regions and the most under-dense voids. Moreover, the mass-distribution maps via weak lensing that will also be obtained with the survey will provide a unique and presumably more direct probe of the underlying distribution of dark matter, making it possible to map the relationship between the dark matter 'scaffolding' and luminous galaxies, up to the largest scales. In addition to providing invaluable constraints for galaxy formation models and simulations, this is a crucial for cosmological tests, such as the BAO (described Section 5.2.3), which rely on certain assumptions about 'bias' — how galaxies trace the underlying density field.

In summary, the contribution that the NEW WFIRST wide-field galaxy surveys will make to the study of galaxy evolution is enormous, comparable to the impact the SDSS has had on understanding the properties of today's galaxies.

5.5.2 Evolution of clusters of Galaxies

By imaging over 10,000 galaxy clusters with 0.11 arcsec resolution, including some of the most massive clusters in the early ($1.5 < z < 3$) universe, a NEW WFIRST survey would greatly enhance our knowledge of the evolution of cosmic structure and test competing cosmological models. Studies of high-redshift rich clusters have already begun with HST+WFC3 (see Figure 11); the wide field surveys that would come from NEW WFIRST would vastly expand the power of this approach.

In our standard cosmological model, structure formation proceeds from the smallest to the largest units. We believe that stars formed when the universe was less than 100 million years old but that another ~100 million years passed before they began to assemble into the first proto-galaxies. Clusters of galaxies, the largest gravitationally bound objects to have formed so far, started to emerge only 1 Gyr later, and continued to assemble over a broad range in redshifts. Observationally, clusters and proto-clusters in formation have now been identified at $1.5 < z < 4.1$ (Miley et al. 2004, 2006; Stanford et al. 2006; Gobat et al. 2010).

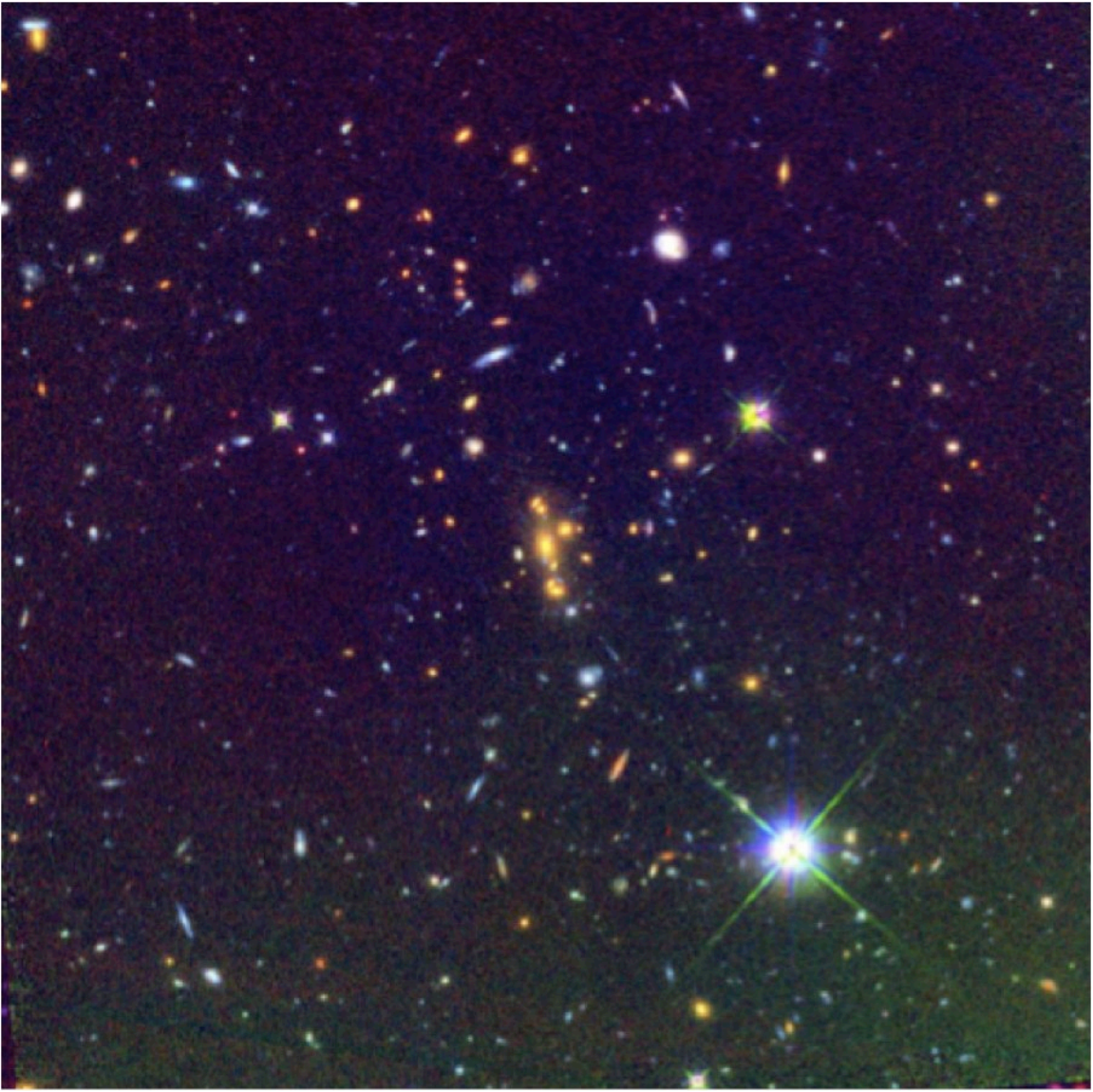

*Figure 11: One of the most distant massive clusters discovered to date at z = 1.4 with mass $M_{vir} \approx 8 \cdot 10^{14}$ M⊙ determined from both lensing and X-ray data. The galaxy population is dominated by old, red spheroids, similar to low redshift clusters. The image is a color composite from HST/ACS (i+z), WFC3 (J), and VLT Ks, covering 2 arcmin (corresponding to ~1 Mpc) across.*

With evolutionary histories that are highly sensitive to the cosmological parameters and the physical processes responsible for structure formation, galaxy clusters are a sensitive probe for dark energy (DE) evolution and non-canonical cosmologies. The latter point is of great interest: distinguishing a time-varying DE component from the cosmological constant is one of the most pressing questions in modern theoretical and observational astronomy, with profound implications for fundamental physics.

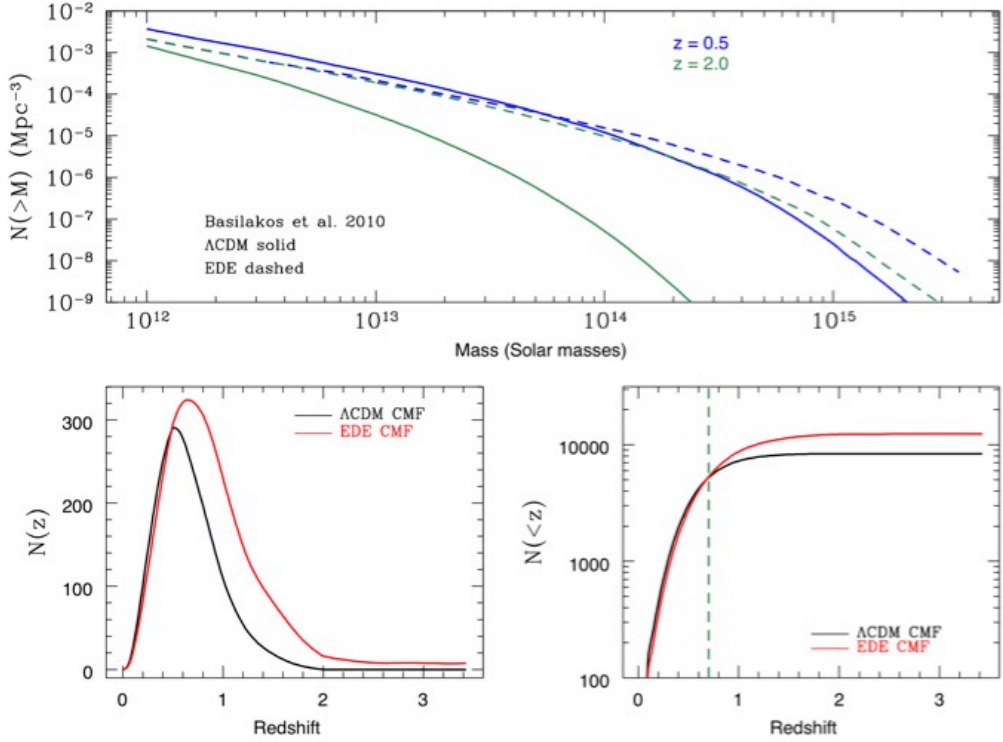

*Figure 12: (top) Cluster Mass Function for the standard ΛCDM (blue) and a proposed Early Dark Energy model (red line). (lower left and right) The predicted differential and cumulative number of clusters, respectively, in the two scenarios for the NEW WFIRST high latitude survey. CMFs are normalized to the observed cluster abundance at z < 1, and then extrapolated according to the respective cosmological models. A 3,400 deg² survey area, and a cluster mass selection function that scales as the evolution of the characteristic L\* (H-band) magnitude, are assumed.*

In this context, *massive* ($M > 10^{14}$ $M_\odot$) high-redshift ($z > 1$) clusters are particularly important (Jee et al. 2009; Rosati et al. 2009; Brodwin et al. 2010; Mortonson et al. 2010; Holz & Perlmutter 2012). The reasons are conceptually simple: for instance, in Early Dark Energy (EDE) models (Grossi and Springel 2009; Basilakos, Plionis, and Lima 2010) the contribution of DE to the total energy density of the universe does not vanish at high redshift, unlike in standard ΛCDM models or other simple scenarios (such as "quintessence"). The presence of a non-negligible DE at *z > 2* would require accelerated structure formation in order to match the cluster abundance observed at z ~ 0. EDE models thus predict that massive clusters form significantly earlier than they would in a ΛCDM cosmology (see Figure 12). Similarly, deviations from gaussianity of the primordial density fluctuations would lead to an enhanced probability of forming massive collapsed structures at high redshifts (Sartoris et al. 2010). As Figure 12 shows, the cluster densities predicted by these cosmological models differ most dramatically for massive ($> 10^{14}$ $M_\odot$) clusters at $z > 1.5$. *This is precisely the regime NEW WFIRST will have the unique capability of exploring.*

Detecting the *intrinsically rare, massive* clusters at $z > 1.5$ is a formidable challenge, requiring large areas of the sky to be surveyed in the NIR at depths sufficient to detect a significant number of cluster members. A 3,400 deg$^2$ survey to a limiting H-band magnitude of 27 will meet the challenge: by extrapolating the evolution of L* (the luminosity of a "typical" galaxy (De Propris et al. 2010), the NEW WFIRST high-latitude survey will probe galaxies at least 3 magnitudes fainter than L* in clusters as distant as $z \sim 2.5$, making the clusters "stand out" against the sea of more nearby galaxies. (For comparison, even the brightest cluster members in such distant clusters will not be detected by VIKING, the deepest of VISTA's wide survey.)

The NEW WFIRST cluster sample will be a great resource that can be mined for follow-up observations. Lensing, dynamical, and/or Sunyaev–Zel'dovich cluster masses will be measured with JWST, GSMT, and/or future sub-millimeter and X-ray facilities; the redshift evolution of the resulting cluster mass function (CMF) will provide sound statistical constraints on cosmological models. In a ΛCDM scenario, of the 10,000 clusters that we expect to detect, only one or two, if any, will be $z > 2$, whereas 1000 times as many are expected if a non-negligible DE is present at high redshifts (Figure 12). For clusters more massive than $5 \times 10^{14}$ M☉, no clusters are expected at $z > 2$ in the ΛCDM scenario, but ~130 are expected in the EDE scenario (Sartoris et al. 2010).

We conclude by pointing out that although NEW WFIRST is unique in its ability to identify massive, distant clusters, it will also detect a multitude of more common, less massive clusters, down to $5 \times 10^{13}$ M☉, in the critical range of $1 < z < 3$ (i.e., at a time when the star formation rate peaks). NEW WFIRST will therefore play a fundamental role in advancing our understanding, not only of the formation and evolution of galaxy clusters (by witnessing cluster assembly and the emergence of well-known scaling properties in clusters), but also of massive early-type galaxies, which are a fundamental tracer of the mass assembly history of the baryonic matter in the universe.

5.5.3 Co-Evolution of Galaxies and AGN.

NEW WFIRST will map the co-evolution of galaxies and AGN out to high redshifts, covering the peak epoch of star formation and AGN activity in the universe at $1 < z < 2$. Bright, unobscured sources will allow for redshift and black hole mass measurements from their (slitless) grism spectra, while obscured quasars will have spectroscopic redshifts (for bright sources) and improved photometric redshifts (for fainter sources). NEW WFIRST will enable unprecedented studies of AGN host galaxy properties, including measuring their morphologies as well as their dark matter halo properties from clustering and weak lensing studies. Such analyses will be split by AGN type (e.g., obscured vs. unobscured; luminosity class) as well as AGN selection method. For instance, recent work, based on samples of approximately 300-500 AGN, has shown that while radio-selected AGN tend to reside in massive, early-type galaxies, optical and mid-infrared selected quasars tend to reside in "blue cloud" galaxies, e.g., lower mass, late-type galaxies (Coil et al. 2007; Hickox et al. 2009; Griffith and Stern 2010). X-ray selected AGN samples tend to straddle these distributions. NEW WFIRST will extend such studies with samples many orders of magnitude larger, allowing evolutionary studies split by AGN luminosity.

Clustering and weak-lensing analyses of matched obscured and unobscured populations will allow the study of AGN halo properties and thus test unification schemes. The standard torus model (e.g., Urry and Padovani 1995) posits that orientation is the only parameter that differentiates obscured and unobscured AGN, thereby implying identical halo properties. In contrast, dynamical models (e.g., Hopkins et al. 2005) posit that the obscuration comes from a post-merger evolutionary phase, implying that halo properties could depend on obscuration. Large samples will also allow studies of how obscuration depends on luminosity. For example, the physically motivated "receding torus model" (e.g., Simpson 2005) implies that higher luminosity quasars impact their obscuring torus, providing more sightlines to the central engine, such that the fraction of unobscured quasars increases with increasing AGN luminosity (e.g., Treister et al. 2008; Assef et al. 2012). NEW WFIRST will provide data sufficient to test these models.

## 5.6 Studying Cosmic Dawn

The three top-level science goals given by the *NWNH* decadal survey report are "Cosmic Dawn," "New Worlds," and the "Physics of the Universe." WFIRST already addresses the Physics of the Universe through its dark energy surveys, and New Worlds through its microlensing survey. A NEW WFIRST mission could capably address the central question of Cosmic Dawn, by characterizing the epoch, speed, and inhomogeneity of cosmological reionization, and also characterizing the sources responsible for reionization. All of this can be done with the instrumentation planned for the dark energy experiments, and much of it via the dark energy surveys and deep field observations that can be incorporated into regular calibration efforts.

### 5.6.1 Quasars in the Epoch of Reionization

The high latitude extragalactic survey will identify unprecedented numbers of quasars at very high redshift, probing the epoch(s) of reionization, mapping the first structures to form in the universe, and providing unique probes of the intergalactic medium. Quasars are among the most luminous objects in the universe, observable to the earliest cosmic epochs. The discovery of a Gunn-Peterson (1965) trough in the spectra of several quasars at redshift $z > 6$ implies that the universe completed reionization near that redshift (e.g., Fan et al. 2001), though poor sample statistics, the lack of higher redshift quasars, and the coarseness of the Gunn-Peterson test make that inference somewhat uncertain. Currently, the most distant confirmed quasar is at z = 7.1 (Mortlock et al. 2011), and only a dozen quasars have been confirmed at z > 6. Ambitious surveys will push this number to around 100 in the next few years, likely identifying one or two quasars at z ~ 8.

NEW WFIRST will fundamentally change the landscape of early universe investigations. Willott et al. (2010) provide the best measurement to date of the high-redshift quasar luminosity function. Their measurement includes a robust sample of quasars at z ~ 6 and reasonably extrapolates the decrease in quasar space densities seen at z > 3 to higher redshifts. Using this luminosity function, the baseline wide-area NEW WFIRST survey (10,000 deg$^2$ to AB ~ 26) should contain 3400 quasars at z > 7, 900 at z > 8, 240 at z > 9, 65 at z > 10, and 5 at z > 12 (albeit with increasing uncertainty at the highest redshifts). If the increased capabilities of the NEW hardware were spent on depth rather than area, instead reaching AB ~ 26.6 over 3400 deg$^2$, these numbers would drop by a factor of ~ 2 at all redshifts. For comparison, Euclid is expected to identify 1370 quasars at z > 7 and 20 quasars at z > 10 using the same luminosity function (see Table 3).

Table 3: The First QSO's with NEW WFIRST

| Survey | Area (deg$^2$) | Depth (5-sigma, AB) | z>7 QSO's | z>10 QSO's |
|---|---|---|---|---|
| UKIDSS-LAS | 4000 | Ks=20.3 | 8 | - |
| VISTA-VHS | 20,000 | H=20.6 | 40 | - |
| VISTA-VIKING | 1500 | H=21.5 | 11 | - |
| VISTA-VIDEO | 12 | H=24.0 | 1 | - |
| Euclid | 15,000 | H=24.0 | 1359 | 22 |
| WFIRST SDT DRM1 | 3400 | K=26.0 | 1304 | 25 |
| NEW WFIRST | 10,000 | H=26.0 | 3440 | 65 |

*Table 3: Number of very high-redshift quasars predicted for various ground- and space-based near-infrared surveys, based on the quasar luminosity function of Willott et al. (2010). Note: 6.6% of the Euclid wide survey will have two or fewer dithered exposures per sky position. Such data will lack the redundancy required to robustly identify rare objects, and thus have been omitted in the above calculations.*

Such discoveries will directly measure the first epoch of supermassive black hole formation in the universe, probe the earliest phases of structure formation, and provide unique probes of the intergalactic medium along our line of sight to these distant, luminous sources. The large numbers of quasars identified in the first Gyr after the Big Bang will enable clustering analyses of their spatial distribution (e.g., Coil et al. 2007; Myers et al. 2007), thereby probing the mass of the dark matter halos in which these early sources reside. Spectroscopic follow-up of such a sample would yield another probe of reionization by studying the sizes of their surrounding ionized bubbles, or Strömgren spheres. Studies of a large sample of very high redshift quasars, rather than the single object studies done to date (e.g., Bolton et al. 2011; Mortlock et al. 2011), would allow us to study the "topology of reionization" (e.g., Furlanetto et al. 2004). Optical surveys are not capable of identifying quasars above $z \sim 6.5$ since their optical light is completely suppressed by the redshifted Lyman break and Lyman-$\alpha$ forests. Though JWST will be extremely sensitive at near-infrared wavelengths, it will not survey nearly enough sky to find the rarest, most distant, luminous quasars, which have a surface density of just a few per 10,000 deg$^2$.

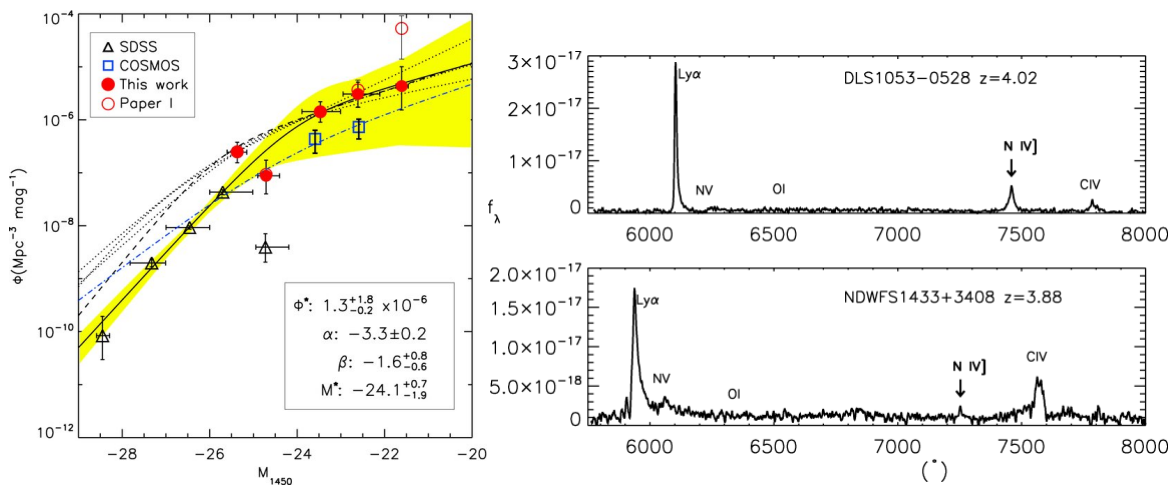

*Figure 13: (left, from Glikman et al. 2007) QSO luminosity function (QLF) at $z \sim 4$, combining data from SDSS (triangles), COSMOS (blue squares; Ikeda et al. 2011) and the Bootes field (filled red circles; Glikman et al. 2011). Note the similar faint end slopes but differing normalizations for the latter two. The solid line shows the best-fit double power law; parameters listed in lower right. (right, from Glikman et al. 2011) Keck-LRIS spectra of two faint QSOs at $z \sim 4$ with unusually strong N IV] 1486 emission. This emission line is very rare in bright QSO samples such as in the SDSS, but appears in ~9% of faint quasars at $z \sim 4$ in the study of Glikman et al. (2007).*

NEW WFIRST will also uniquely probe the faint end of the quasar luminosity function (QLF) at high redshift. Such studies are important for understanding the sources responsible for reionization, the role of AGN feedback in early galaxy formation, as well as the accretion history of black holes in the universe. Recent studies of the faint end of the QLF at $z \sim 4$ have found similar faint end slopes, $\beta \sim$

–1.6 (e.g., Glikman et al. 2011, Masters et al. 2012), albeit it with different normalizations (Figure 13, left). This work implies that quasars could account for about half the radiation needed to ionize the intergalactic medium at this redshift. Work to date also suggests that new AGN populations appear at the faint end of the QLF. Specifically, Glikman et al. (2007) found unusually strong NIV] 1486 emission in ~9% of their faint z ~ 4 quasars (Figure 13, right). Though not unprecedented, NIV] 1486 emission is extremely rare in the SDSS quasar sample (e.g., Bentz et al. 2004). The Lynx arc, at z = 3.357, has this same emission line, interpreted by Fosbury et al. (2003) as being powered by photoionization by very massive young stars and a top-heavy initial mass function (IMF; e.g., dominated by massive stars). This suggests an early co-evolution of galaxies and their supermassive black holes.

Finding both the brightest quasars at the highest redshifts as well as lower luminosity quasars over a wide range of redshifts, NEW WFIRST will be the definitive probe of the early phase of supermassive black hole growth in the universe.

5.6.2 Finding Gravitationally Lensed Galaxies to Probe the Early Universe

By imaging ~3,400 deg$^2$ to depth of 27.5 AB mag in 4 NIR passbands, the NEW WFIRST high-latitude survey will potentially contain more than 300,000 z > 9 galaxies in the field and more than 5,000 highly magnified z > 9 galaxies behind strongly lensing clusters. A complementary IR survey to 26 AB mag at wavelengths longer than 2$\mu$m will be required to unambiguously measure the photometric redshifts of these very young galaxies. The large number of such objects identified by the NEW WFIRST telescope will provide stringent constraints on the early star formation rate density and on the nature of the sources responsible for re-ionization of the universe.

Mapping the formation of cosmic structure in the first 1 billion years after the Big Bang is essential for achieving a comprehensive understanding of star and galaxy formation. Little is known about this critical epoch when the first galaxies formed. Determining the abundance and properties of such early objects is critical in order to understand (1) how the first galaxies were formed through a hierarchical merging process; (2) how the chemical elements were generated and redistributed through the galaxies; (3) how the central black holes exerted influence over the galaxy formation; and (4) how these objects contributed to the end of the "dark ages."

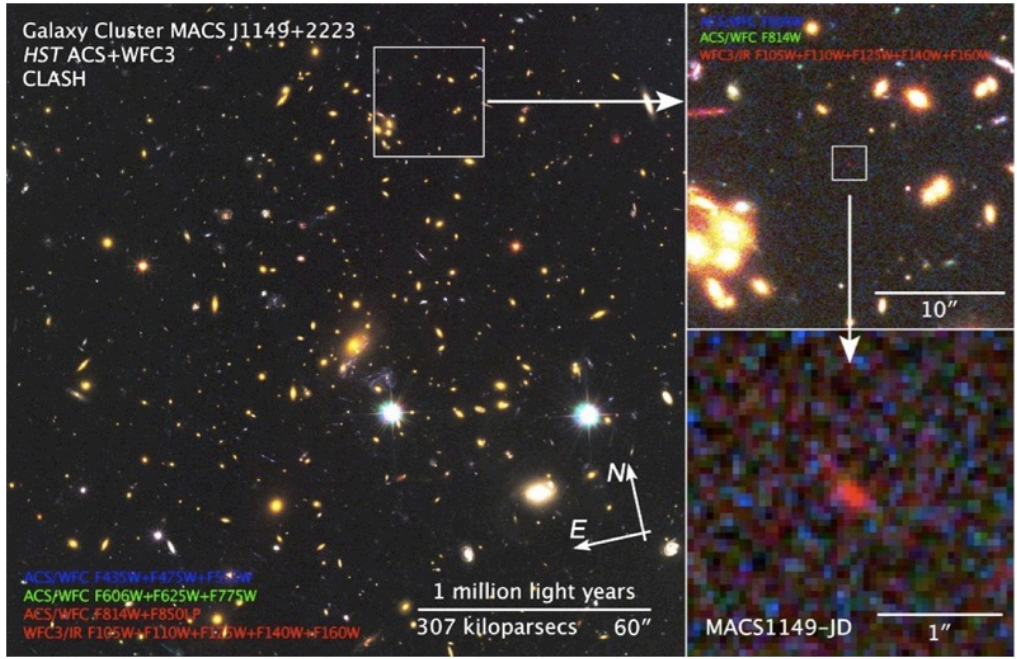

*Figure 14: A candidate for a galaxy at z = 9.6, magnified by a factor of ~15 by the foreground cluster MACS J1149+2223 (z = 0.54). The object was found in an HST survey using the WFC3IR camera (Zheng et al. 2012). This young object is seen when the universe is only ~500 million years old.*

Finding candidate z > 7 galaxies has been pursued in two ways — from moderate area but very deep imaging surveys (e.g., UDF, Subaru Deep Field, CANDELS) and from images of strongly lensing clusters of galaxies (e.g., Bradley et al. 2008; Postman et al. 2012). Recent observations have identified at least 3 candidate objects at z > 9 (Bouwens et al. 2011a; Zheng et al. 2012; Coe et al. 2012). Two of these candidates were found in a survey of massive galaxy clusters. One of the lensed candidates is shown in Figure 14. While unlensed sources at these redshifts are expected to be extremely faint, with total AB magnitudes greater than 28, lensed sources can be 10 – 30 times brighter. This is particularly important because it puts some of these sources within the reach of the spectrographs on the James Webb Space Telescope and from very large ground-based telescopes. Luminous z > 7 galaxies are extremely valuable as their spectra can be used to determine the epoch of the IGM reionization. This is because only a tiny fraction of neutral hydrogen is needed to produce the high opacity of Lyα observed at z ~ 6. The damped Lyα absorption profile that results from a completely neutral IGM (Miralda-Escude 1998) can be measured even at low-spectral resolution. Furthermore, direct measurements of the early star formation rate (via Lyα and Hα emission, Iye et al. 2006) can be derived from the spectra of bright high-z galaxies.

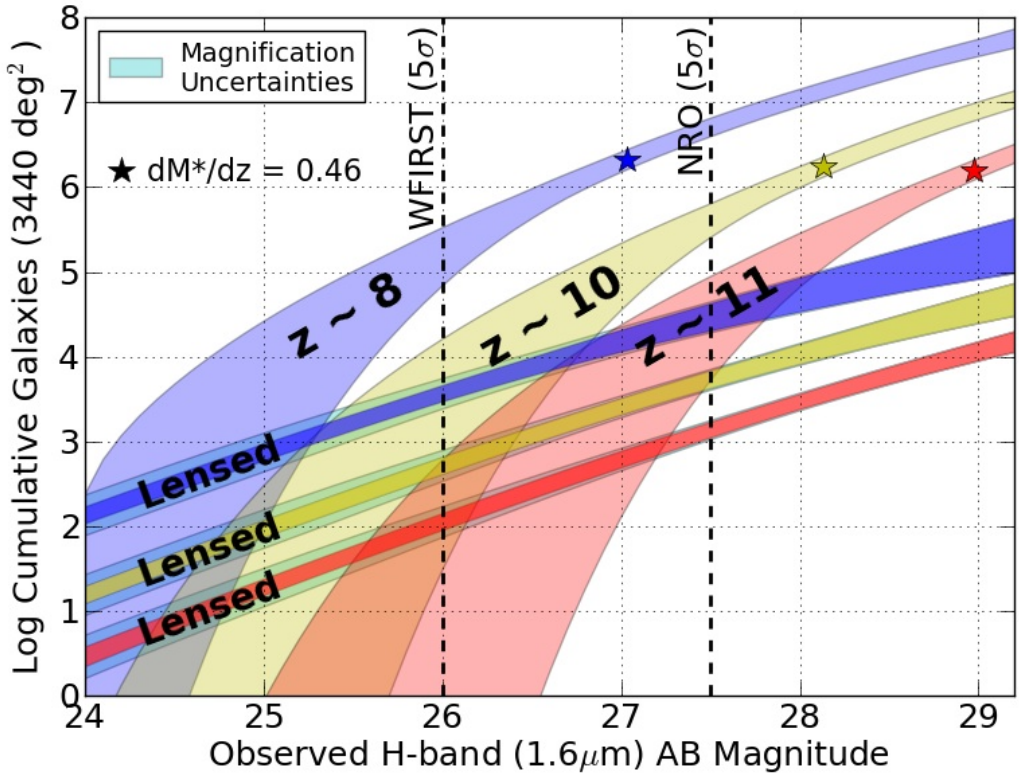

*Figure 15: The cumulative number of lensed and unlensed z = 8, 10, and 11 galaxies predicted in the NEW WFIRST HLS. Darker curves correspond to the lensed number counts. Lensing is advantageous primarily for finding bright (H < 25.5 AB mag) z > 9 sources. The unlensed field will produce an overall higher number of z > 8 sources, however.*

The high-latitude survey (HLS) performed by the NEW WFIRST telescope as part of its weak lensing program will also be superb for finding and studying objects in the early universe. A key advantage that a 2.4 meter aperture provides is that, in the same amount of time, the NEW WFIRST HLS will reach 1.5 magnitudes fainter than the corresponding DRM2 HLS. The final 5-sigma limiting depth of the NEW WFIRST HLS is estimated to be H = 27.5 AB mag. Figure 15 shows the predicted number of z = 8, 10, and 11 objects that would be found as a function of limiting H-band magnitude for both gravitationally-lensed and unlensed regions in a 3,440 deg$^2$ survey. The predicted number at z = 8 is constrained by the BoRG survey z ~ 8 luminosity function results (Bradley et al. 2012). The predicted counts for z = 10 and z = 11 sources are based on extrapolations of the z = 8 results, assuming that the characteristic magnitude of galaxies evolves as dM*/dz = +0.46. The lensed source count predictions assume ~1 strongly lensing cluster per square degree. The mass models were adopted from the CLASH Multi-cycle treasury program (Postman et al. 2012). Figure 16 shows the observed SED of the z = 9.6 candidate from Zheng et al. (2012). Reaching a limiting magnitude of H = 27.5 makes a significant improvement in the ability to detect such sources.

A sample of upwards of 5,000 luminous lensed galaxies will place very stringent limits on the star-formation-rate density, on the amount of ionizing radiation per unit volume, and on the physical properties of early galactic structures. An NRO telescope's 0.1 – 0.2 arcsecond resolution, combined with lens magnifications in the 10 – 30 range, will allow the mapping of structures on scales of 20 to 50 parsecs, thanks to the boost in spatial resolution provided by the cluster lenses. Such a survey will certainly herald a remarkable era in probing the first 1 billion years of cosmic history.

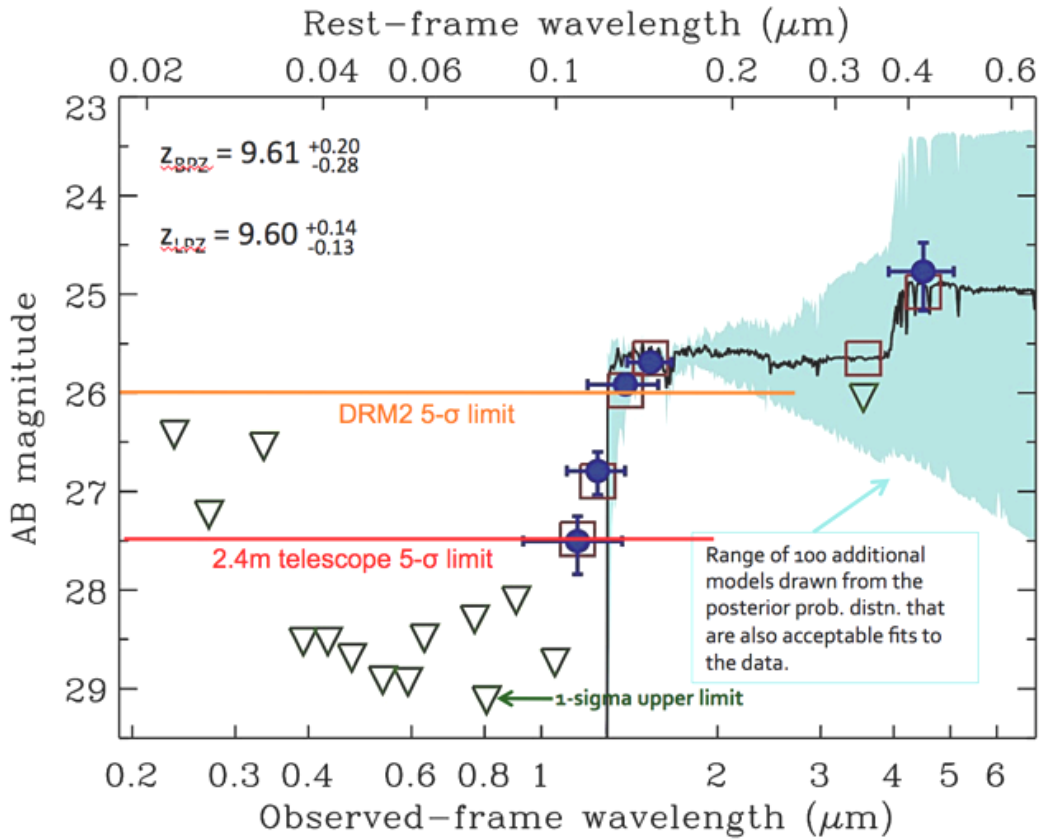

*Figure 16: The observed (blue points, open triangles) and best-fit (black curve) SEDs for the z = 9.6 galaxy in MACS J1149+2223.*

5.6.3 Ly$\alpha$ emitters as probes of reionization

Ly$\alpha$ lines from high redshift galaxies should be suppressed when these galaxies are surrounded by neutral inter-galactic medium (IGM), due to resonant scattering of the Ly$\alpha$ photons by neutral hydrogen. After accounting for galactic winds, ionized bubbles around galaxies, and other mitigating factors, it is still expected that Ly$\alpha$ line emission is suppressed by a factor of three or more (e.g., Santos 2004) for an IGM that is ~50% neutral. The simplest test involving Ly$\alpha$ galaxies is to compare the luminosity functions of Ly$\alpha$ galaxies before and after the redshift of reionization (e.g., Malhotra & Rhoads 2004; Stern et al. 2005). Recent data indicate a less than 10% neutral fraction at z = 6.5 (Ouchi et al. 2010). This is a local test of whether the IGM is neutral at the 10-50% level. It thus complements the more global CMB polarization tests, and the Gunn-Peterson test, which probes lower neutral fractions (~1%).

Higher order tests of reionization measure clustering of Ly$\alpha$ sources (e.g., Furlanetto et al. 2006; McQuinn et al. 2007). In the intermediate phases of reionization, when the ionized bubbles have not yet overlapped, Ly$\alpha$ galaxies should show enhanced angular clustering. This test has the advantage of being insensitive to number or luminosity evolution of the Ly$\alpha$ galaxy population. It also gives extra information about the preferred scale of ionized bubbles. NEW WFIRST, with its NIR coverage and wide-field spectroscopy, can measure the luminosity function of Ly$\alpha$ galaxies between redshifts z = 6 – 12. It has a tremendous advantage in NIR sensitivity over ground-based instruments, which are hampered by the earth's atmosphere. It wins over JWST by its FOV, which is 150 and 268 times the area covered by NIRSPEC and NIRISS. It wins over Euclid by having 4 times the aperture.

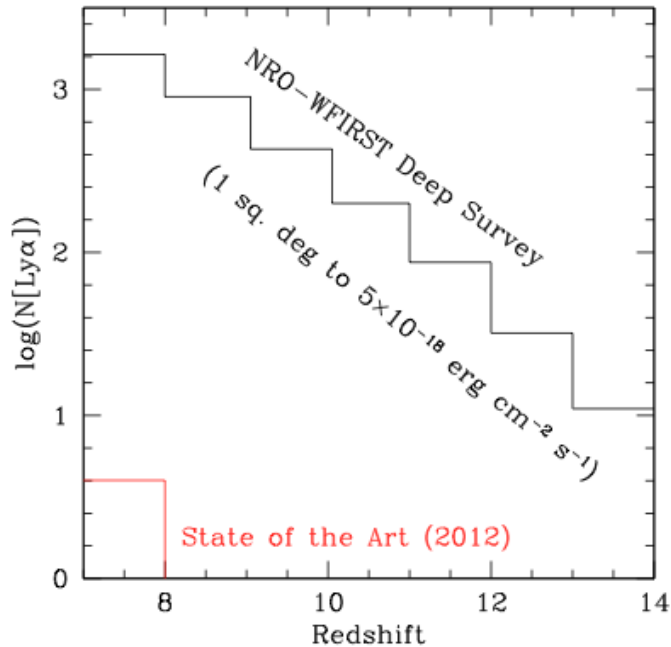

*Figure 17: Number of Ly$\alpha$ emitting galaxies that would be expected in an NEW WFIRST "Cosmic Dawn" field. The numbers are based on the z = 6.5 Ly$\alpha$ luminosity function from Ouchi et al. (2010), together with an assumed grism sensitivity limit of $5\times10^{-18}$ erg cm$^{-2}$ s$^{-1}$. Reionization would be manifested as a break in the number-redshift curve near the redshift where the neutral fraction exceeds 50%. Also shown, for comparison, is an approximation of the current observational situation in 2012, where we have a sample of about 5 known Ly$\alpha$ emitting galaxies at 7 < z < 8, and no robust detections beyond that.*

We now consider a strawman reionization test using Ly-α emitters. To get the best statistics, one needs to get deep enough to reach L* in the Lyα luminosity function. At redshift z = 8.5, that implies a sensitivity to $5 \times 10^{-18}$ ergs cm$^{-2}$ s$^{-1}$. A $10^6$ s exposure with the slitless spectrograph should be sufficient to reach that level (scaling from Euclid sensitivities). With that we should be able to observe 900 Lyα emitters per square degree per unit redshift at z ~ 8.5 (see Figure 17), and about 3000 from z = 7 – 12. Three NEW WFIRST fields distributed over the sky would suffice. These could, for example, be fields that are visited for short exposures periodically in order to keep track of absolute calibration drift over time. These statistics and areal coverage should be sufficient to apply the clustering test.

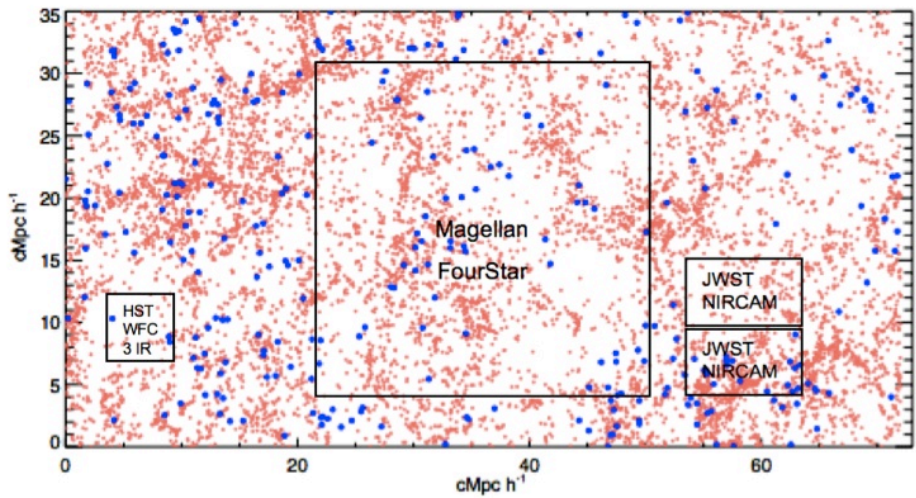

*Figure 18: Excerpt from a cosmological simulation of Lyα galaxy formation and growth (Tilvi et al. 2009). The field plotted corresponds approximately to 38' x 18' of sky coverage for the simulated redshift (z = 6.5) and cosmology (h=0.72). Red dots show the location of collapsed dark matter halos, and blue dots those that currently host a Lyα emitting galaxy. Field sizes of deep near-IR survey instruments are overplotted: WFC3-IR on HST (2.3'x2.3'); NIRCAM on JWST (4.6' x 2.3'); and FourStar on the 6.5m Magellan telescopes (11' x 11'. Two adjacent NIRCAM fields are shown to illustrate the potential impact of field-to-field variations: One contains no Lyα galaxies, the other, about 15. The larger FourStar field of view provides some robustness to local density variations, but lacks the depth offered by space-based near-IR imaging. The field of the nominal NEW WFIRST instrument is not marked, being about 1.5 times larger than the entire plotted area.*

Slitless spectroscopy enables a uniform and unbiased search for both Lyα emitters and Lyman break galaxies at the redshifts where the Lyα line or break (1216 Å) lies in the wavelength coverage, with no pre-selection on color, etc. Also, the area covered by even three deep NEW WFIRST fields gives a robust statistical sample of high-redshift galaxies, on scales sufficient to avoid strong cosmic variance effects. Figure 18 shows Lyα galaxies in a cosmological simulation by Tilvi et al.

(2009). The area of WFIRST (0.375 deg² = 119 cMpc at z=8 on the side) samples a range of regimes at that redshift. The smallest ionized bubble that will allow Lyα to escape is about 1 pMpc (Malhotra and Rhoads 2006), and the bubble size distribution should peak at about 1.5 pMpc (Furlanetto et al. 2004). Thus, a single field of view of WFIRST will contain several hundred bubbles between z = 7.5 – 8.5, assuming an ionized fraction around 50%.

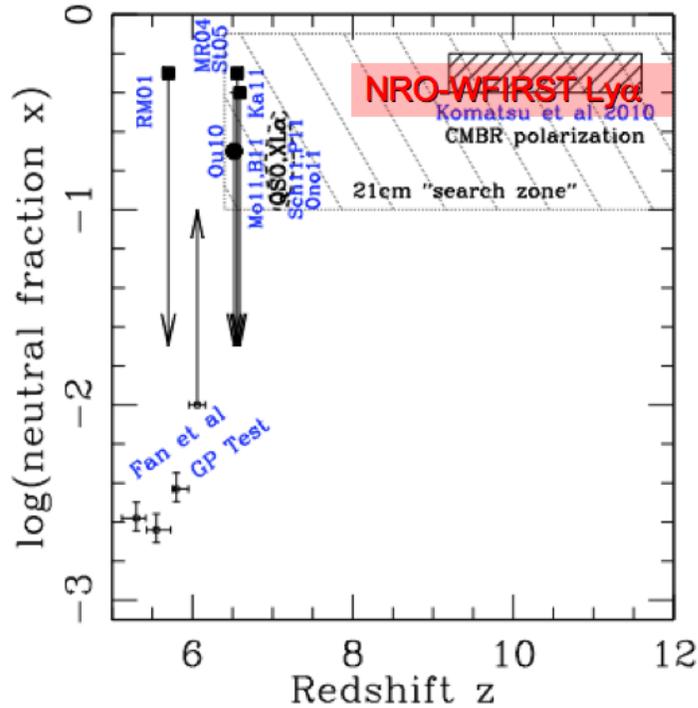

*Figure 19: Reionization history— the IGM neutral fraction as a function of redshift, as constrained by several present and future methods. The measurements at z < 6 and lower limit at z = 6.0 come from quasar Gunn-Peterson test results (Becker et al. 2004, Fan et al. 2006). Lyα emission studies provide complementary upper limits at 5.7 < z < 7 (RM01: Rhoads and Malhotra 2001, MR04: Malhotra and Rhoads 2004; St05: Stern et al. 2005; Ou10: Ouchi et al 2010; Ka11: Kashikawa et al. 2011). Lyα equivalent width studies (Sch11: Schenker et al. 2011; Ono11: Ono et al. 2011; P11: Pentericci et al. 2011) and the near zone studies of the z=7.08 quasar (Mo11: Mortlock et al. 2011; B11: Bolton et al 2011) provide recent evidence for neutral gas at z=7, as do some Lyα searches (Iye et al. 2006). At higher redshift, microwave background polarization (Komatsu et al. 2010) gives an integral constraint on the free electron column density and tells us that the neutral fraction was about 50% somewhere between z = 9 and 11.5. Lyα studies using NEW WFIRST will probe this range of neutral fractions and redshifts, enabling us to pin down the history of reionization at the crucial epoch and in great detail. They will provide a crucial second method to compare with future 21cm radio studies, which are a very promising method but whose application depends on successful mitigation of extremely bright foreground signals.*

The role of WFIRST in Ly$\alpha$ based reionization tests is summarized in the context of our present knowledge in Figure 19. Such tests could be accomplished within the NEW WFIRST wide area survey, provided that 1-2% of the total survey time is spent revisiting about 1 deg$^2$ of calibration fields. The benefit of such fields to the wide area survey would be long term monitoring of slitless spectroscopy performance, which would help with the calibration and data uniformity required for precision cosmology. If such repeated fields are not ultimately included in the dark energy surveys, the reionization science could be done through a dedicated program on the scale of 1 or 2 x 10$^6$ s. In other words, we could determine the history of reionization for an investment of telescope time comparable to the Hubble Ultra Deep Field or a similar treasury-scale HST GO program.

5.6.4 Lyman break galaxy samples from WFIRST

NEW WFIRST can offer an essentially complete census of Lyman break galaxies at redshift z = 8. Consider the combined data from the full supernova survey, as described above. Six square degrees, three filters, and an aggregate 0.45 years of telescope time translates to just under 300 ksec of integration time per field per filter. Scaling from z = 8 Lyman break studies by Bouwens et al. (2011), based on WFC3-IR data, this should achieve 5$\sigma$ sensitivities of 30.3 AB magnitude, while L* is between 27 and 28 mag. Presuming clean selection of LBGs down to a 10$\sigma$ flux level, the Bouwens et al. LF suggests that 6 deg$^2$ should include about 10$^5$ Lyman break galaxies per unit redshift around z = 8 – 9. With such large numbers, estimates of the ionizing photon production enter the "systematic dominated" regime, where uncertainties in the ionizing photon production (due to incomplete knowledge of stellar populations), the extrapolation of the luminosity function to fainter levels, and the ionizing photon escape fraction become dominant. While progress on some of these issues may require JWST follow-up spectroscopy, some may be partially addressed using NEW WFIRST itself, if deep spectroscopy accompanies the supernova imaging program. The inclusion of a filter wheel with filters beyond the basic Y-J-H trio could also provide for more flexible Lyman break surveys. In particular, inclusion of a deep $K_s$ imaging capability would allow the J-drop Lyman break galaxies to be confidently selected.

If NEW WFIRST were to search for supernovae through *spectroscopic* monitoring, using an R ~ 75 slitless grism, the aggregate deep field data would yield spectroscopic Lyman break confirmations for about 3000 galaxies down to AB = 27 over 6 deg$^2$. For reasonable assumptions, the most distant confirmations from such a spectroscopic survey might be as distant as z = 12 (conservatively) to 14 (optimistically).

The above Lyman break sample numbers come for free with the dark energy supernova survey. If considered as a GO program, a more modest survey area (1

deg$^2$) and depth (29 mag AB) would be achievable in 250 ksec of telescope time, corresponding to a medium-sized HST GO program, and would yield a sample of order 2000 Lyman break galaxies at z ~ 8.

5.6.5 Gamma Ray Burst Afterglows

Finally, NEW WFIRST could address reionization science through Target-of-Opportunity follow-up of high redshift gamma ray burst afterglows. Spectra of GRB afterglows prior to reionization will show the damping wing of Ly$\alpha$ absorption from the surrounding neutral IGM (e.g., Totani et al. 2006; Mesinger and Furlanetto 2008). Sensitive IR spectroscopy at R ~ 600, unimpeded by Earth's atmosphere, should detect this signature. The main issues to consider will be (a) response time, (b) field of regard, and (c) availability of events.

5.7 Summary: How does NEW WFIRST stack up compared to SDT DRM1?

WFIRST as envisioned in *NWNH* was anchored by two core programs. However, the *NWNH* stressed that the wealth of data for general astrophysics programs from the dark energy surveys and GO-led surveys, and from GO targeted observations, were the ***most important components*** of the WFIRST scientific program. The above discussion has made clear that, at this early stage of analysis, NEW WFIRST will perform the dark energy surveys and the microlensing planet search planned for SDT DRM1 at an equivalent and in most cases superior level. Particularly in the case of the dark energy program, "known-unknown" issues, most notably the stability and degree-of-knowledge of the PSF, are crucial to the success of the weak-lensing program, and these matters will require considerable development and understanding of the NEW WFIRST system. This, and the photometric stability issue discussed above for the supernova program, are the only potential barriers that would challenge the ability of NEW WFIRST to meet or exceed the performance of the SDT DRM1 in the two core programs.

We have many few examples of the kinds of GO science, including surveys and pointed observations, that demonstrate the breadth of exciting, science that either the SDT DRM1 or NEW WFIRST could achieve, greatly beyond the capability of any existing or planned facility. These examples are but a taste of the contribution NEW WFIRST could make, over and above the DRM1 design. This is because the increased aperture and sharper PSF take this version of WFIRST into the Hubble regime, that is, the power of a 2.4m diffraction-limited space telescope. HST is arguably the most successful science tool in history. When one considers how much imaging alone contributed to the wealth of Hubble science, and how a factor of 100 increase in field size — such as NRO realization of WFIRST provides — can open the door to programs even Hubble couldn't accomplish, it is clear that NEW WFIRST has the potential to be a facility that can be placed in the same class as the Hubble Space Telescope, something that would be unprecedented for an investment of this size.

## 6.0 Value Added: Direct-imaging Exoplanet Science with NEW WFIRST

6.1 Overview

In addition to dark energy, *NWNH* placed a high priority on Exoplanet discovery and characterization. The growth in interest and the rate of discoveries in the exoplanet field has been explosive over the last decade. Although gravitational microlensing with the NRO telescope offers the potential for a significant statistical characterization of exoplanet populations, that technique measures only planet masses and orbital parameters, not their composition. However, unlike the baseline WFIRST, the NEW WFIRST could also be equipped to **directly** detect nearby exoplanets, resolving them from their parent star, allowing for spectroscopic characterization that can provide unique insights into planetary properties, history, and formation. The power of this technique has recently been shown by extensive studies of the young giant planets in the HR8799 system (see Figure 20).

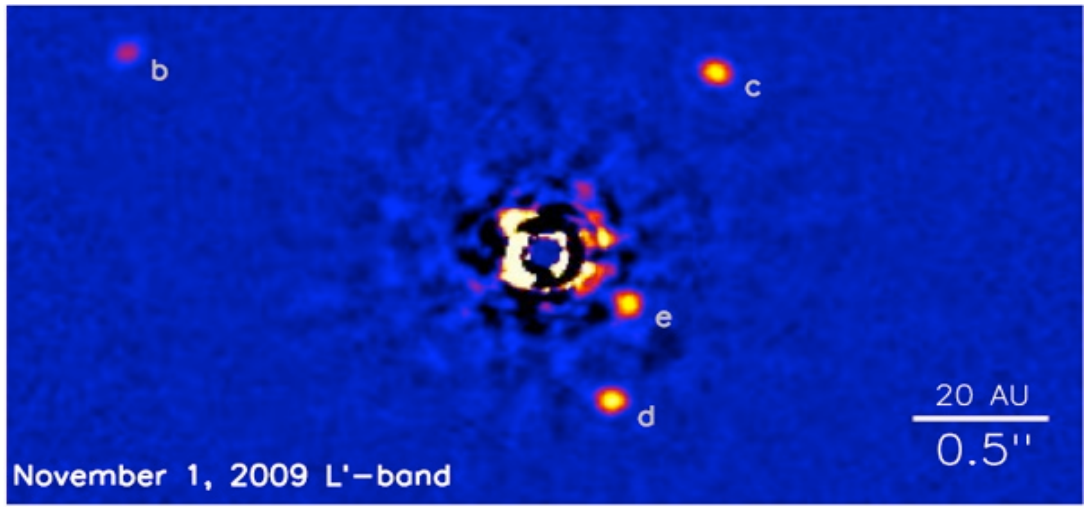

*Figure 20: Near-infrared image of four giant planets orbiting the young star HR8799, taken with the Keck adaptive optics system (Marois et al. 2010).*

Although a NEW WFIRST coronagraph would be less capable than one on an optimized mission with a 2.4m aperture, advances in technology would give it orders of magnitude more sensitivity than any instruments on HST, and it would be an important complement to ground-based facilities. This would have significant scientific value, particularly if the instrument meets at least one of the following requirements — sensitivity to planet-to-star contrast ratios of $10^{-9}$, a near-infrared channel for studying self-luminous planets, and/or a moderate resolution combined coronagraph/spectrometer. It would also serve to demonstrate the technology for future missions that would one day detect and characterize Earthlike planets.

6.2 High-Contrast Imaging Technology

While the NRO telescopes were originally designed for wide-field applications, and the WFIRST science is enabled by that capability, and NRO telescope also make an excellent platform for including a narrow-field camera (and potentially a spectrometer). The telescope's reported performance of 60 nm RMS wavefront error suggests that an instrument could be diffraction-limited for wavelengths longer than 840 nm (using the rule of thumb that diffraction-limited imaging requires $\lambda/14$ RMS wavefront error). The performance at the on-axis field point is likely to be better than 60 nm, allowing for diffraction-limited imaging at shorter wavelengths with a narrow-field on-axis instrument. In addition, correction of the static wavefront error at the on-axis field point could be carried out using one or more deformable mirrors if necessary, further improving the performance of an on-axis imager.

Furthermore, any wide-field instruments designed for the system are likely to be off-axis in field angle (such as the designs presented in Section 4). These designs leave room for a pickoff mirror at the on-axis field point, in the image plane of the two-mirror telescope. Three reflections (the primary, secondary, and the pickoff) would be needed to direct the beam into a narrow-field imager. The volume available for instruments is large enough to accommodate multiple instruments. This opens up the possibility of including a coronagraph with wavefront control, or under certain restrictions, flying an accompanying occulter, to perform high-contrast imaging. While cost and volume may dictate only a single narrow-band channel, significant science results and technological development could be achieved. In this section we summarize the state of technology and the feasibility of using the telescope for high-contrast imaging.

6.2.1 Coronagraph Technology

Great advances have been made over the last decade in coronagraph design and wavefront control for high-contrast imaging. There are numerous designs proposed each with different advantages and disadvantages and performance spaces. There are also at least three active laboratories studying wavefront control for high-contrast space-based applications (the high-contrast imaging testbed (HCIT) at the Jet Propulsion Laboratory (JPL), the high-contrast imaging laboratory (HCIL) at Princeton, and the NASA Ames Testbed). Contrasts as high as a few times $10^{-10}$ have been achieved at JPL in 10% broadband light. Numerous concepts were submitted to the last decadal survey using telescopes of similar apertures to the NRO telescope that utilized coronagraphs to image and characterize earth-sized exoplanets around the nearest stars. NEW WFIRST provides a unique opportunity to advance the technology for coronagraphic imaging and achieve unique science by including a narrow field coronagraph. Not only could interesting science be accomplished, but testing components and algorithms is a crucial step toward a future large flagship exoplanet mission.

The NRO telescope does present some unique challenges to implementing a coronagraph. In particular, because of the central obscuration and the six supporting struts, it possesses a complicated pupil geometry that rules out many of the standard coronagraphs. Fortunately, much progress has been made in the last few years on developing coronagraphs for on-axis telescopes. These include new shaped pupil designs that incorporate central obstructions and spiders, new Lyot approaches with phase varying pupil stops for masking out the secondary, the new PIAA-CMC that uses an achromatic focal plane mask, and the Visible Nulling Coronagraph (VNC) that can mask out the central obstruction and segments by careful control of the pupil shear.

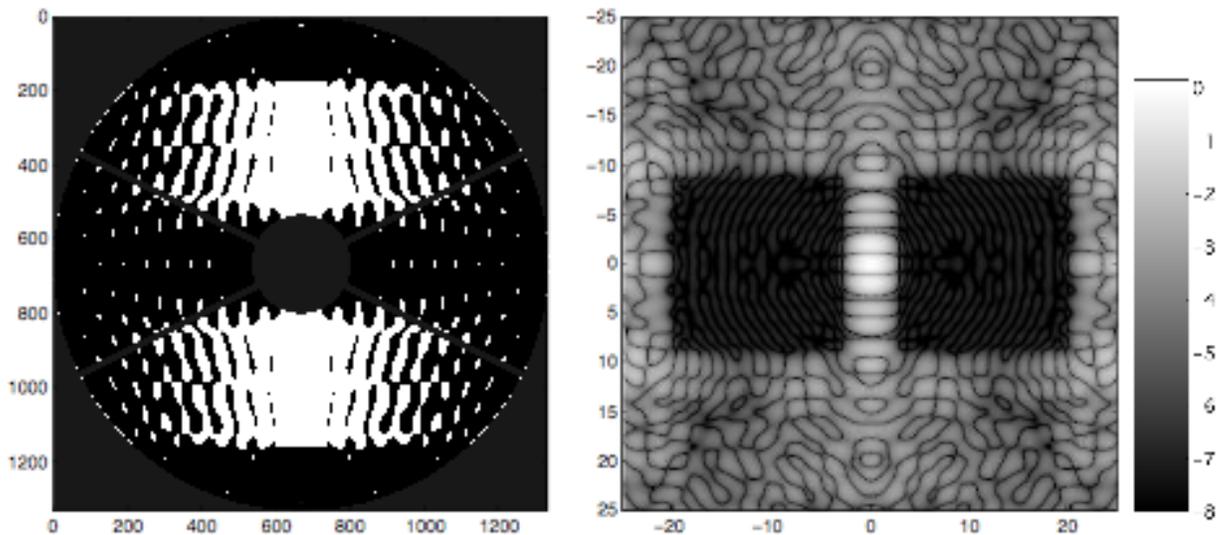

*Figure 21: A shaped pupil (left) designed for a pupil similar to the NRO telescope with central obstruction and (right) the corresponding high-contrast point spread function. Dark holes are $10^{-8}$ contrast with an inner working angle of 3 $\lambda/D$ (0.2 arcsec at 800 nm) and an outer working angle of 21 $\lambda/D$.*

For illustration, Figure 21 shows a shaped pupil designed for a pupil similar to the NRO telescope pupil that generates two dark holes with an inner working angle of 3 $\lambda/D$, an outer working angle of 21 $\lambda/D$, 23% throughput, and a contrast of $10^{-8}$. Other designs are also possible with higher (or lower) contrast and smaller inner working angles that trade with the size of the discovery space and the system throughput. Similar masks to these have been manufactured and verified in the laboratory, demonstrating the maturity of the technology for on-axis telescopes. Figure 22 shows a dark hole formed using a shaped pupil mask at HCIT, demonstrating the maturity of wavefront control. The mask was used in conjunction with a wavefront control system to reduce scatter in the dark hole to a contrast of $2.4 \times 10^{-9}$ in a 10% optical bandwidth (Belikov et al. 2007).

It is also possible to modify the shaped pupil mask to directly account for measured phase errors in the telescope. For instance, Carolotti et al. (2012) show how a mask was designed for JWST that created a single-sided dark hole in the presence of the nominal phase error without any wavefront correction. This sort of unified design approach ensures that it is possible to combine wavefront control with existing deformable mirror technology with the wavefront quality of the NRO telescope "as is."

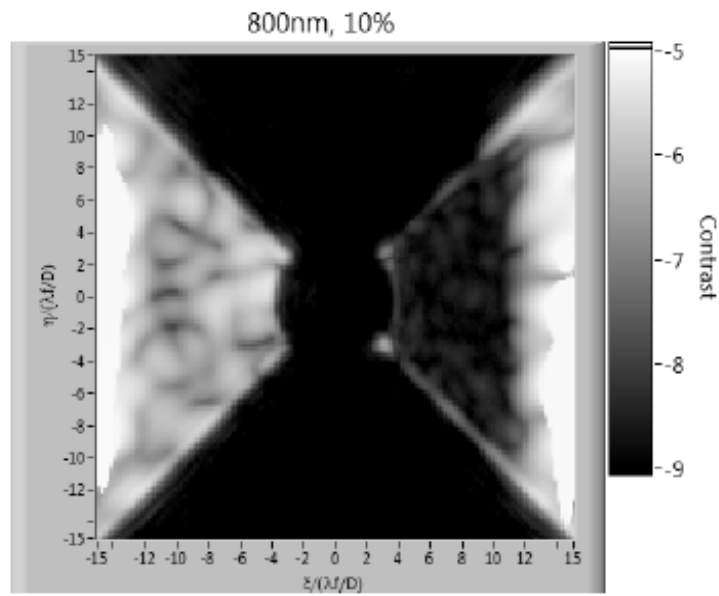

*Figure 22: The dark trapezoidal region on the right side of the image is a dark hole formed in a laboratory testbed.*

The largest unknown affecting the achievable contrast is the expected stability of the telescope. Because of the long integration times needed to perform correction, the telescope needs to be stable during the several hours that might be necessary to take an image. For very high-contrast ($10^{-9}$ or better) this can place extremely taxing requirements on telescope stability. Fortunately, the NRO telescope is equipped for active thermal control and active control of the secondary which may make contrasts up to $10^{-8}$ or better possible, though $10^{-10}$ is likely impossible. This needs to be studied carefully. However, it is important to note that placing a coronagraph on NEW WFIRST provides an unparalleled opportunity to study coronagraph performance in space and gain direct knowledge regarding the interaction of the telescope short- and long-term stability with achievable contrast.

6.2.2 Occulters

An occulter mission employs a large free-flying starshade to block the light of the parent star from the telescope. This is an attractive alternative to coronagraphs for high-contrast science in space. Their biggest advantage is that they provide the high-contrast starlight rejection so that a conventional telescope — without especially challenging stability or optical quality requirements — can image exoplanets. Further, occulters decouple the angular resolution from telescope size: even with a 2.4m telescope, very small inner working angles can be achieved and a sizeable number of Earthlike planets could be discovered. The telescope instrumentation consists only of a simple camera (and possibly a spectrograph). Since no starlight enters the telescope, no sophisticated wavefront control system is needed. Figure 23 shows the precision petal that was recently completed at JPL. This petal is consistent with star to planet contrast of $10^{-10}$ (Kasdin et al. 2012).

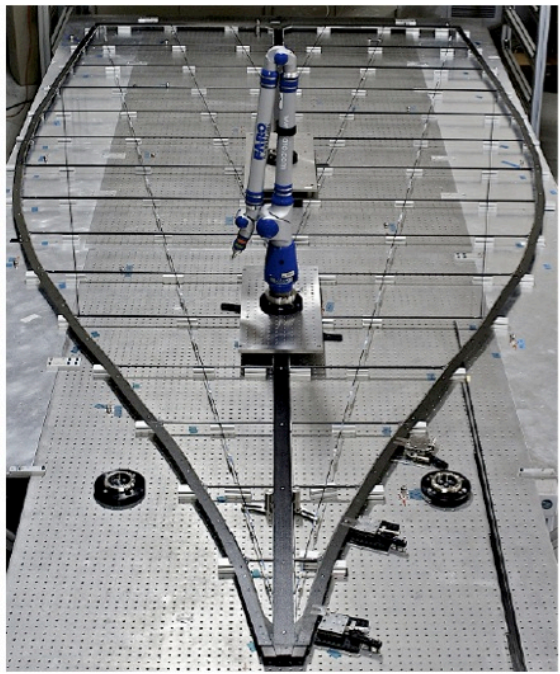

*Figure 23: Full scale precision starshade. The overall shape is consistent with $10^{-10}$ imaging. The petal includes the precision optical edge but is missing the non-structural opaque blanket.*

Of course, these advantages come at a cost. A separate spacecraft for the starshade is needed that would likely cost many hundreds of millions of dollars. More importantly, it is extremely difficult, if not impossible, to architect an occulter mission in Earth orbit. Avoiding interference from the Sun, Earth and Moon make finding geometrically favorable viewing angles challenging and the large gravity gradient makes 'stationkeeping' difficult and costly with existing thrusters. One could consider a NEW WFIRST mission as an opportunity to do technology verification of the deployment and stationkeeping of an occulter, but the more challenging environment in Earth orbit may make this less representative of a real mission and the cost may be too high for a mission with little or no science. For a reasonable science return, an occulter mission should be in deep space, either Earth-trailing or an L2 orbit, possibly too costly for the first NRO telescope. Of course, if the telescope is flown at L2 for other scientific reasons, then a follow-on mission in which an occulter flies into formation with the telescope should be seriously considered, as it would be capable of achieving the very high contrast and small inner working angle needed to image exo-earths.

6.3 Potential Science with an NRO Coronagraph

6.3.1 Extrasolar planets

We estimate that a NEW WFIRST coronagraph could conservatively achieve a detectable planet:star contrast of $10^{-8}$ with an optimistic bound of $10^{-9}$ at an inner working angle of ~0.2 arcsec. For comparison, the contrast between Earth and the sun is ~$10^{-10}$; terrestrial planets would be beyond the reach of any plausible NRO telescope capability. However, a NEW WFIRST coronagraph would have significant capability to detect and characterize giant planets — photometrically or spectroscopically. These techniques will be most powerful for planets 1-10 AU from their parent star, a region of phase space almost inaccessible to transits. Characterization is especially valuable. The massive statistical harvest from Kepler and Doppler techniques provides a robust census of planets within 1 AU of their parent star, but very little information about their nature — for example, are rocky planets common, or are most planets ice and hydrogen? Different models of planet formation make different predictions about the nature of other solar systems. Studying the distribution of giant planets across the "snow line" at 3 AU, and their composition, will provide a unique window into the planet formation process.

Ground-based, high-contrast facilities such as the Gemini Planet Imager (GPI, Macintosh et al. 2012) and its ESO counterpart Spectro-Polarimetric High-contrast Exoplanet Research instrument (SPHERE. Beuzit et al. 2010) will of course also help with this puzzle, but NEW WFIRST would be highly competitive, particularly if it can reach a contrast of $10^{-9}$ — GPI and SPHERE achieve contrasts of only $10^{-7}$ to $10^{-6}$ on typical targets. To detect giant planets, these instruments observe at near-IR wavelengths, where young (< 500 Myr) giant planets remain detectable from thermal radiation of their initial energy of formation. GPI and SPHERE have very limited sensitivity at shorter wavelengths and are essentially blind to mature solar systems such as our own. A coronagraph on NEW WFIRST could see exoplanets of any age, allowing us to study their evolution over periods similar to the age of our solar system. For systems detected by both NRO and GPI, the availability of photometry at visible wavelengths and at near-infrared wavelengths inaccessible from the ground (e.g., water absorption lines) will help strongly constrain the planet's properties, such as radius and atmospheric composition, placing it on evolutionary tracks that in turn allow determination of its formation process (e.g., Fortney et al. 2008). Self-luminous planet detection such as by GPI also strongly biases the sample to higher mass; NEW WFIRST would be capable in some circumstances of directly imaging Neptune-mass planets.

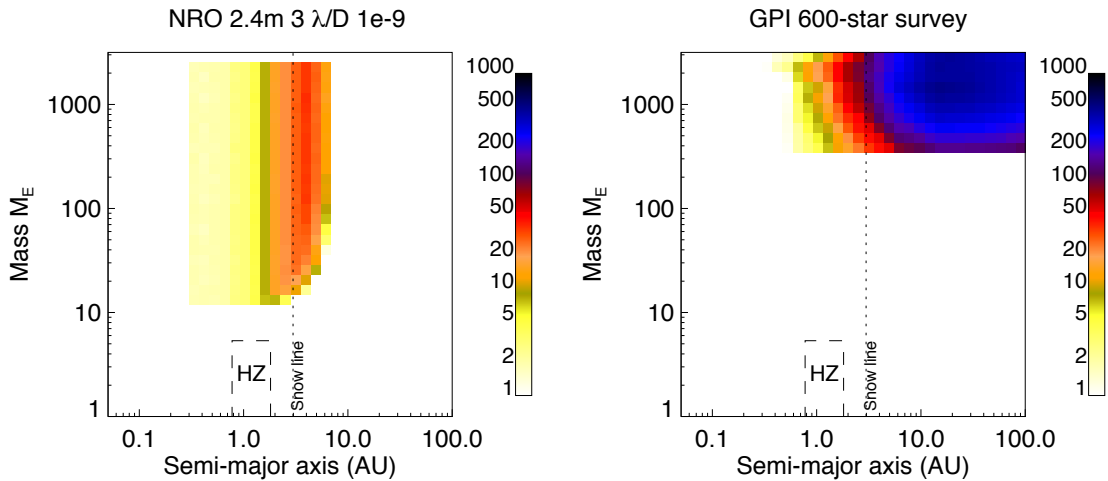

*Figure 24: Sensitivity comparison between a 2.4m space coronagraph and the Gemini Planet Imager for a notional survey of nearby stars with mass 0.5-1.5 solar masses. The quantity plotted is the ExoPTF "depth of search" (Lunine et al. 2008). This is the sum of the completeness of a survey for all targets — it can be thought of as the number of planets of a given mass and semi-major axis that would be discovered if all targets had such a companion. GPI has a much larger potential target sample, but since it detects self-luminous planets, whose brightness declines steeply with mass and age and it is only sensitive to high-mass planets.*

Figure 24 shows the science reach of an NRO coronagraph capable of achieving $10^{-9}$ contrast – roughly sufficient to detect a Jupiter analog. Poorer performance (such as a contrast of only $10^{-8}$) would significantly restrict the ability to detect mature planets. However, this could be mitigated by including a near-infrared coronagraph channel that could detect younger self-luminous planets. Finally, since spectroscopic characterization of even a small number of planets is very valuable, a coronagraphic moderate-resolution spectroscopic capability would provide unique insights into planetary composition and formation history.

JWST will also incorporate coronagraphic capabilities, but the design of these, and the uncorrected aberrations on JWST's mirror segments, limits their performance. JWST, however, can observe at longer IR wavelengths where thermal emission from giant planets is strong. High-performance visible-light coronagraphy with the NEW WFIRST telescope, when combined with near-IR on the ground and mid-IR on JWST, would be an extremely powerful combination for completely characterizing nearby giant planets.

6.3.2 Circumstellar dust

Planets are of course not the only component of other solar systems; seen from outside, the brightest thing in our solar system (after the Sun) is not Jupiter but the disk of zodiacal dust, remnants of comets and asteroids in the inner solar system.

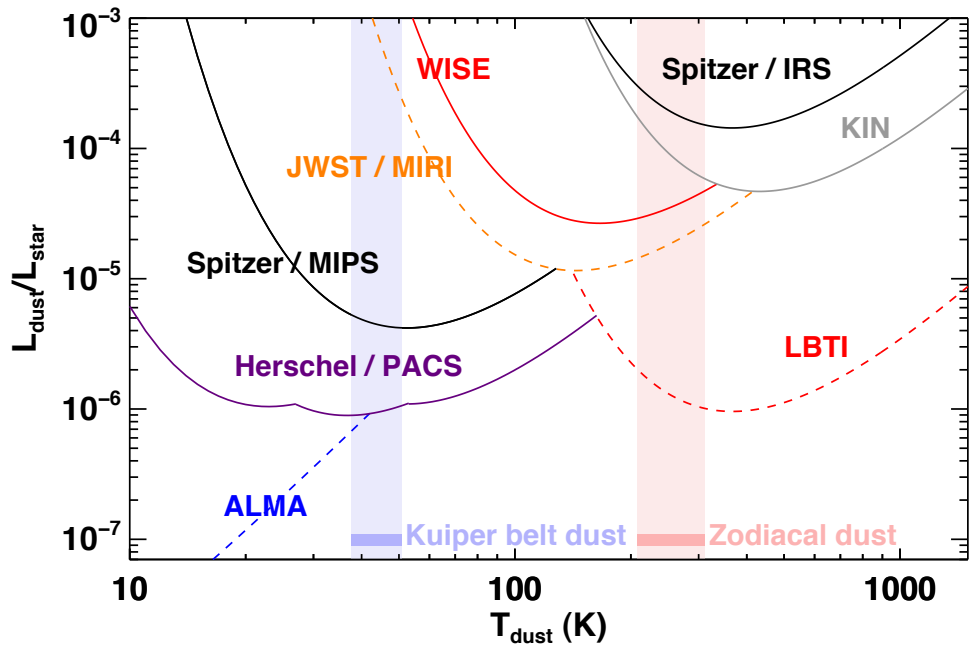

*Figure 25: Taken from Fig. 3 of Roberge et al. (2012). Sensitivity limits for detection of debris dust around nearby Sun-like stars, for various recent (Spitzer, WISE), current (KIN, Herschel) and near-term facilities (LBTI, JWST, ALMA). The curves show 3σ detection limits in terms of the fractional dust luminosity ($L_{dust}/L_{star}$) versus its temperature. Recent and current facilities are plotted with solid lines, near-term ones with dashed lines. The temperature ranges for dust in two zones around the Sun are shown with vertical bars, calculated assuming blackbody grains. The habitable zone (0.8 AU – 1.8 AU) is shown in pink, the Kuiper belt (30 AU – 55 AU) is shown in light blue. The modeled $L_{dust}/L_{star}$ values for the Solar System's Kuiper belt dust are marked with horizontal light blue and pink bars. An optimized coronagraph on the NRO telescope could be sensitive to $Ldust/L* \sim 10^{-6}$ at $T_{dust} \sim 200$ K.*

This zodiacal dust is both a tracer of undetectable planets and a possible barrier to future detection of other Earths. See Roberge et al. (2012) for an extensive discussion of this "exozodi." Planets cause structures (gaps, resonances, leading and trailing concentrations) in the dust that can be directly resolved by a telescope, allowing the possibility of inferring the presence of planets with masses as small as Earth. If the dust disk is massive, however, the starlight it scatters may obscure otherwise-detectable planets. Simulations show that a disk 5-10 times the brightness of our own would significantly decrease the ability of a future flagship mission to detect and characterize habitable planets. This is a major source of planning uncertainty for such missions – the actual amount of zodiacal dust around nearby Sun-like stars is almost completely unknown: if our system is atypical, and most systems are dustier, only a very large (8-m class) mission could be capable to detecting other Earths.

The best current limits are at the level of ~100 times the luminosity of our solar system (see Figure 25); the LBTI may improve this limit by a factor of ten but that is unproven and will not be able to map structure in the disks. An optimized coronagraph on NEW WFIRST could significantly advance this field. Precision PSF subtraction or polarimetry could allow the NRO telescope to be sensitive to zodiacal dust at the level of ~5 times the solar zodiacal cloud at separations of ~2AU — a factor of twenty better than current limits and significantly better than any other near-term facility.

6.4 Other Exoplanet Observation Capabilities -- Astrometry

There may be other applications of a NEW WFIRST telescope to high priority *NWNH* goals for exoplanet discovery and characterization. One is to search for planetary systems around nearby stars through sub-micro-arcsecond ($\mu$as) astrometry. Such precision was the basis of the Space Interferometry Mission (SIMLite), which was ranked scientifically as high priority in the *NWNH* Decadal Survey, but judged to be too expensive in the proposed implementation. The basic concept is to equip an NRO telescope with a diffractive pupil to calibrate non-systematic distortions: this would enable astrometry at the sub-micro-arcsecond level. As with SIMLite, an NRO telescope with this feature would be capable of detecting and characterizing earth-like planet detection and estimating their masses around nearby stars (~10pc). The ability to track telescope distortions might also enhance other types of observation, such as by improving the accuracy of galaxy weak-lensing measurements.

A specific implementation would be to imprint a 2-D grid of regularly spaced small dark spots on the coating of the primary mirror that will generate faint stellar diffraction spikes (see Figure 26). This concept is simple and robust, with no moving parts or complexity, requiring only a different coating. Once the coating is applied on the ground, performance can be fully tested and characterized prior to launch, minimizing the risk and cost. Furthermore, the implementation is flexible with regard to the pattern of diffraction spikes; spacing between the spikes and total light in the spikes can be realized by choice of the diffractive pupil geometry. As a result, the technique can be designed to avoid compromising other observations. For example, in a baseline design for a 1.4m telescope, the additional light introduced by the dots on the primary mirror is less than 1% of the zodiacal background level over 95% of the field (Guyon et al. 2012a). A trade study is needed to consider different science cases in order to find the best compromise between astrometric accuracy and diffracted light.

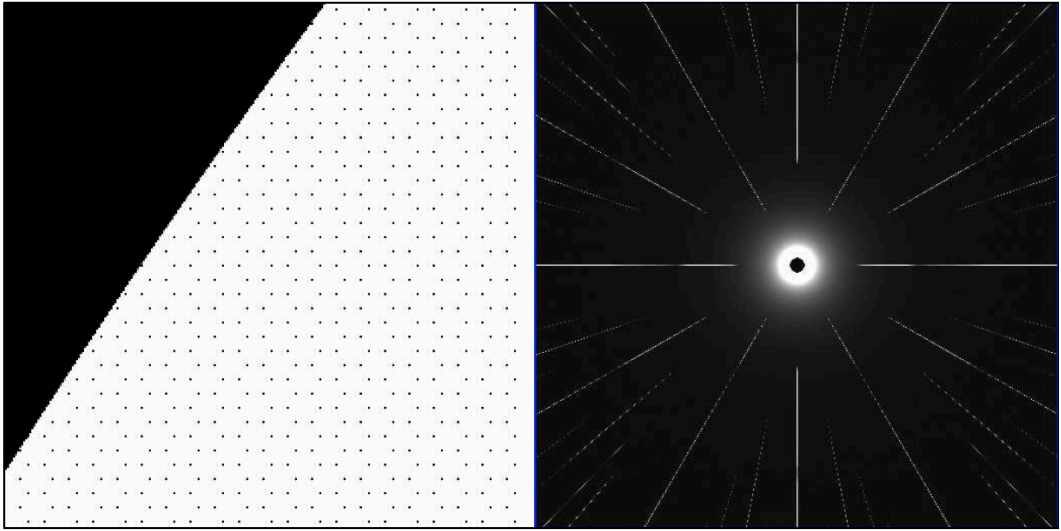

*Figure 26. Dots on the telescope primary mirror (left) and corresponding on-axis PSF in the wide field astrometric camera (right). The primary mirror area shown is 3% of the pupil diameter across and is located at the edge of the pupil. The spacing between the diffraction spikes, their extent in the focal plane, and their overall luminosity can be chosen by an appropriate design of the dot pattern on the pupil. The surface brightness of the spikes shown is a factor of $10^9$ (22.5 magnitudes) dimmer than the central star.*

In this approach, the astrometric motion of the host star is measured by comparing the position of its diffraction spikes to the background field stars (Guyon et al. 2012b). This concept can work at visible and IR wavelengths, does not require a new camera or instrument, and can be implemented with the telescope "as is," since it only requires recoating the primary mirror. Simulations show that the NRO telescope, with an aperture of 2.4m, a 0.3 $deg^2$ FOV, and a hexagonal dot pattern covering 1% of the pupil, could in principle achieve 0.1µas (1σ) astrometric accuracy in a single visit, assuming a 2-day integration observing at the galactic pole. Guyon et al. (2012a) estimate that such data would be capable of measuring the masses of Earth-like planets to a precision of 10% around nearby stars. This result applies for visible or NIR imaging without relying on the accurate pointing, external metrology technique, or high stability hardware required with previous high precision astrometry concepts (Guyon et al. 2012b).

Compatibility of a diffractive pupil telescope and a high-performance coronagraph has been tested, showing that the diffracted light does not contaminate the inner working angle of the coronagraph (Bendek et al. 2012). These two capabilities together offer tremendous synergy to solve for orbital parameters and masses of planetary systems.

# 7. Programmatic Issues

7.0 Overview

Designs of a WFIRST mission using an NRO 2.4m telescope are based on previous work done for other WFIRST configurations studied by the WFIRST Project office and community Science Definition Team (SDT Final report: http://arXiv.org/pdf/1208.4012.pdf ) They involved telescopes of aperture size 1.1 - 1.3m and focal planes of tens of HgCdTe detectors. The concepts were found to be technically feasible using hardware that is current available with a level of high technical readiness. Costs have been estimated in the $1.0 - 1.6B range.

7.1 Needed elements for using the NRO telescope

The NEW WFIRST designs can leverage off this previous work, but several missing pieces need to be filled in.

1) The first need is to obtain the detailed optical, mechanical and thermal design documents for the NRO telescopes. Some information has been released, allowing the above studies to be done. However, the detailed design of the mission will require full information to be available.

2) The second need is to form a project and community team with a short-term objective of producing a report on the scientific and technical aspects of the mission.

3) The third need is to have adequate investment in 4k x 4k H4RG detector development. Compared to previous designs with smaller mirrors, the finer pixel scale enabled by the larger 2.4m telescope brings a need for more pixels to maintain a wide-field capability. For example, a large 0.5 deg$^2$ field of view sampled at 0.11 arcsec/pixel requires 32 H4RG detectors. Development is underway on H4RGs, and some detectors are already available, but more work is needed to improve yields and performance.

## 7.2 Cost comparison with comparable and incomparable missions

Missions that have technologies ready to go and funded for rapid development can be done at a relatively low cost. An excellent example is Kepler —a 1.4m telescope with a focal plane of 32 CCDs built with a full cost in 2009 of approximately $0.6B. Explorer missions such as WMAP, Swift and WISE provide examples of space science missions executed for costs in the range of $100M – $200M. HST, Chandra and JWST with their multi-billion-dollar costs are not good models as they were technologically challenging and built with stringent requirements of long life. The technical envelopes were pushed to the maximum for optical precision and (in JWST's case) cryogenic operations. The use of an existing space qualified 2.4m optical telescope, used "as is" to the greatest extent possible, combined with using low risk technical and engineering approaches for the instrumentation and spacecraft, opens up new opportunities to make NEW WFIRST an cost-effective option.

## 7.3 Streamlining integration and testing

An advantage of the NRO telescopes is that they were designed to specifications that are a good match to WFIRST. They have active control of the mirror orientations and metrology that will provide the needed stability of the point spread function for the WFIRST weak lensing and microlensing science.

Another advantage of these telescopes is that they have already undergone extensive testing at the component and subsystem level. A full thermo-vacuum test of the full system was planned, but not completed. This test could be done in the early phases of a NEW WFIRST mission to retire risk early. A well-phased and streamlined integration and test plan for NEW WFIRST, combined with a highly modular approach to the spacecraft systems and instrument, has the potential to greatly simply the overall integration-and-testing process and hence reduce mission cost.

## 7.4 Launch vehicles and orbits

There are several orbit options for NEW WFIRST. These include (1) L2, (2) geosynchronous and (3) elliptical high orbits. Each has particular advantages. L2 is highly stable and constantly far from earth. Geosynchronous has launch-sharing options and a good orbit for possible robotic servicing. It also provides continuous, inexpensive data down-link. Elliptical orbits can be synchronized for continuous down-link and have lower trapped-electron levels than geosynchronous.

There are two currently-operating launch vehicles capable of delivering NEW WFIRST to any of the above orbits, the Delta-IV and Atlas V. Both are well matched to the NEW WFIRST-NRO requirements and both have excellent reliability records. Other attractive options may become available in time for NEW WFIRST. The NASA SLS launcher currently in development has more than enough lift capability for NEW WFIRST. A ride on an early launch might be possible, or a shared launch to synchronous orbit. A suitable Falcon rocket could be available by 2020 with low cost.

7.5 Operations

Restricting the number of instruments, and especially the number of operating modes (a feature of spending the majority of observing time on survey science as compared to GO pointed observations) could significantly lower the operations cost for the NEW WFIRST mission. In addition, use of a geosynchronous or elliptical high orbit, and a single ground-station, could further simplify the ground-operations. Further savings could be made in the science and data operations by leveraging off the infrastructure of existing science operations centers.

7.6 A new approach enabled by NEW WFIRST

The cost of astrophysics missions has been rising in the last two decades. Certainly one reason for this is greater scientific aspiration — technical complexity and sophistication is a natural result of tackling ever-more-challenging scientific questions. In addition, because of their increased scale and complexity, large astrophysics missions programs have become progressively more risk-averse: achieving a very high level of confidence drives cost and schedule. There is concern within the science community that increasing costs for the most challenging missions combined with a decreasing budget (in real terms) for space astrophysics could have serious consequences for our ability to realize some of the high-priority-science programs that were described in *NWNH*.

NEW WFIRST could be a unique opportunity to break free of this dilemma. NEW WFIRST would be a large and powerful space telescope, but it would require essentially no technical innovation or development. One of the longest-lead-time components — the telescope itself — arrives in an advanced state of near-completion, having successfully passed testing at the component and subsystem level, and basically ready for a system thermo-vacuum test. Not only does this mean that a significant part of the design, development, and construction cost of NEW WFIRST has already been invested, but that the risk associated with this phase is nearly retired. Also, quire crucially, the designs for other components — mainly instruments and spacecraft — are tightly constrained by the pre-determined, well documented properties of the existing NRO telescope. The risk is further reduced for this particular project because the core instrument, a multi-detector near-IR camera, can leverage off the considerable experience gained building cameras for HST, Kepler and JWST. In addition, it is very likely the spacecraft can be based on previous, mature designs. Finally, the very existence of two NRO telescopes at a similar level of completion means that a carefully constructed program could generate most of a backup mission should the worst happen. These are all aspects of a potential NEW WFIRST that offer a lower risk profile than is usually attached to such a large mission.

We believe there is an opportunity with NEW WFIRST to change the existing science mission paradigm, by challenging the project team and science community to explore ways to keep the costs in the $1B range. This will require — at a minimum — investments to be made up front on IR detectors and for preliminary designs, and the scientific requirements to be defined early. Crucial steps are building on the capabilities of the existing hardware, and then having the discipline to hold to these requirements rigidly. Keeping NEW WFIRST at moderate cost is predicated on using the existing NRO hardware with no or few modifications, and having a rapid development schedule with a suitable funding profile.

## 8. Summary and Conclusions

The present decade is a time of unprecedented opportunity in astrophysics research, but also a time when funding for new space initiatives is the lowest in decades. The *NWNH* Decadal Survey proposed a mission, WFIRST, to be an effective response to this situation. WFIRST would achieve a genuinely new and powerful capability and support a wide range of science goals from stellar populations and galaxy evolution to exoplanets and cosmology. By keeping lifecycle cost well below $2B, WFIRST would achieve a "science-per-dollar" ratio among the best of previous space astrophysics missions. Even so, the increased cost for JWST and the reduced overall funding envelope for NASA Astrophysics since the release of the *NWNH* Decadal Survey has made it problematical to develop WFIRST for a launch in the early 2020s — when its contributions to dark energy and exoplanet research, and synergies with European missions such as Euclid, and mainline new facilities such as JWST, ALMA, LSST, and planned ELTs are vitally needed. The unanticipated availability of two Hubble-sized telescopes from the NRO has the potential to be a game changer at this critical point in time.

The present discussion is only a preliminary assessment of how well an NRO telescope could perform the WFIRST mission. At this level of investigation, we find some examples where the better PSF and larger collecting area does not offer a great advantage compared to the SDT DRM1 version of WFIRST, because in these cases deeper and sharper images may not improve the execution of the program. In most cases, however, we find that a NEW WFIRST would either do a better job, or do the same job more quickly, than DRM1. This is especially true for much of the GO program, where surveying large fractions of the sky is not the Figure-of-Merit for accomplishing the science program. While DRM1 may be well matched to the dark energy and exoplanet microlensing goals of WFIRST as proposed in *NWNH*, the possibility of a near-IR Hubble with 100 times larger entendue enables a vastly larger range of astrophysics programs, as exemplified by the Hubble itself.

An increase in capability comes with a cost. Although the imaging capability is substantially better, it is more difficult to achieve the same FOV as with a smaller telescope like DRM1, and the number of detector pixels required to cover that field — while taking advantage of the better image quality — increases significantly. The spacecraft that is required must handle a larger, more massive telescope with greater power requirement, and the launch vehicle needs to be substantially larger as well. The latter may lead to constraints on orbits and operations that will lead to further science tradeoffs.

There is healthy and justifiable skepticism that the use of a larger, more capable telescope for the WFIRST mission can be done without increasing a cost that already exceeds identifiable resources. However, the possibility of better execution of the WFIRST program, as outlined in the Decadal Survey and developed by the SDT, and perhaps adding capability for high-priority science, such as the coronagraph for the direct imaging of exoplanets discussed here, is a strong motivation to find an approach that can build NEW WFIRST within the $1.6B cost estimated for DRM1. Accomplishing this will require changes in approach to mission definition —for example, adapting our science to existing hardware, as we have discussed here. We believe there can be changes in the execution of the program, and in particular the management of risks, that will lower cost by departing from traditional practices for large astrophysics missions. Moreover, we believe the starting point of near-complete 2.4m telescopes with no further technical innovation required for other components is exactly the kind of program where such a paradigm shift can be made. In addition, although it is hard to evaluate the magnitude of the effect, the re-use of an existing 2.4m "Hubble-like" telescope for astronomy could put some extra wind behind the sails of the WFIRST mission, with respect to governmental --- and perhaps also public — support and interest.

However, there is no way to demonstrate the feasibility of using NEW WFIRST — an NRO telescope used to accomplish the *NWNH* goals for WFIRST, and more — without far more detailed studies than those described here. Now is an ideal time to engage in that process, as vigorously as possible.

# References

Albrecht, A., et al. 2006, arXiv:astro-ph/0609591

Assef, R.J., et al. 2012, ApJ, submitted (arXiv:1209.6055)

Basilakos, S., Plionis, M., and Lima, J. A. S. 2006 arXiv:1006:3418  Bentz, M., Hall, P.B. &

Osmer, P.S. 2004, AJ, 128, 561

Belokurov, V. et al. 2007, ApJ, 654, 897

Bendek, E., Belikov, R. Guyon, O, and Pluzhnik, E.  2012, Submitted PASP, 2012.

Bennett, D. P., Anderson, J., & Gaudi, B. S. 2007, ApJ, 660, 781

Beuzit, J. L., Feldt, M., Mouillet, D., et al. 2010, In "The Spirit of Lyot 2010: Direct Detection of Exoplanets and Circumstellar Disks"

Bolton, J. S., et al. 2011, MNRAS 416, L70

Bouwens, R. J., et al., 2011a, Nature, 469, 504.

Bouwens, R., et al. 2011b, ApJ 737, 90

Bradley, L. D., Bouwens, R. J., Ford, H. C. et al., 2008, ApJ, 678, 647.

Bradley, L. D., Trenti, M., Oesch, P. A., et al. 2012, arXiv:1204.3641 (submitted to ApJ).

Brodwin, M., et al. 2010 ApJ, 721, 90

Bruzual, G. and Charlot S.  2003, MNRAS, 344, 1000

Cassan, A. et al.  2012, Nature, 481, 167

Carolotti, et al. 2012, Proc. SPIE 8442 (193)

Coe, D., Zitrin, A., Carrasco, M., et al. 2012, ApJ, submitted

Coil, A. L., et al. 2007, ApJ, 654, 115

Dalcanton, J., et al. 2012a,  ApJS, 198, 6

Dalcanton, J. et al. 2012b, ApJS, 200, 18

Davidge, T. J., Olsen, K. A. G., Blum, R., Stephens, A. W., and Rigaut, F. 2005, AJ, 129, 201

Dotter, A., Chaboyer, B., Jevremovic, D., Baron, E., Ferguson, J. W., Sarajedini, A., and Anderson, J. 2007, AJ, 134, 376

Fan, X., et al. 2001, AJ, 122, 2833

Fortney, J. J., Marley, M. S., Saumon, D., and Lodders, K. 2008, ApJ, 683, 1104

Fosbury, R., et al. 2003, ApJ, 586, 797

Furlanetto, S. R., Zaldarriaga, M., and Hernquist, L. 2004, ApJ 613, 1

Furlanetto, S. R., et al 2006, MNRAS 365, 1012

Glikman, E., et al. 2007, ApJ, 663, 73

Glikman, E., et al. 2011, ApJ, 728, 26

Griffith, R., & Stern, D. 2010, AJ, 140, 533

Green, J., et al. 2012, arXiv:1208.4012

Gobat, R., et al. 2010, A&A, in press (arXiv1011.1837)

Grossi, M., and Springel, V. 2009, MNRAS, 394, 1559

Guhathakurta, P. et al. 2005, arXiv:astro-ph/0502366

Gunn, J. E. & Peterson, B.A 1965, ApJ, 142, 1633

Guyon, O., et al. 2012a, ApJ, in press

Guyon, O., et al. 2012, ApJS, 200, 11

Han, C. & Gould, A. 2003, ApJ, 592, 172

Harris, H. C., et al. 2006, AJ, 131, 571

Herwig, F. 2005, ARAA, 43, 435

Hickox, R., et al. 2009, ApJ, 696, 891

Holz, D. E., & Perlmutter, S. 2012, ApJ, 755, 36

Hopkins, P. F., et al. 2005, ApJ, 630, 705

Ibata, I., et al. 2007, ApJ, 671, 1591

Iye, M., Ota, K., Kashikawa, N., et al. 2006, Nature, 443, 186

Jee, M. J., et al. 2009, ApJ, 704, 672

Kalirai, J., et al. 2012, AJ, 143, 11

Karachentsev, I., et al. 2004, AJ, 127, 2031

Kasdin, N. J., et al. 2012, Proc. SPIE 8442

Laureijs, R., et al. 2011, *Euclid Definition Study Report* (the "Red Book"), ESA/SRE(2011)12

Lunine, J. I., et al. 2008, arxiv:0808.2754

Macintosh, B., et al. 2012, Proc. SPIE, in press

Malhotra, S., and Rhoads, J. E. 2004, ApJ, 617, L5

Malhotra, S., and Rhoads, J. E. 2006, ApJ, 647, L95

Maraston, C., et al. 2006, ApJ, 652, 85

Marois, C., Zuckerman, B., Konopacky, Q.M., Macintosh, B. A., & Barman, T. 2010, Nature, 468, 1080

Masters, D., et al. 2012, ApJ, 755, 169

Mortlock, D.J., et al. 2011, Nature, 474, 616

Myers, A.D., et al. 2007, ApJ, 658, 85

McQuinn et al. 2007, MNRAS, 381, 75

McBride, J., Graham, J. R., Macintosh, B., Beckwith, S. V. W., Marois, C., Poyneer, L. A., and Wiktorowicz, S. J. 2011, arXiv:1103.6085

McConnachie, A., et al. 2009, Nature, 461, 66

Mesinger, A., & Furlanetto, S. 2008, MNRAS 385, 1348

Melbourne, J., et al. 2010, ApJ, 712, 469

Melbourne, J., et al. 2012, ApJ, 748, 47

Miley, G. K., et al. 2004, Nature, 427, 47

Miley, G. K., et al. 2006, ApJ, 650, L29

Miralda-Escude, J. 1998, ApJ, 501, 15.

Mortlock, D. J., et al. 2011, Nature, 474, 616

Mortonson, M. J., Hu, W., & Huterer, D. 2010, Phys.Rev.D, in press (arXiv:1011.0004)

Mouhcine, M., Ibata, R., & Rejkuba, M. 2011, MNRAS, 415, 993

Nikolaev, S., and Weinberg, M. D. 2000, ApJ, 542, 804

Olsen, K. A. G., Blum, R. D., Stephens, A. W., Davidge, T. J., Massey, P., Strom, S. E., and Rigaut, F. 2006, AJ, 132, 271

Ouchi, M., et al. 2010, ApJ 723, 869

Postman, M., Coe, D., Benitez, N., et al. 2012, ApJS, 199, 25

Sartoris, B., Borgani, S., Fedeli, C., Matarrese, S., Moscardini, L., Rosati, P., Weller, J. 2010, MNRAS, 407, 2339

Simpson, C. 2005, MNRAS, 360, 565

Stanford, S. A., et al. 2006, ApJ, 646, 13

Stern, D., et al 2005, ApJ 619, 12

Radburn-Smith, D., et al. 2011, ApJS, 195, 18

Roberge, A., et al. 2012, arXiv:1204.0025

Rosati, P., et al. 2009, A&A, 508, 583

Tilvi, V. S. et al. 2009, ApJ 704, 724

Tollerud, E. J., Bullock, J. S., Strigari, L. E., and Willman, B. 2008, ApJ, 688, 277

Totani, T. et al. 2006, PASJ 58, 485

Treister, E., Krolik, J. H. & Dullemond, C. 2008, ApJ, 679, 140

Urry, C. M. & Padovani, P. 1995, PASP, 107, 803

van der Wel, A., et al. 2006, ApJ, 652, 97

Weinberg, D. H., Mortonson, M. J., Eisenstein, D. J., Hirata, C., Riess, A. G., Rozo, E. 2012, arXiv:1201.2434

Willott, C. J., et al. 2010, AJ, 139, 906

Woitke, P. 2006a, A&A, 452, 537

Woitke, P. 2006b, A&A, 460, L9

Zheng, W., Postman, M., Zitrin A., et al. 2012, Nature, 489, 406

# Acknowledgements

The authors thank NASA for undertaking, on behalf of the astronomical community, the long and challenging process of making the NRO telescopes available for scientific research. We appreciate as well the participation of all the attendees of the Princeton Workshop, whose engagement in this exciting challenge of using an available 2.4m NRO space telescope for the WFIRST program added substantially to the content of this paper. Portions of this work performed under the auspices of the U.S. Department of Energy by Lawrence Livermore National Laboratory under Contract DE-AC52-07NA27344.